\documentclass[iop]{emulateapj}
\usepackage{apjfonts}

\usepackage{amsmath,amssymb}
\usepackage{xspace}
\usepackage{afterpage}

\newcommand{\sbzks}{sBzKs\xspace}
\newcommand{\pbzk}{pBzK\xspace}
\newcommand{\pbzks}{pBzKs\xspace}

\newcommand{\ha}{H$\alpha$}
\newcommand{\hasp}{H$\alpha$\xspace}

\newcommand{\hbsp}{H$\beta$\xspace}

\newcommand{\msun}{\ensuremath{M_{\odot}}}

\newcommand{\kms}{\ensuremath{\,\text{km}~\text{s}^{-1}}}
\newcommand{\cahk}{\ensuremath{\text{\ion{Ca}{2}}\;\text{{H+K}}}\xspace}

\newcommand{\hdf}{H$\delta_\text{F}$}

\newcommand{\re}{\ensuremath{r_\text{e}}}

\newcommand{\hst}{\textit{HST}}
\newcommand{\spitzer}{\textit{Spitzer}}

\newcommand{\ppxf}{pPXF\xspace}

\newcommand{\sersic}{S\'ersic\xspace}

\newcommand{\starlight}{STARLIGHT\xspace}

\slugcomment{Accepted for publication in The Astrophysical Journal}

\shorttitle{Near-IR Spectroscopy of $\lowercase{z} \gtrsim 1.4$ Passive Galaxies}
\shortauthors{Onodera et al.}

\begin{document}

\title{
  Deep Near-infrared spectroscopy of passively evolving galaxies at $\lowercase{z} \gtrsim 1.4$\,\altaffilmark{1}
}

\author{
  M.~Onodera\altaffilmark{2}, 
  A.~Renzini\altaffilmark{3},
  M.~Carollo\altaffilmark{2}, 
  M.~Cappellari\altaffilmark{4},
  C.~Mancini\altaffilmark{3},
  V.~Strazzullo\altaffilmark{5},
  E.~Daddi\altaffilmark{5},
  N.~Arimoto\altaffilmark{6,7,8},
  R.~Gobat\altaffilmark{5},
  Y.~Yamada\altaffilmark{6},
  H.~J.~McCracken\altaffilmark{9},
  O.~Ilbert\altaffilmark{10},
  P.~Capak\altaffilmark{11}, 
  A.~Cimatti\altaffilmark{12}, 
  M.~Giavalisco\altaffilmark{13}, 
  A.~M.~Koekemoer\altaffilmark{14},
  X.~Kong\altaffilmark{15}, 
  S.~Lilly\altaffilmark{2}, 
  K.~Motohara\altaffilmark{16}, 
  K.~Ohta\altaffilmark{17}, 
  D.~B.~Sanders\altaffilmark{18}, 
  N.~Scoville\altaffilmark{19}, 
  N.~Tamura\altaffilmark{8}, 
  and
  Y.~Taniguchi\altaffilmark{20}
}

\altaffiltext{1}{
  Based on data collected at the Subaru telescope, 
  which is operated by the National Astronomical Observatory of Japan. 
  (Proposal IDs: S09A-043 and S10A-058).
}

\affil{$^{2}$Institute for Astronomy, ETH Z\"urich, Wolfgang-Pauli-strasse 27, 8093 Z\"urich, Switzerland}
\affil{$^{3}$INAF-Osservatorio Astronomico di Padova, Vicolo dell'Osservatorio 5, I-35122, Padova, Italy}
\affil{$^{4}$Sub-Department of Astrophysics, University of Oxford, Denys Wilkinson Building, Keble Road, Oxford OX1 3RH, United Kingdom}
\affil{$^{5}$CEA, Laboratoire AIM-CNRS-Universit\'e Paris Diderot, Irfu/SAp, Orme des Merisiers, F-91191 Gif-sur-Yvette, France}
\affil{$^{6}$National Astronomical Observatory of Japan, 2-21-1 Osawa, Mitaka, Tokyo, Japan}
\affil{$^{7}$Graduate University for Advanced Studies, 2-21-1 Osawa, Mitaka, Tokyo, Japan}
\affil{$^{8}$Subaru Telescope, 650 North A'ohoku Place, Hilo, Hawaii 96720, USA}
\affil{$^{9}$Institut d'Astrophysique de Paris, UMR7095, Universit{\'e} Pierre et Marie Curie, 98 bis Boulevard Arago, 75014 Paris, France}
\affil{$^{10}$Laboratoire d'Astrophysique de Marseille, BP 8, Traverse du Siphon, 13376 Marseille Cedex 12, France}
\affil{$^{11}$Spitzer Science Center, California Institute of Technology 220-06, Pasadena, CA 91125, USA}
\affil{$^{12}$Dipartimento di Astronomia, Universit{\`a} di Bologna, Via Ranzani 1, 40127 Bologna, Italy}
\affil{$^{13}$Department of Astronomy, University of Massachusetts, Amherst, MA, USA}
\affil{$^{14}$Space Telescope Science Institute, 3700 San Martin Dr., Baltimore, MD, 21218, USA}
\affil{$^{15}$Center for Astrophysics, University of Science and Technology of China, Hefei, Anhui 230026, China}
\affil{$^{16}$Institute of Astronomy, University of Tokyo, 2-21-1 Osawa, Tokyo, Japan}
\affil{$^{17}$Department of Astronomy, Kyoto University, Kyoto 606-8502, Japan}
\affil{$^{18}$Institute for Astronomy, University of Hawaii, 2680 Woodlawn Drive, Honolulu, HI 96822, USA}
\affil{$^{19}$California Institute of Technology, MC 105-24, 1200 East California Boulevard, Pasadena, CA 91125, USA}
\affil{$^{20}$Research Center for Space and Cosmic Evolution, Ehime University, 2-5 Bunkyo-cho, Matsuyama 790-8577, Japan}

\email{monodera@phys.ethz.ch}
\begin{abstract}

  We present the results of new near-IR spectroscopic observations of
  passive galaxies at $z\gtrsim 1.4$ in a concentration of
  \textit{BzK}-selected galaxies in the COSMOS field.  The
  observations have been conducted with Subaru/MOIRCS, and have
  resulted in absorption lines and/or continuum detection for 18 out
  of 34 objects. This allows us to measure spectroscopic redshifts for
  a sample that is almost complete to
  $\mathit{K}_\text{AB}=21$. COSMOS photometric redshifts are found in
  fair agreement overall with the spectroscopic redshifts, with a
  standard deviation of $\sim 0.05$; however, $\sim 30\%$ of objects
  have photometric redshifts systematically underestimated by up to
  $\sim 25\%$.  We show that these systematic offsets in photometric
  redshifts can be removed by using these objects as a training set.  
  All galaxies fall in four distinct redshift spikes 
  at $z=1.43$, $1.53$, $1.67$ and $1.82$, with this
  latter one including 7 galaxies. SED fits to broad-band fluxes
  indicate stellar masses in the range of $\sim 4$--$40\times
  10^{10}\, M_\odot$ and that star formation was quenched $\sim 1$ Gyr
  before the cosmic epoch at which they are observed.  The spectra of
  several individual galaxies have allowed us to measure their \hdf{}
  indices and the strengths of the 4000 \AA{} break, which confirms
  their identification as passive galaxies, as does a composite
  spectrum resulting from the coaddition of 17 individual spectra.
  The effective radii of the galaxies have been measured on the COSMOS
  \hst/ACS $i_{\rm F814W}$-band image, confirming the coexistence at
  these redshifts of passive galaxies which are substantially more
  compact than their local counterparts with others that follow the
  local effective radius-stellar mass relation. For the galaxy with
  best S/N spectrum we were able to measure a velocity dispersion of
  $270 \pm 105$ km s$^{-1}$ (error bar including systematic errors),
  indicating that this galaxy lies closely on the virial relation given its
  stellar mass and effective radius.
\end{abstract}


\section{Introduction}

Galaxy formation and evolution in the cosmological context is one of
the leading themes in observational cosmology today, with much effort
being currently dedicated using all facilities both on the ground and
in space. The goal is to map the evolution of the population of
galaxies as a function of redshift (cosmic epoch) and environment,
with each population being described by distribution functions of
mass, star formation rate (SFR), morphology, AGN activity, etc. This
is accomplished through a full multiwavelength approach, covering the
whole electromagnetic spectrum, so to ensure the most robust
characterization of each individual galaxy.  This growing body of
observational evidence has started to provide enough data for a fully
empirical reconstruction of the main trends in the evolution of
galaxies, such as their mass growth via star formation and merging,
and the quenching of their star formation activity, turning them into
passively-evolving, early type galaxies
\citep[e.g.,][]{peng:2010,peng:2011}. In parallel, these observational
constraints are continuously fed into theoretical simulations, forcing
them to match closer and closer the picture that is progressively
emerging from observations.

Quenching of star formation is perhaps the most dramatic event in the
evolution of galaxies, turning them into passively-evolving galaxies
(PEGs). Hence, properly mapping this transformation across cosmic times
and in a range of environments must be central in our attempt to
understand galaxy evolution.  Indeed, massive PEGs represent
the culmination of the galaxy evolution process and are known to be
the most strongly clustered galaxies locally as well as at $z\sim 2$
\citep{kong:2006,mccracken:2010}. Besides marking the highest
density peaks, PEGs (ellipticals) are extremely interesting on their
own. In the local Universe PEGs (including bulges) encompass almost
60\%{} of the total stellar mass \citep{baldry:2004}, and therefore the
formation and evolution of this type of galaxies are central to the
broader problem of galaxy evolution in general. 

A most extended and coherent effort to map galaxy evolution up to
$z\sim 3$ is certainly the COSMOS project \citep{scoville:2007} with
full multiwavelength imaging coverage of an equatorial 2 square
degree field.  Besides the imaging campaign extending
from the X-rays to radio, a major spectroscopic effort over this field
has been conducted with the Very Large Telescope (VLT), targeting
galaxies all the way to $z\sim3$ \citep[the zCOSMOS project,][and in preparation]{lilly:2007,lilly:2009}. 
As a result, spectra and redshifts have been obtained for $\sim 20,000$ galaxies up
to $z\simeq 1.2$ over the full COSMOS field (zCOSMOS-Bright) and for $\sim 5,000$
galaxies at $1.4\lesssim z\lesssim 3$ over its central square degree (zCOSMOS-Deep).

Thus, mapping the evolution of PEGs as a function of mass, redshift
and environment has been fairly well accomplished up to $z\sim 1$,
especially as part of the zCOSMOS-Bright project
\citep[e.g.,][]{scarlata:2007,bolzonella:2010,pozzetti:2010}, and
there is a claim that the most massive PEGs are virtually all in place
by $z\sim 1$ \citep{cimatti:2006,pozzetti:2010}.  However, their
number density rapidly falls at higher redshifts as suggested by the
number density of PEGs at $\langle z\rangle=1.7$ being only $\sim
20\%$ of that of local ellipticals of similar mass \citep[][see also
\citealt{daddi:2005:pbzk, brammer:2009,brammer:2011, taylor:2009,
  williams:2009, ilbert:2010, cassata:2011}]{kong:2006}.

However, much of the current statistics of $z\gtrsim 1.4$ PEGs relies
on photometric redshifts, which are not accurate enough to properly
map the density field at these high redshifts.  This is especially the
case for the highest density regions, where environmental effects are
presumably strongest, and likely to be preferentially inhabited by
PEGs which are the most strongly clustered population at $1.4\lesssim
z\lesssim 3$ \citep{kong:2006,mccracken:2010}. As a consequence, it is
not currently possible to measure the fraction of red (quenched)
galaxies at $z\gtrsim1 .4$ as a function of overdensity, as done at
lower redshifts by \citet{peng:2010}.  In fact, due to instrumental
limitations the high-redshift component of the zCOSMOS survey has been
designed to observe only star-forming galaxies (down to $B_\text{AB}
\lesssim 25$), and therefore none of the $\sim 4,000$ $z>1.4$
passively evolving galaxies brighter than $K_{\rm AB}\sim23$ was
included among the targets. Thus, with zCOSMOS-Deep alone one cannot
study this important galaxy population, hence one cannot properly map
the highest peaks of the density field at these redshifts.

Beyond $z\simeq1.4$, only a handful of such galaxies have been
observed spectroscopically \citep[e.g.,][see \citealt{damjanov:2011}
for a recent
compilation]{cimatti:2004,cimatti:2008,mccarthy:2004,daddi:2005:pbzk},
and their stellar populations dated.  Except for such rare cases, the
stellar population properties of high-$z$ PEGs have been explored only
via their broad-band spectral energy distribution (SED), or with
intermediate-width passbands \citep[e.g.,][]{whitaker:2011}.  Besides
locating individual PEGs relative to the density field, full
spectroscopic coverage is instead essential to alleviate the
age--metallicity degeneracy and to get dynamical information, in
particular in the rest-frame optical where features are extensively
used and calibrated for this purpose in local elliptical galaxies.

At redshifts below $\sim 1.4$ the strong spectral features that allow
one to measure the redshift of PEGs are the \ion{Ca}{2} H\&K lines and
the 4000 \AA\ break, but by $z=1.4$ these features  move to the
near-IR.  The only useful spectral feature for optical spectroscopy
then remains the complex of rest-frame UV absorption lines of neutral
and ionized iron and magnesium in the
$2650<\lambda_{\text{rest}}(\text{\AA})<2850$ spectral range, where
the continuum of PEGs is very faint.  Nevertheless, this feature has
been used to derive redshifts and stellar ages for most of the few
passive galaxies at $z>1.4$ with spectroscopic information (see
references above).  However, reaching the extremely faint UV continuum
of a passive galaxy requires extremely long integrations, which in the
case of \citet{cimatti:2008} ranged from $\sim30$ to 60 hours per object. 

Alternatively, one can try to follow the \ion{Ca}{2} H\&K lines and
the 4000 \AA\ break in the near-IR, where the continuum flux is much
higher (but also the background).  This was attempted with long-slit
spectroscopy with GNIRS on Gemini by
\citet{kriek:2006:passive,kriek:2008:survey,kriek:2009} resulting in
17 spectroscopic redshifts of PEGs at $z\simeq2.3$ judged from the
absence of emission lines.  Multi-object spectroscopy in the near-IR
offers decisive advantages over optical or single-channel near-IR
spectroscopy of high-redshift PEGs, and the Multi-Object InfraRed
Camera and Spectrograph
\citep[MOIRCS;][]{ichikawa:2006:moircs,suzuki:2008:moircs} mounted on
the Subaru telescope offered an unique opportunity amongst 8--10m
class telescopes.  In 2006 we then started a project exploiting this
opportunity and, after a few unsuccessful runs, useful data have
started to accumulate from 2009 on and the results are presented in
this paper.

The paper is organized as follows.  In Section \ref{sec:obs} we
describe our dataset; redshift identifications are presented and
compared to photometric redshifts in Section \ref{sec:speczz}, whereas
the presence of several PEG redshift spikes is discussed in Section
\ref{sec:overdensity}.  In Section \ref{sec:stacking} a composite
spectrum corresponding to $\sim 140$ hours of integration is
presented. The properties of the stellar populations of the program
galaxies are investigated in detail in Section
\ref{sec:stellarpopulation}.  In Section \ref{sec:size_sigma} the
structural and kinematical properties of these high-$z$ PEGs are
derived and compared to those of other high-$z$ PEGs with
spectroscopic redshifts and of local early-type galaxies.  The number
and mass densities of our PEGs are discussed in Section
\ref{sec:number}.  Finally, we summarize our main results and
conclusions in Section \ref{sec:summary}.

Throughout this paper, we assume a flat cosmology with
$H_0=70\kms\,\text{Mpc}^{-1}$, $\Omega_\text{M}=0.3$, and
$\Omega_\Lambda=0.7$.  Photometric magnitudes are expressed in the AB
system \citep{oke:1983}.

\section{Sample, Observations, and Data Reduction}
\label{sec:obs}

\subsection{Sample Selection}
\label{sec:sampleselection}
Our MOIRCS spectroscopic targets are primarily \textit{BzK}-selected
candidate PEGs \citep[\pbzks,][]{daddi:2004:bzk} extracted from the
\textit{K}-band selected catalog for the COSMOS field \citep{mccracken:2010}.
Among its $\sim 4,000$ \pbzks down to $K<23$, 
we selected 34 \pbzks with $19.4<K<21.9$ 
among the brightest objects that satisfied as many of the following additional criteria as possible:

\begin{itemize}
\item no detection at $24$ \micron{} in the COSMOS \spitzer/MIPS data
  \citep{sanders:2007,lefloch:2009} down to $80\,\mu\text{Jy}$,
  corresponding to a $\sim 3\sigma$ detection limit, to exclude 
  star-forming galaxies and AGNs.  About 3,100 of 4,000 \pbzks satisfy this criterion. 
  At $z\gtrsim1.4$, where the \pbzks are expected to be located, this
  flux limit corresponds to a SFR of roughly
  $50\msun\,\text{yr}^{-1}$, which still fairly high. 
  As an exception to this criterion, one \pbzk object with higher $24$
  \micron{} flux was also included in the target list;
\item a higher $4.5$ \micron{} flux compared to that at $3.6$ \micron{} in the
  \spitzer/IRAC bands, i.e., $m_{3.6}-m_{4.5}>0$, which is satisfied by 
  about 2,900 \pbzks. 
  The IRAC photometry is taken from the publicly available S-COSMOS catalog \citep{sanders:2007}.
  A red $m_{3.6}-m_{4.5}$ color makes likely  that the object is at
  $z>1.4$, hence enabling us to access major rest-frame optical
  absorption features.;
\item photometric redshift \citep[from][]{ilbert:2009} $z_\text{phot}>1.4$ 
  to detect 4000 \AA{} break in the observed wavelength range. 
  About 2,400 \pbzks have $z_\text{phot}>1.4$. 
  Several \pbzk{} objects with $z_\text{phot}$ slightly below 1.4 were also included 
  to account for photometric redshift errors. 
  About 30\%{} of our selection have $z_\text{phot}<1.4$;
\item high projected concentration of \pbzk galaxies over each MOIRCS
  field of view (FoV), to make sure that the highest possible number of
  suitable targets is observed for each telescope pointing, thus
  exploiting the multiplex of the instrument. Bright \pbzks are
  strongly clustered whereas their surface number density on the sky
  is relatively low \citep[$r_0\simeq 7$ Mpc and 
    $\Sigma \simeq 0.4$ arcmin$^{-2}$ to the $K=22$ limit,
    respectively;][]{kong:2006,mccracken:2010}. Besides maximizing the
  yield, targeting \pbzk concentrations allows us to check whether such
  concentrations are mere statistical fluctuations in surface density,
  or real physical structures. Thanks to the wide 2 square degree
  field of COSMOS, several concentrations with at least 15 \pbzks with $K\lesssim22$ per
  MOIRCS field of view were identified;
\item regions including some of the 12 brightest \pbzks studied by \citet{mancini:2010}, 
  hence ensuring the brightest possible continuum emission and the
  detection of absorption features with as high a S/N as possible. 
\end{itemize}

Considering the above constraints, we selected a FoV containing the
object 254025 at $z=1.82$ \citep{onodera:2010:pbzk} which is one of
the 12 brightest galaxies in \citet{mancini:2010}, and observed it in
2009 finding a possible overdensity of \pbzks with spectroscopic
redshift at $z=1.82$ (see Section \ref{sec:overdensity}).  Motivated
by this finding, we then selected two more adjacent FoVs to cover a
possible large scale structure around it as well as to maximize the
number of the brightest \pbzks in each mask. Thus, the three selected
MOIRCS FoVs cluster around a concentration of bright \pbzks, but the
selected fields lie outside the central square degree covered by
zCOSMOS-Deep \citep{lilly:2007}.

\begin{figure}[htbp]
  \begin{center}
    \includegraphics[width=0.95\linewidth]{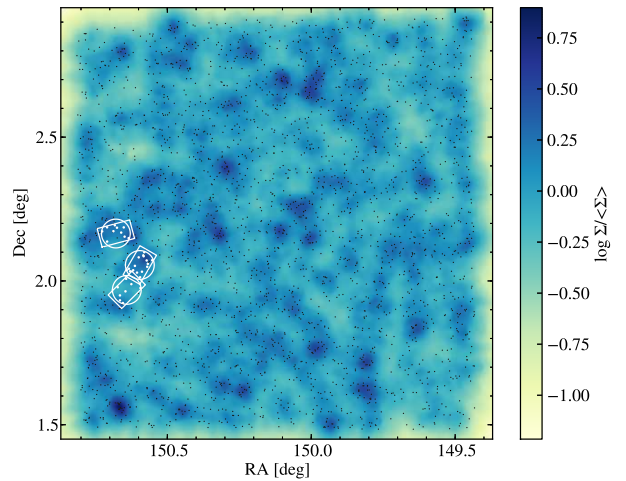}
  \end{center}
  \caption{
    The density map of the \pbzk galaxies in the COSMOS 2 square
    degree field covered by the \textit{K}-band image from \citet{mccracken:2010}. 
    Black and white dots show the location of the $\sim 2800$ \pbzks
    down to $\mathit{K}=22$ and the spectroscopically observed \pbzks, respectively.
    White rectangles and circles show the FoVs covered by MOIRCS
    observations (the mask IDs are B, A, and C from north to south, respectively). 
    Slits were placed in the overlapping regions between a circle and the corresponding rectangle. 
    Background shows the surface density of \pbzks as calculated from 10-th nearest neighbour 
    and normalized by an average surface density of 0.41 arcmin$^{-2}$.
  }
  \label{fig:densitymap}
\end{figure}

Figure \ref{fig:densitymap} shows the projected density map of \pbzks
in the COSMOS field with $K<22$, along with the position of the three
MOIRCS pointings for which the data presented here have been taken.
Besides 254025, two other bright \pbzks (namely 217431 and 307881)
were also included in the sample of 12 ultra-massive PEGs whose
structural parameters have been measured by
\citet{mancini:2010}. Together with 254025, these are the three
brightest objects in our sample, and therefore are listed at the top
of the tables in this paper.

\begin{figure}[htbp]
  \begin{center}
    \includegraphics[width=0.95\linewidth]{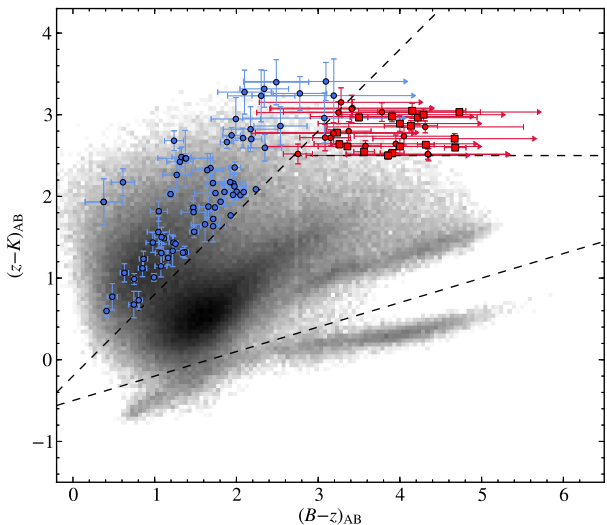}
  \end{center}
  \caption{ \textit The \textit{BzK} 2-color diagram for objects in
    the COSMOS field.  The \textit{K}-selected objects are shown in
    gray scale as a background and the \pbzks in our spectroscopic
    sample are shown in red.  Red squares and circles show \pbzks with
    and without continuum detection, respectively.  Blue circles
    represent star-forming \textit{BzK}-selected galaxies that were
    also included in the MOIRCS masks which will be described in a
    forthcoming paper.  }
  \label{fig:bzkdiagram_pbzks}
\end{figure}

\begin{figure}[htbp]
  \begin{center}
    \includegraphics[width=0.95\linewidth]{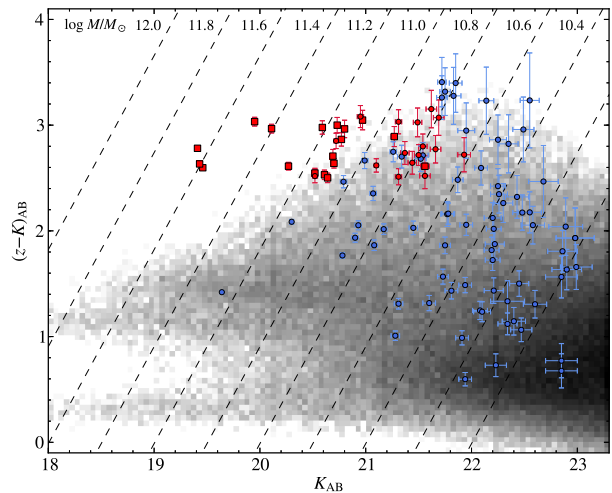}
  \end{center}
  \caption{ The $(\mathit{z}-\mathit{K})$--\textit{K} color--magnitude
    diagram for objects in the COSMOS field.  Symbols are the same as
    in Figure \ref{fig:bzkdiagram_pbzks}.  Dashed lines show constant
    stellar mass derived by using Equation (6) and (7) of
    \citet{daddi:2004:bzk}, i.e.,
    $\log(M_*/10^{11}\,M_\odot=-0.4(\mathit{K}^\text{tot}-21.38)+0.218(\mathit{z}-\mathit{K}-2.29)$,
    converted to the Chabrier IMF adopted in this paper.  }
  \label{fig:cmd_pbzks}
\end{figure}

One MIPS 24 \micron{} detected object (313880, see Section
\ref{sec:mips}) with $f_{24}=130\,\mu\text{Jy}$ was included because
of geometrical constraints there is no alternative \pbzk object that
could be placed in the MOIRCS mask. The list of the 34 \pbzk targets
along with their photometric redshifts and magnitudes are reported in
Table \ref{tab:propphot}.  Note that all the \pbzks{} down to $K=21$
included in the MOIRCS explored area have been included in the target
list.  Figure \ref{fig:bzkdiagram_pbzks} displays the \textit{BzK}
plot for the COSMOS field (adapted from \citealt{mccracken:2010}) with
the observed \pbzk targets shown with red symbols. These same objects
are also shown in the $(\mathit{z}-\mathit{K})-\mathit{K}$
color-magnitude diagram displayed in Figure \ref{fig:cmd_pbzks}.

After including in the MOIRCS masks the highest possible number of
\pbzks, the residual fibres were used to observe star-forming
\textit{BzK}-selected galaxies (\sbzks, shown as blue symbols in
Figures \ref{fig:bzkdiagram_pbzks} and \ref{fig:cmd_pbzks}). The
results for these \sbzks will be presented and discussed in a future
paper.

%
%
\begin{deluxetable*}{ccccccccccc}[htbp]
  \tablewidth{0pt}
  \tablecolumns{11}
  \tablecaption{Photometric properties of observed \pbzks \label{tab:propphot}}
  \tablehead{
    \colhead{ID} &
    \colhead{RA} &
    \colhead{Dec} &
    \colhead{$B$} &
    \colhead{$z$} &
    \colhead{$K$} &
    \colhead{$B-z$} &
    \colhead{$z-K$} &
    \colhead{$z_\text{phot}$} &
    \colhead{Mask ID\tablenotemark{a}} &
    \colhead{Exp. Time}
    \\
    \colhead{} &
    \colhead{(deg)} &
    \colhead{(deg)} &
    \colhead{(mag)} &
    \colhead{(mag)} &
    \colhead{(mag)} &
    \colhead{(mag)} &
    \colhead{(mag)} &
    \colhead{}&
    \colhead{}&
    \colhead{(min)}
  }
  \startdata
  \cutinhead{\textit{Identified}}
  254025 & 150.6187115 & 2.0371363 & $25.15 \pm 0.03$ & $22.19 \pm 0.02$ & $19.41 \pm 0.01$ & $3.23 \pm 0.04$ & $2.78 \pm 0.02$ & 1.7407 & A & 490 \\
  217431 & 150.6646939 & 1.9497545 & $26.11 \pm 0.10$ & $22.07 \pm 0.03$ & $19.43 \pm 0.01$ & $4.32 \pm 0.10$ & $2.63 \pm 0.03$ & 1.2457 & C & 450 \\
  307881 & 150.6484873 & 2.1539903 & $26.47 \pm 0.13$ & $22.06 \pm 0.03$ & $19.46 \pm 0.01$ & $4.68 \pm 0.13$ & $2.60 \pm 0.03$ & 1.4201 & B & 550 \\
  233838 & 150.6251048 & 1.9889180 & $27.44 \pm 0.29$ & $22.98 \pm 0.05$ & $19.95 \pm 0.01$ & $4.73 \pm 0.29$ & $3.03 \pm 0.05$ & 1.8428 & C & 450 \\
  277491 & 150.5833512 & 2.0890266 & $26.31 \pm 0.10$ & $23.08 \pm 0.06$ & $20.11 \pm 0.01$ & $3.50 \pm 0.12$ & $2.97 \pm 0.06$ & 1.7936 & A & 490 \\
  313880 & 150.6603061 & 2.1681129 & $25.97 \pm 0.07$ & $22.88 \pm 0.05$ & $20.27 \pm 0.02$ & $3.36 \pm 0.09$ & $2.61 \pm 0.05$ & 1.3430 & B & 550 \\
  250093 & 150.6053729 & 2.0288998 & $27.21 \pm 0.25$ & $23.57 \pm 0.09$ & $20.59 \pm 0.02$ & $3.90 \pm 0.27$ & $2.98 \pm 0.09$ & 1.4234 & A & 490 \\
  263508 & 150.5677283 & 2.0594318 & $26.77 \pm 0.14$ & $23.14 \pm 0.06$ & $20.61 \pm 0.02$ & $3.90 \pm 0.15$ & $2.53 \pm 0.06$ & 1.5658 & A & 490 \\
  269286 & 150.5718552 & 2.0712204 & $26.73 \pm 0.14$ & $23.14 \pm 0.06$ & $20.64 \pm 0.02$ & $3.85 \pm 0.15$ & $2.50 \pm 0.06$ & 1.6654 & A & 490 \\
  240892 & 150.6432950 & 2.0073169 & $>27.90        $ & $23.40 \pm 0.08$ & $20.69 \pm 0.02$ & $>4.67        $ & $2.71 \pm 0.08$ & 1.5474 & C & 450 \\
  205612 & 150.6542714 & 1.9233323 & $26.32 \pm 0.09$ & $23.33 \pm 0.06$ & $20.70 \pm 0.02$ & $3.26 \pm 0.11$ & $2.64 \pm 0.06$ & 1.2932 & C & 450 \\
  251833 & 150.6293675 & 2.0336620 & $26.36 \pm 0.12$ & $23.07 \pm 0.06$ & $20.52 \pm 0.02$ & $3.56 \pm 0.13$ & $2.55 \pm 0.06$ & 1.1874 & A & 490 \\
  228121 & 150.5936156 & 1.9754018 & $27.75 \pm 0.38$ & $23.73 \pm 0.10$ & $20.73 \pm 0.02$ & $4.29 \pm 0.39$ & $3.00 \pm 0.10$ & 1.7198 & C & 450 \\
  321998 & 150.7093826 & 2.1863891 & $27.50 \pm 0.32$ & $23.63 \pm 0.09$ & $20.77 \pm 0.02$ & $4.13 \pm 0.33$ & $2.87 \pm 0.09$ & 1.4320 & B & 550 \\
  299038 & 150.7091894 & 2.1369001 & $>27.90        $ & $24.02 \pm 0.13$ & $20.97 \pm 0.02$ & $>4.15        $ & $3.05 \pm 0.13$ & 1.7667 & B & 550 \\
  209501 & 150.6645174 & 1.9325604 & $>27.90        $ & $24.17 \pm 0.14$ & $21.56 \pm 0.04$ & $>4.00        $ & $2.61 \pm 0.15$ & 1.3413 & C & 450 \\
  253431 & 150.6408360 & 2.0378335 & $>27.90        $ & $24.17 \pm 0.13$ & $21.27 \pm 0.03$ & $>4.00        $ & $2.89 \pm 0.13$ & 1.5583 & A & 490 \\
  275414 & 150.5822420 & 2.0857211 & $>27.90        $ & $23.76 \pm 0.12$ & $20.80 \pm 0.02$ & $>4.22        $ & $2.96 \pm 0.12$ & 1.4332 & A & 490 \\
  \cutinhead{\textit{No detection}}
  222961 & 150.6455001 & 1.9637953 & $>27.90        $ & $24.76 \pm 0.25$ & $21.69 \pm 0.05$ & $>3.41        $ & $3.07 \pm 0.25$ & 1.4755 & C & 450 \\
  229536 & 150.6724126 & 1.9780668 & $26.58 \pm 0.14$ & $24.08 \pm 0.16$ & $21.56 \pm 0.05$ & $2.75 \pm 0.21$ & $2.52 \pm 0.17$ & 1.6066 & C & 450 \\
  243138 & 150.5947944 & 2.0130046 & $27.46 \pm 0.27$ & $24.65 \pm 0.23$ & $21.93 \pm 0.06$ & $3.08 \pm 0.35$ & $2.72 \pm 0.24$ & 1.9387 & A & 280 \\
  251051 & 150.5881329 & 2.0320317 & $27.10 \pm 0.20$ & $24.21 \pm 0.14$ & $21.50 \pm 0.04$ & $3.15 \pm 0.24$ & $2.72 \pm 0.15$ & 1.4514 & A & 280 \\
  255465 & 150.6336861 & 2.0423782 & $27.35 \pm 0.24$ & $24.43 \pm 0.16$ & $21.66 \pm 0.05$ & $3.19 \pm 0.29$ & $2.77 \pm 0.17$ & 1.8805 & A & 280 \\
  258867 & 150.5698729 & 2.0491841 & $>27.90        $ & $24.77 \pm 0.25$ & $21.62 \pm 0.05$ & $>3.28        $ & $3.15 \pm 0.25$ & 2.3654 & A & 280 \\
  260120 & 150.6126504 & 2.0523735 & $>27.90        $ & $23.82 \pm 0.10$ & $21.31 \pm 0.03$ & $>4.35        $ & $2.51 \pm 0.10$ & 0.9670 & A & 280 \\
  268884 & 150.5525002 & 2.0719597 & $>27.90        $ & $24.34 \pm 0.17$ & $21.31 \pm 0.03$ & $>3.78        $ & $3.03 \pm 0.17$ & 1.6244 & A & 280 \\
  273534 & 150.5994479 & 2.0819904 & $27.50 \pm 0.27$ & $24.52 \pm 0.20$ & $21.49 \pm 0.04$ & $3.25 \pm 0.34$ & $3.03 \pm 0.20$ & 1.5801 & A & 280 \\
  281751 & 150.5802008 & 2.1000786 & $27.03 \pm 0.18$ & $23.72 \pm 0.09$ & $21.10 \pm 0.02$ & $3.58 \pm 0.20$ & $2.62 \pm 0.09$ & 1.1811 & A & 280 \\
  305677 & 150.6343972 & 2.1488364 & $27.11 \pm 0.33$ & $23.04 \pm 0.09$ & $20.52 \pm 0.02$ & $4.34 \pm 0.34$ & $2.52 \pm 0.09$ & 1.3697 & B & 340 \\
  315704 & 150.6874049 & 2.1729448 & $>27.90        $ & $24.10 \pm 0.14$ & $21.37 \pm 0.04$ & $>4.05        $ & $2.74 \pm 0.15$ & 2.5197 & B & 550 \\
  316338 & 150.7264016 & 2.1721785 & $>27.90        $ & $24.08 \pm 0.15$ & $21.44 \pm 0.04$ & $>3.95        $ & $2.64 \pm 0.16$ & 1.1895 & B & 340 \\
  321193 & 150.6749498 & 2.1850247 & $27.45 \pm 0.25$ & $24.34 \pm 0.16$ & $21.54 \pm 0.04$ & $3.37 \pm 0.30$ & $2.80 \pm 0.16$ & 1.5988 & B & 550 \\
  322048 & 150.6514093 & 2.1861803 & $27.18 \pm 0.27$ & $24.03 \pm 0.14$ & $20.95 \pm 0.03$ & $3.42 \pm 0.30$ & $3.08 \pm 0.14$ & 1.5022 & B & 340 \\
  325564 & 150.6824812 & 2.1934154 & $>27.90        $ & $23.57 \pm 0.10$ & $20.72 \pm 0.02$ & $>4.31        $ & $2.85 \pm 0.10$ & 1.3220 & B & 550 
  \enddata
  \tablenotetext{a}{The mask IDs are given in Table \ref{tab:obslog}.}
  \tablecomments{
    Upper limits are $3\sigma$. 
  }
\end{deluxetable*}

\subsection{Observations}
The near-IR multi-object spectroscopic observations have been carried
out with Subaru/MOIRCS.  The imaging FoV of MOIRCS is $7\times4$
arcmin$^{2}$ with 2 detectors (channel 1 and 2, respectively) and
slits can be placed within the intersection with the 6 arcmin diameter
circular region as illustrated in Figure \ref{fig:densitymap}. The
low resolution zJ500 grism was used with $0.7$ arcsec width slits,
which covers about 9000--17500 \AA{} with a resolution $R\simeq500$.
The slit length ranges from 8 to 14 arcsec, adequate to subtract the
sky background.

Six masks in total, two masks per FoV, were observed during six
separate observing runs as reported in Table \ref{tab:obslog}.  We
replaced targets with alternative ones when neither continuum emission
nor emission lines were seen in the first observing runs, but the
majority of the targets are common between the masks targeting the
same FoVs (cf.\ Table \ref{tab:obslog}). Sequences of 600--1200 sec
integrations were made in a standard two-position ``AB'' dithering
pattern separated by 2 arcsec. At the beginning and/or end of the
nights A0V-type standard stars were observed for flux calibration
(i.e., atmospheric absorption and instrument response) with the
identical instrumental setup as the targets and with similar airmass
as for the COSMOS field observations.

In 2009, observations were affected by cloudy weather and $\sim 1.2$
arcsec seeing whereas in 2010 we took advantage of clear nights and
$\sim 0.4$--$1.0$ arcsec seeing conditions. Each FoV was integrated
for 7--9 hours (about half of this if targets were replaced between
runs).  Table \ref{tab:obslog} summarizes our observations.

%
%
\begin{deluxetable*}{crccc}[htbp]
  \tabletypesize{\scriptsize}
  \tablewidth{0pt}
  \tablecolumns{5}
  \tablecaption{Observation Logs\label{tab:obslog}}
  \tablehead{
    \colhead{Mask ID} &
    \colhead{Date} &
    \colhead{Exp Time} &
    \colhead{Number of \pbzks\tablenotemark{a}} &
    \colhead{Number of \sbzks\tablenotemark{a}} 
    \\
    \colhead{} &
    \colhead{} &
    \colhead{(min)} &
    \colhead{} &
    \colhead{}
  }
  \startdata
  A  & 13, 14 Mar 2009 & 280 & 16 \phm{(0)} & 16 \phm{(0)} \\
     & 29,    Mar 2010 & 210 &  8 (8)       & 20 (8)       \\
  B  &  6     Feb 2010 & 340 & 10 \phm{(0)} & 17 \phm{(0)} \\
     &  1     Apr 2010 & 210 &  7 (7)       & 15 (7)       \\
  C  &  7     Feb 2010 & 330 &  8 \phm{(0)} & 15 \phm{(0)} \\
     & 21     Feb 2010 & 120 &  8 (8)       & 14 (9)
  \enddata
  \tablenotetext{a}{The number of common objects between two runs with the same MOIRCS pointing are indicated in parentheses. }
  \tablecomments{
    Column 1: ID of the three MOIRCS pointings (masks); 
    Column 2: observing date;
    Column 3: exposure time; 
    Column 4: number of \pbzks in the mask; 
    Column 5: number of \sbzks in the mask
  }
\end{deluxetable*}

\subsection{Data Reduction}

The data were reduced with the MCSMDP
pipeline\footnote{\url{http://www.naoj.org/Observing/DataReduction/}}
\citep{yoshikawa:2010} and with custom scripts.  The data are first
flat-fielded using dome-flat frames taken with the same configuration
of the science frames. Bad pixels were then removed using the
bad-pixel map bundled in MCSMDP and cosmic ray hits were removed using
the pair of images in the dithering. The counts of the rejected
pixels were replaced by linear interpolations from neighboring pixels
along the spatial direction.  After removing bad pixels and cosmic
rays, the sky background was removed by subtracting the ``B'' frame
from the ``A'' frame.  Then the distortion of the detectors was
corrected by using the same coefficients used in the MOIRCS imaging
data reduction\footnote{The distortion coefficients are available as
  an IRAF/geotran format bundled with a MOIRCS imaging data reduction
  pipeline provided by Ichi Tanaka from
  \url{http://www.naoj.org/staff/ichi/MCSRED/mcsred.html}}.

Each frame was then rotated to correct the tilt of the spectra on the
detector by making the edges of each slit almost parallel to the
$x$-direction (dispersion direction) of the detector.  We applied a 
rotation angle of $\sim -0.56$ and $\sim -0.73$ degrees for
the channel 1 and 2, respectively.

Individual spectra were cut from the frames and wavelength calibration
was made for each extracted 2-dimensional spectrum based on the
location of the OH-airglow lines \citep{rousselot:2000} present in the
observed wavelength range. The uncertainty of the wavelength
calibration is typically one half the pixel size, i.e., $\sim 2.5$
\AA.  Then the frame was transformed to align the sky lines along the
$y$-direction (spatial direction) and to make wavelength a linear
function of pixel coordinates. Because sky background residuals are
expected due to the time variation of sky brightness (in particular in
the OH-line strength) we performed an additional residual
sky-subtraction by subtracting at each wavelength a mean of the
spatial pixels outside those occupied by the objects.

The spectra of the A0V-type stars were processed in the same manner as
for the targets, and the 2D frames were co-added. Then the 1D spectra
of the standard stars were extracted by using the \texttt{apall} task
in IRAF. The system response curves including atmospheric, telescope,
and instrumental throughputs were obtained by dividing the observed
spectra of the standard stars by spectra from the stellar spectral
library of \citet{pickles:1998}. After the residual sky-subtraction,
the 2D spectra of the targeted galaxies were flux-calibrated by
dividing them by the system response curve and co-added with
appropriate offsets derived either from the location of reference
stars or bright compact objects to which slits had been assigned or
from the fits header information. In performing these co-additions we
applied weights proportional to the exposure time and inversely
proportional to the seeing size and the square of the S/N of each
individual 2D spectrum.

Finally, for the objects with continuum detection the 1D spectra were
extracted from the stacked 2D frames by using the \texttt{apall} task
in IRAF. The 1D spectra from the different observing runs were
extracted separately, and then the pairs of 1D spectra were stacked
together with a weight to maximize the S/N of the final 1D spectra.
These 1D spectra were calibrated to absolute flux scale by comparing
them with the $J$- and $H$-band fluxes.  The uncertainty of the
absolute flux calibration is $\lesssim 20$\%{} indicated by the
difference of the scaling factor between the $J$ and the $H$ band.

The cutouts from the COSMOS \hst/ACS $i$-band image
\citep{koekemoer:2007} and 1D spectra of the 18 \pbzks with continuum
detection for which the determination of the spectroscopic redshifts
was attempted are shown respectively in the left and middle panels of
Figure \ref{fig:pbzk_hst_spec_sed}.

\begin{figure*}[htbp]
  \begin{center}
    \includegraphics[width=0.95\linewidth]{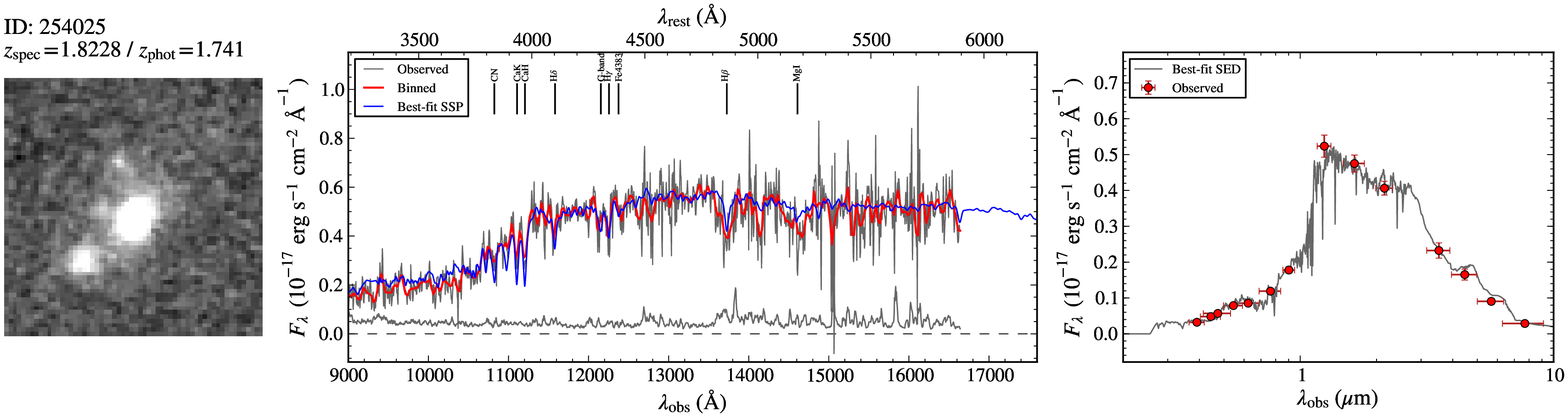}
    \includegraphics[width=0.95\linewidth]{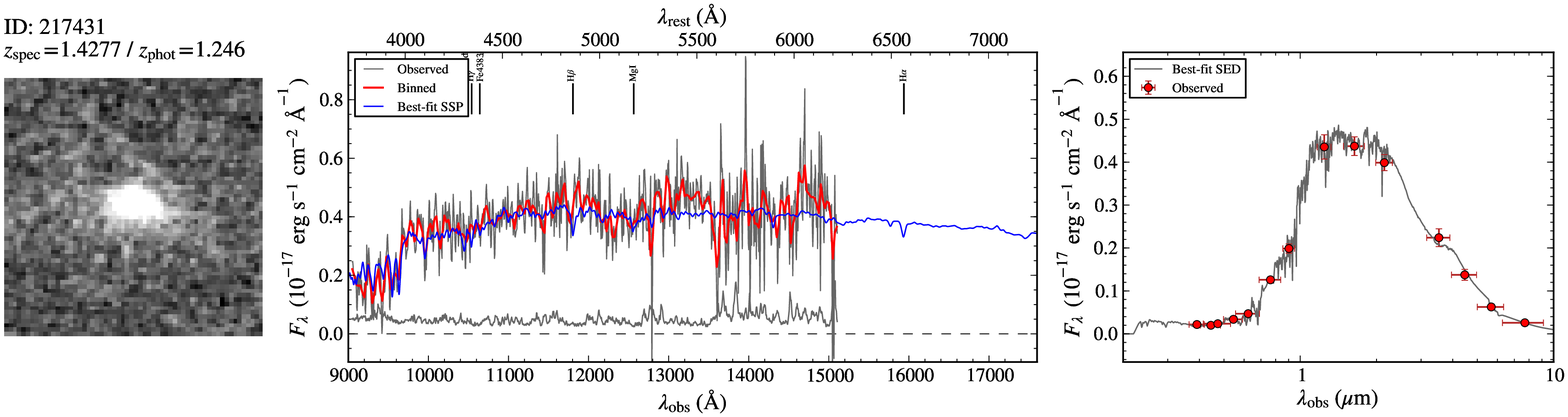}
    \includegraphics[width=0.95\linewidth]{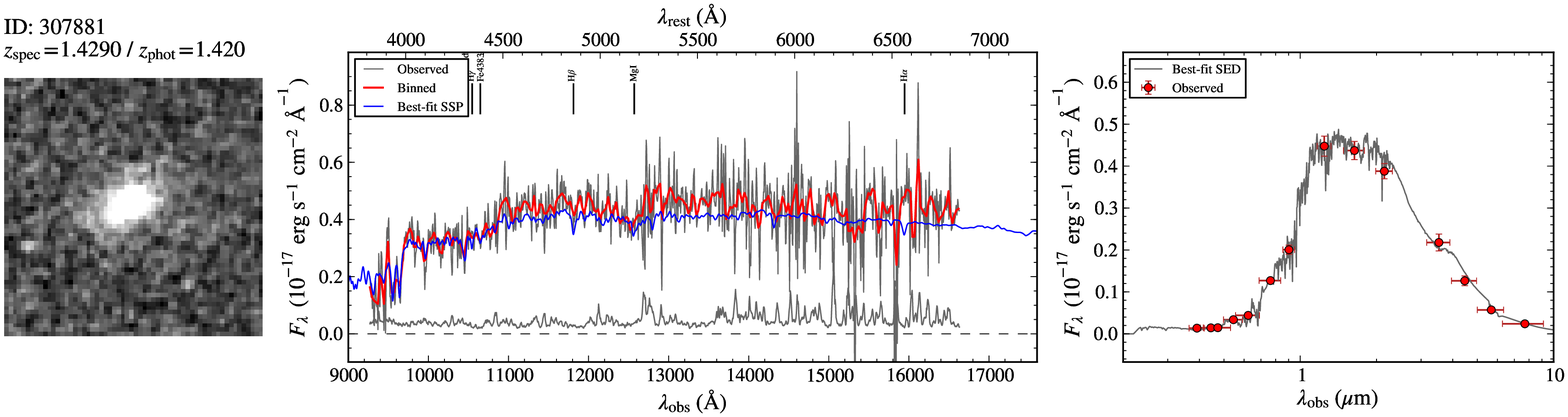}
    \includegraphics[width=0.95\linewidth]{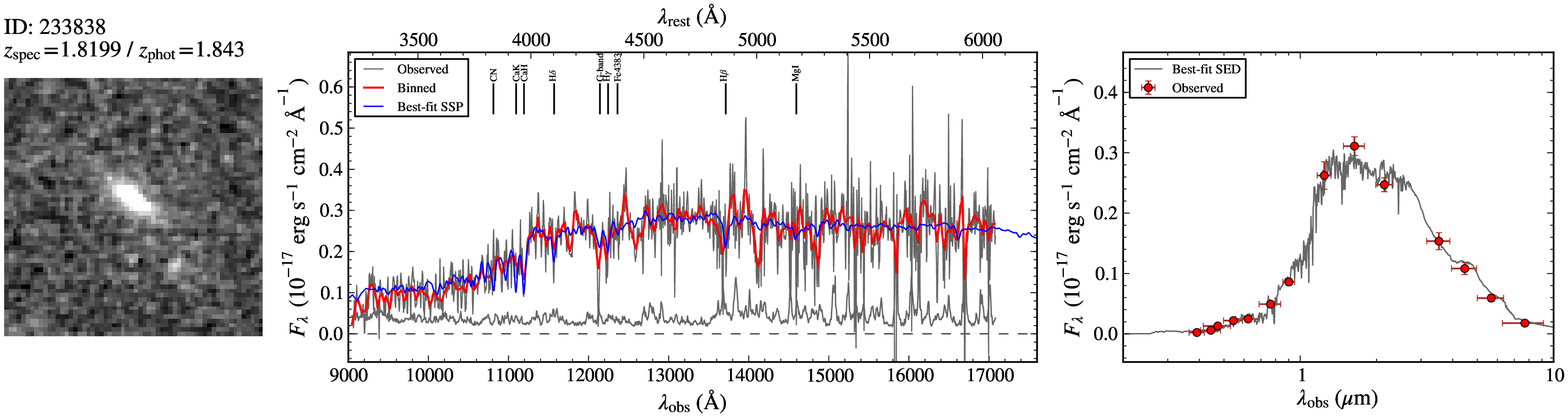}
  \end{center}
  \caption{ \hst/ACS stamps, MOIRCS spectra, and broad-band SEDs of
    the continuum detected \pbzks in this study.  \textit{Left}:
    \hst/ACS cutouts made from the publicly available COSMOS HST-ACS
    Mosaic v2.0 from \citet{koekemoer:2007}.  The images are 3 arcsec
    on a side; \textit{Middle}: MOIRCS spectra. The observed spectrum,
    its 50 \AA{} median smoothed version, and its best-fit SSP
    template from \citet{bruzual:2003} library are shown respectively
    in gray, orange, and blue. The gray spectra at the bottom show the
    $1\sigma$ noise.  The positions of major absorption lines are
    indicated by the vertical lines; \textit{Right}: Broad-band
    SEDs. The red circles show the observed fluxes and the gray line
    shows the best-fit template.  The panels from the top to bottom
    are ordered with increasing total $K$-band magnitude.  The IDs and
    spectroscopic and photometric redshifts are indicated.  }
  \label{fig:pbzk_hst_spec_sed}
\end{figure*}

\addtocounter{figure}{-1}
\begin{figure*}[htbp]
  \begin{center}
    \includegraphics[width=0.95\linewidth]{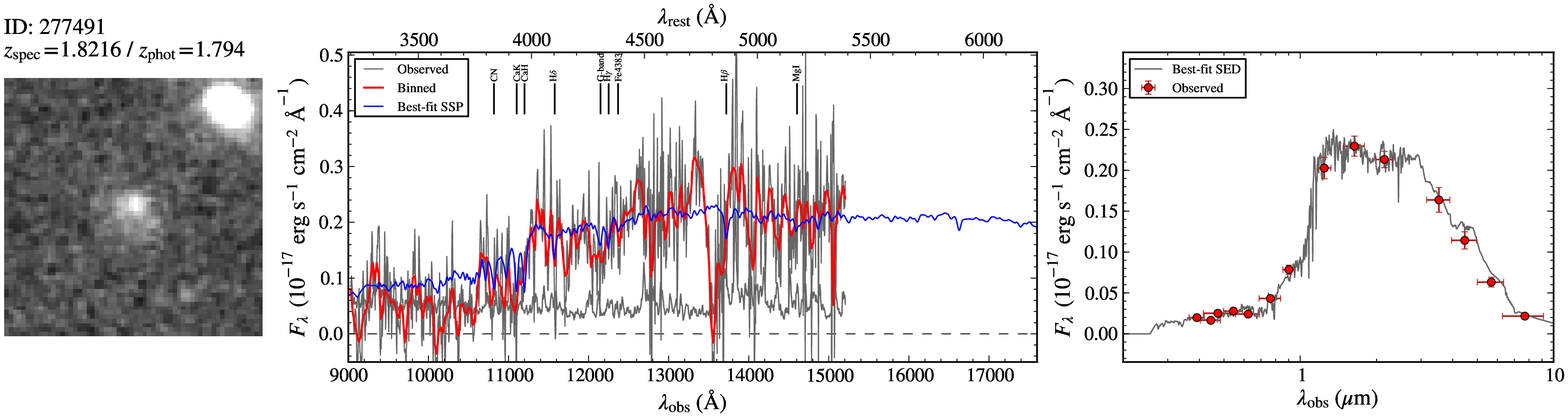}
    \includegraphics[width=0.95\linewidth]{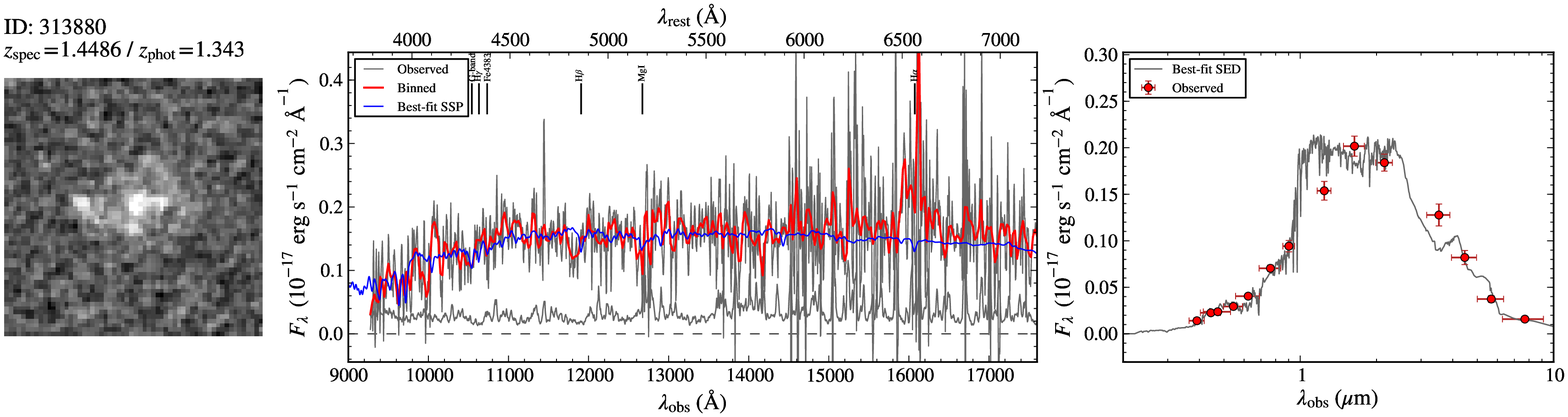}
    \includegraphics[width=0.95\linewidth]{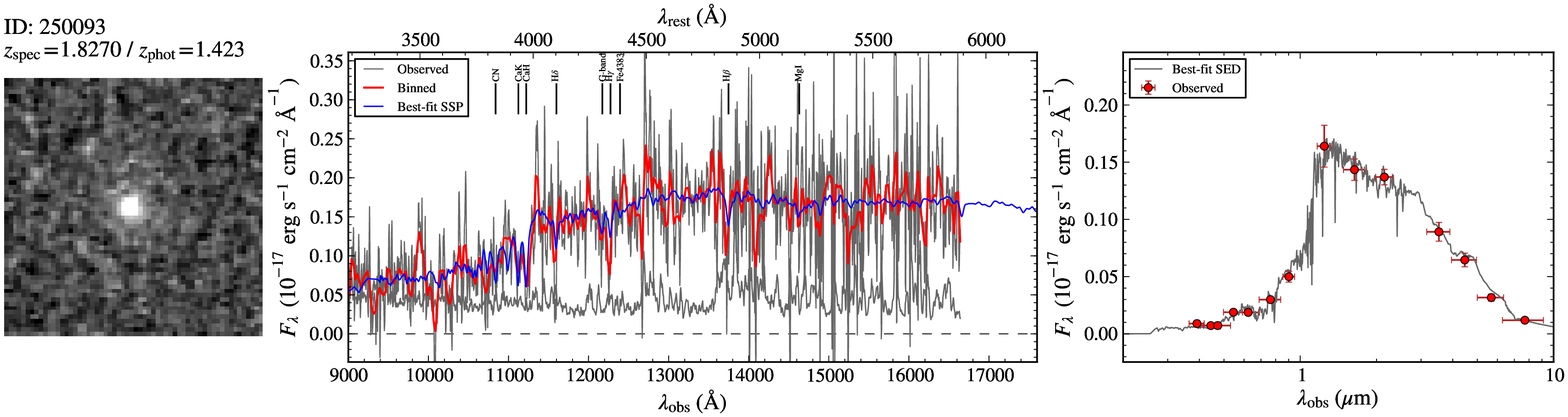}
    \includegraphics[width=0.95\linewidth]{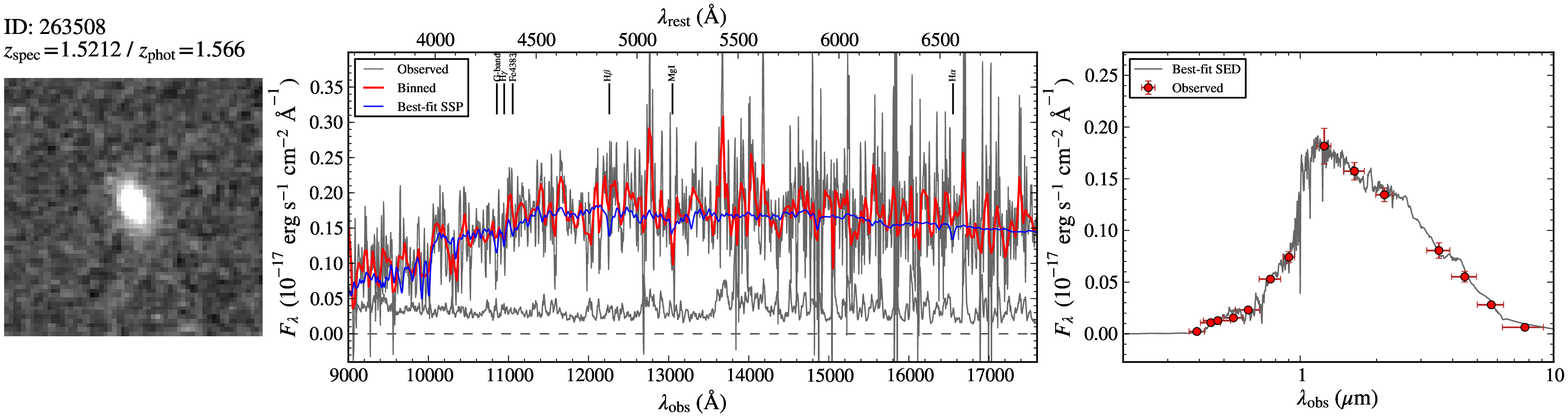}
  \end{center}
  \caption{\textit{Continued.}}
\end{figure*}

\addtocounter{figure}{-1}
\begin{figure*}[htbp]
  \begin{center}
    \includegraphics[width=0.95\linewidth]{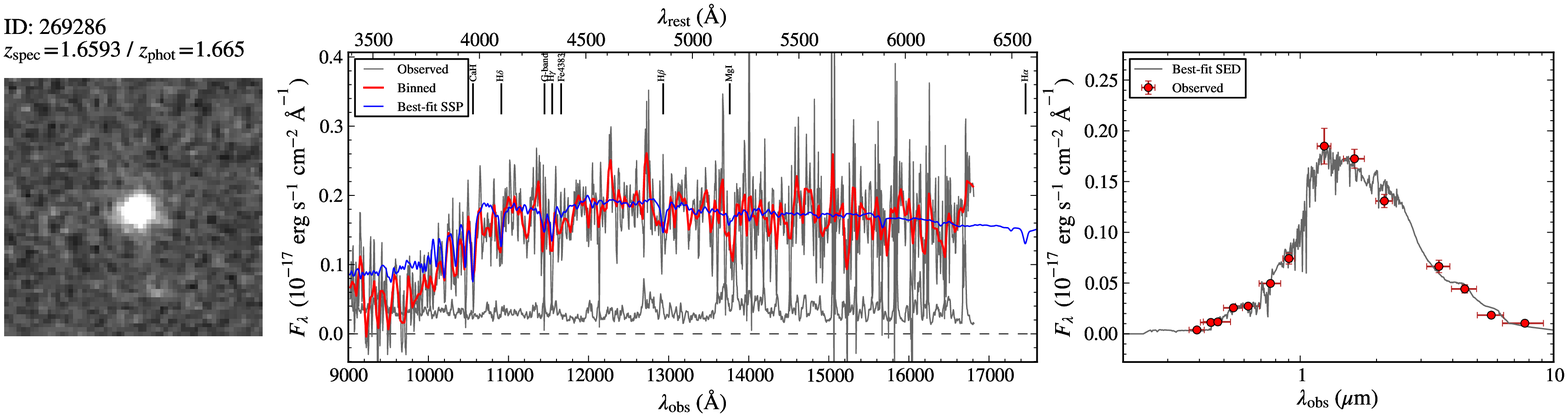}
    \includegraphics[width=0.95\linewidth]{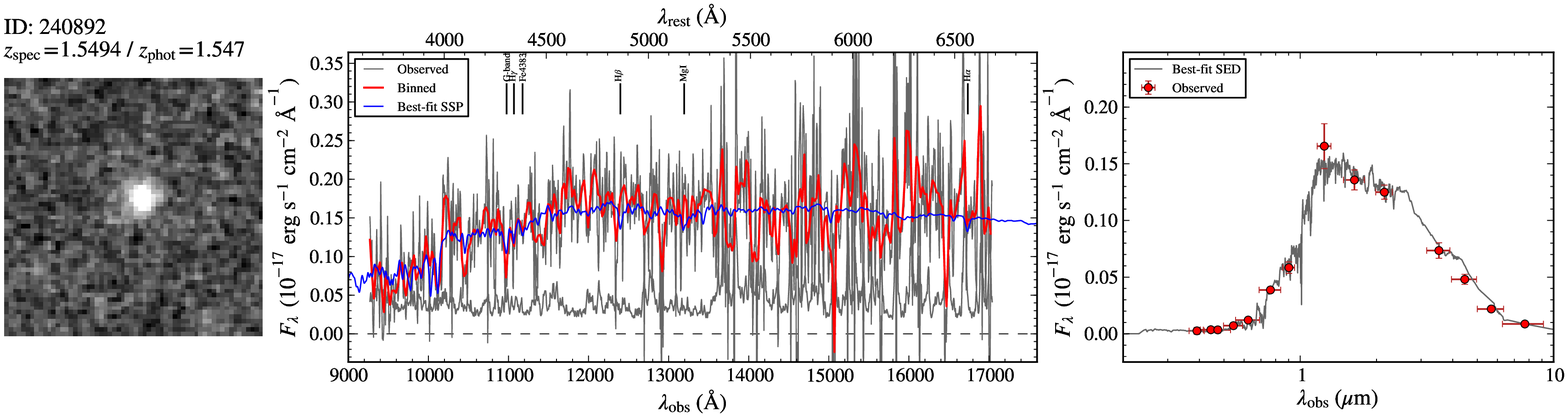}
    \includegraphics[width=0.95\linewidth]{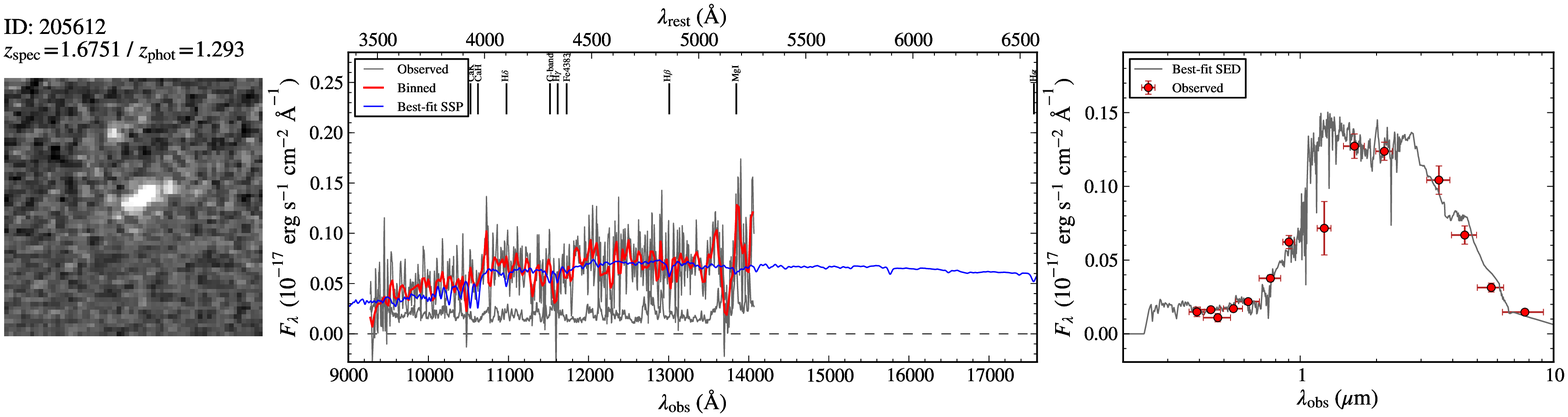}
    \includegraphics[width=0.95\linewidth]{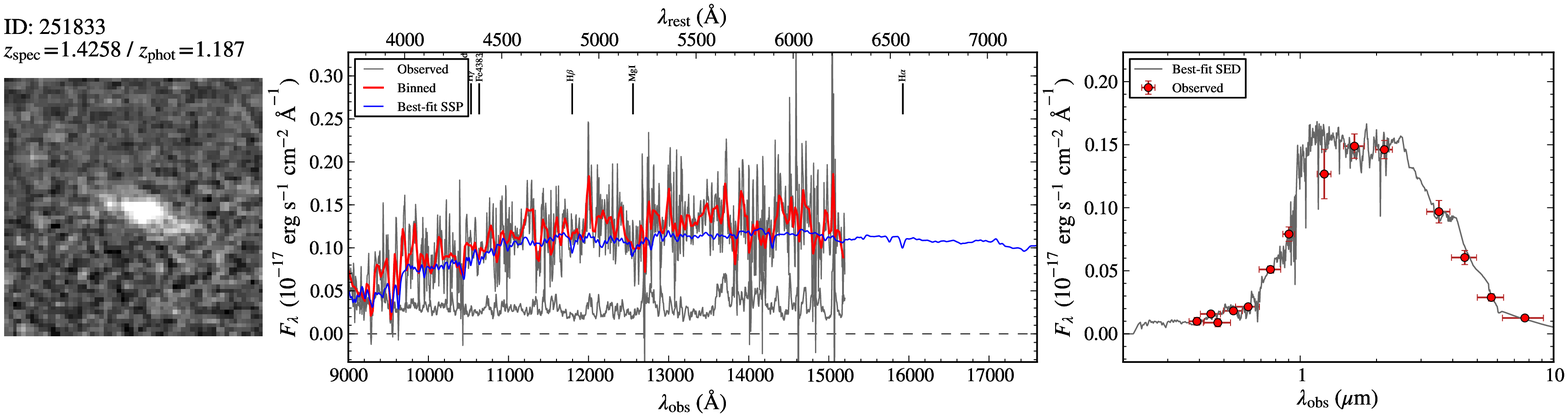}
  \end{center}
  \caption{\textit{Continued.}}
\end{figure*}

\addtocounter{figure}{-1}
\begin{figure*}[htbp]
  \begin{center}
    \includegraphics[width=0.95\linewidth]{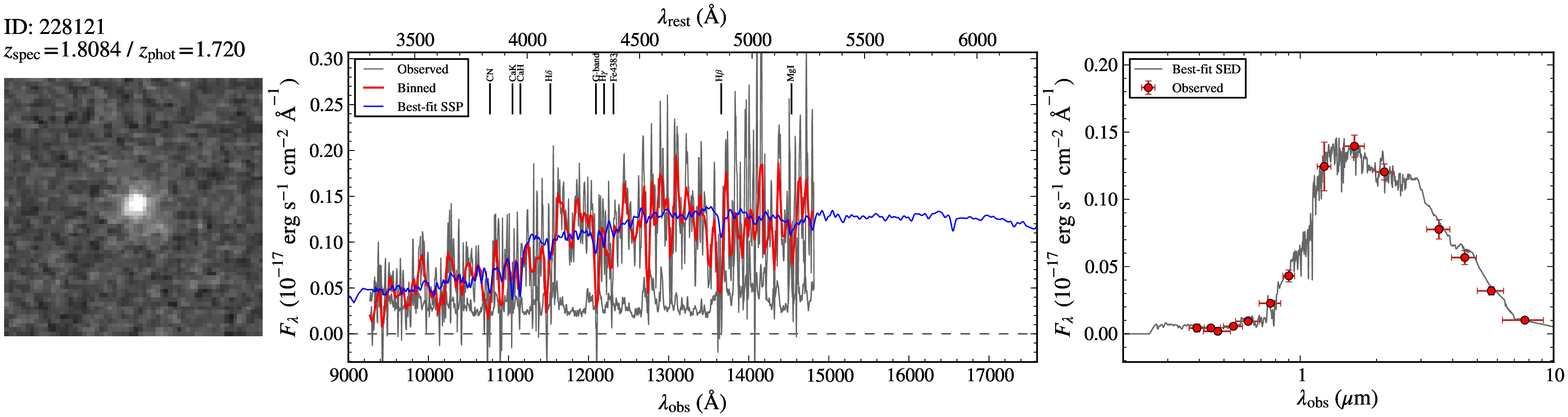}
    \includegraphics[width=0.95\linewidth]{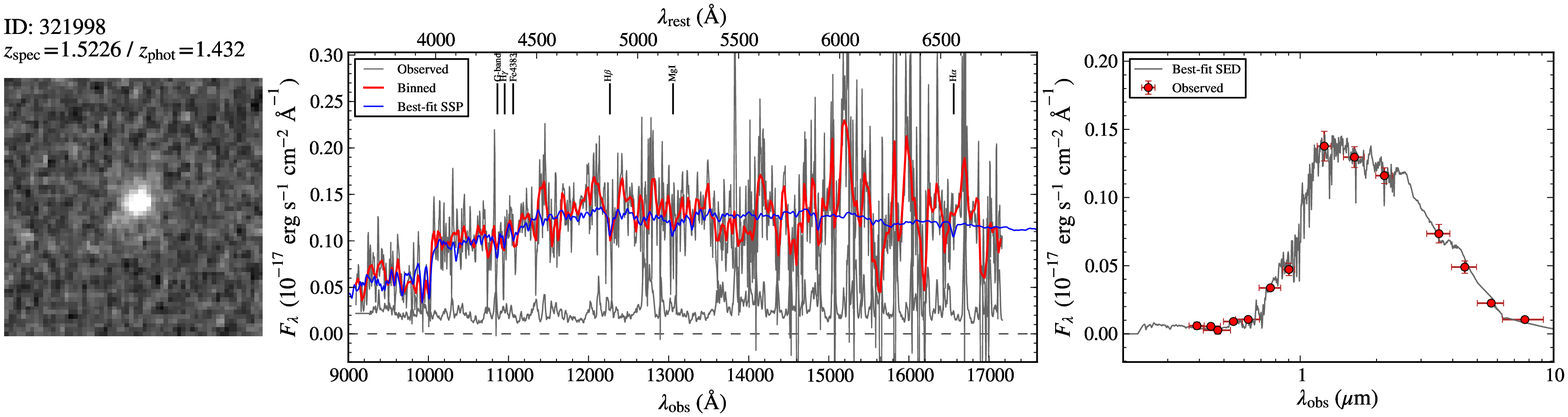}
    \includegraphics[width=0.95\linewidth]{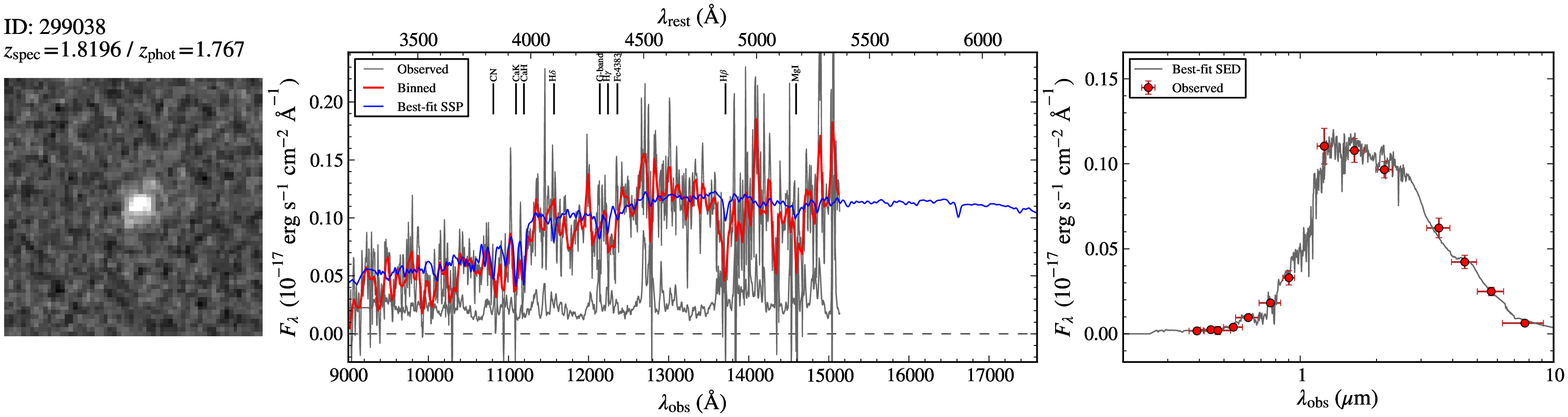}
    \includegraphics[width=0.95\linewidth]{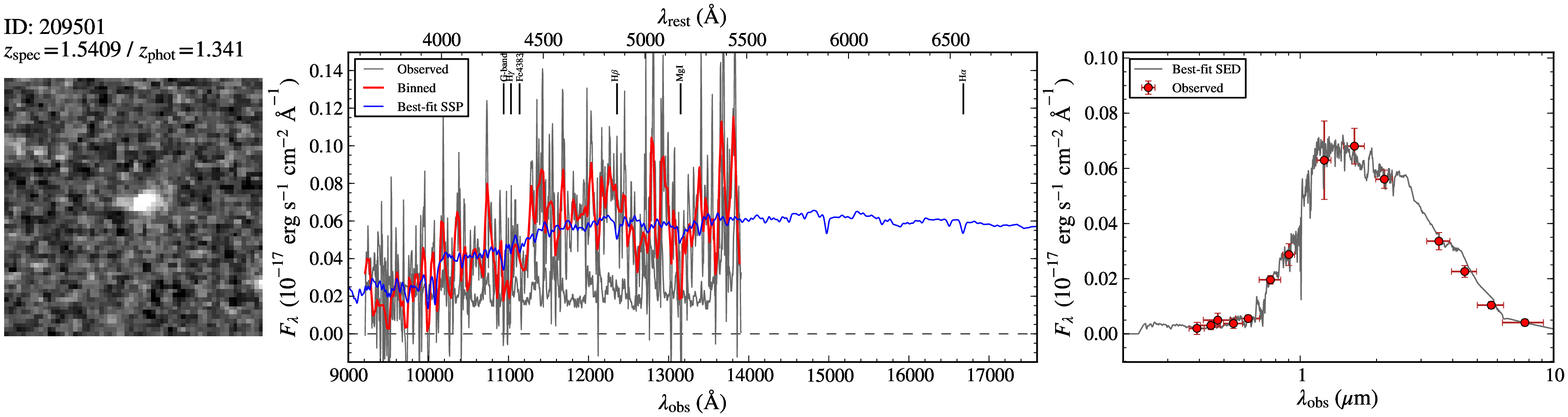}
  \end{center}
  \caption{\textit{Continued.}}
\end{figure*}

\addtocounter{figure}{-1}
\begin{figure*}[htbp]
  \begin{center}
    \includegraphics[width=0.95\linewidth]{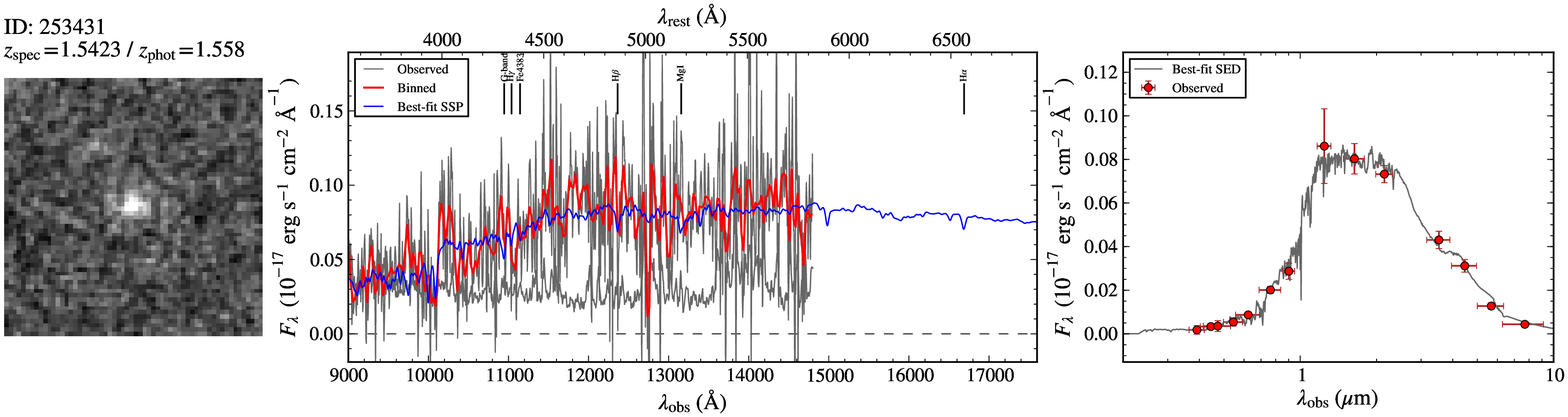}
    \includegraphics[width=0.95\linewidth]{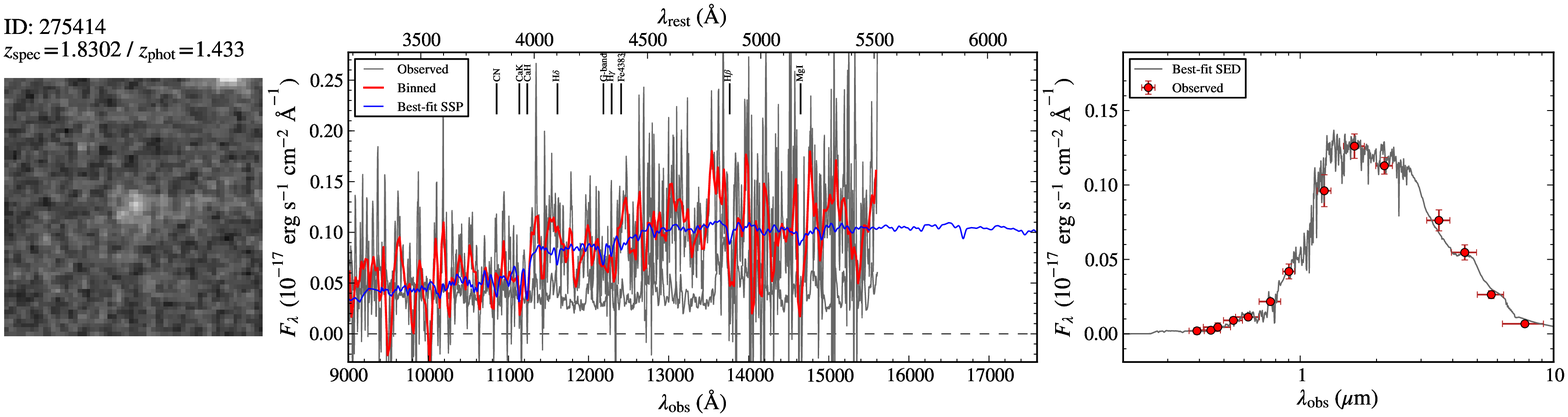}
  \end{center}
  \caption{\textit{Continued.}}
\end{figure*}

\section{Spectroscopic Redshifts}
\label{sec:speczz}

\subsection{Redshift Measurement}
\label{sec:specz}

None of the target objects shows emission lines in the observed
spectral range, which supports their identification as PEGs.  The
measurement of spectroscopic redshifts was therefore attempted from
stellar absorption features such as \cahk, the 4000 \AA{} break and
Balmer lines.  The spectroscopic redshifts were then derived using the
Penalized Pixel-Fitting method \citep[\ppxf;][]{cappellari:2004:ppxf},
using template stellar spectra from the MILES stellar library
\citep{sanchezblazquez:2006:miles}. We have paid special attention to
the error analysis, including systematic effects, in particular those
that could be introduced by the spurious sky residuals and the
correlated noise. Thus, we have adopted the standard bootstrapping
technique of resampling residuals to adjacent groups of spectral
pixels, rather than to the individual pixels, to account for the
correlated noise, then randomly reshuffling the best-fitting residuals
in groups of 50 pixels (300 \kms), allowing for duplication.  These
residuals were added to the galaxy spectrum, then fitting the data
with \ppxf.  To account for the sensitivity of the fits to the
polynomial degree for each noise realization we adopted a random
degree between 0 and 4 for the additive and another random one for the
multiplicative polynomials.  In the new fits we repeated the entire
\ppxf procedure, namely redetermining the set of best fitting MILES
stellar templates thus finding a new set of $\sigma$-clipped
residuals.  This procedure was repeated for 300 random realizations
and $1\sigma$ errors were derived from the distribution of the output
redshifts from the realizations.  The results are reported in Table
\ref{tab:propspec} along with their errors.

As a cross-check, redshifts were also derived by
cross-correlating with simple stellar population (SSP) templates from
\citet[][hereafter BC03]{bruzual:2003} with various ages and
metallicities after broadening to $\sigma=300$ \kms.
For all the objects, the resulting redshifts are consistent within
$1\sigma$ errors with those derived using stellar templates. 

The 18 identified objects are shown with squares in Figure
\ref{fig:cmd_pbzks}.  Since the identification is nearly complete (16
out of 19 objects) at the brighter magnitude, $\mathit{K}\lesssim21$,
the primary factor of the spectroscopic identification in this kind of
observations appears to be the total \textit{K}-band magnitude.
Moreover, since we have selected all \pbzks{} brighter than
$\mathit{K}<21$ in the observed FoVs, the completeness of the
spectroscopic identification is 85\%{} down to this depth.

A redshift confidence class was assigned to each object according to
the following criteria: Class 4 is assigned to the objects which
clearly show both the 4000 \AA{} break and some absorption lines;
Class 3 is assigned to objects with a clear 4000 \AA{} break but
without unambiguously recognizable absorption lines; Class 2 refers to
redshifts derived from the overall shape of the continuum, i.e., the
4000 \AA{} break is not so prominent; Class 1 is for an insecure
redshift. The assigned redshift classes are reported in Table
\ref{tab:propspec}. A redshift class $\ge 1$ was assigned to the 18
objects for which the continuum could be detected, whereas for 16
objects no continuum was detected and they are listed separately in
Table \ref{tab:propphot}.

%
%
\begin{deluxetable}{ccccc}
  \tablewidth{0pt}
  \tablecolumns{4}
  \tablecaption{Spectroscopic Properties\label{tab:propspec}}
  \tablehead{
    \colhead{ID} &
    \colhead{$z_\text{spec}$} &
    \colhead{Class\tablenotemark{a}} &
    \colhead{Dn4000} &
    \colhead{H$\delta_\text{F}$}
    \\
    \colhead{} &
    \colhead{} &
    \colhead{} &
    \colhead{} &
    \colhead{(\AA)}
  }
  \startdata
  254025  & $1.82283    \pm 0.00055    $ & 4 & $1.40 \pm 0.03$ & $ \phm{-}2.84 \pm 1.30$ \\
  217431  & $1.42769    \pm 0.0015\phn $ & 4 & $2.03 \pm 0.10$ & $ \phm{-}6.23 \pm 1.10$ \\
  307881  & $1.42904    \pm 0.00089    $ & 4 & $2.06 \pm 0.08$ & $ \phm{-}5.59 \pm 1.15$ \\
  233838  & $1.8199\phn \pm 0.0016\phn $ & 4 & $1.53 \pm 0.05$ & $ \phm{-}2.95 \pm 1.30$ \\
  277491  & $1.8163\phn \pm 0.0038\phn $ & 3 & $2.74 \pm 0.25$ & $ \phm{-}1.30 \pm 1.51$ \\
  313880  & $1.4486\phn \pm 0.0018\phn $ & 2 & $1.42 \pm 0.09$ & $       -2.66 \pm 1.78$ \\
  250093  & $1.8270\phn \pm 0.0010\phn $ & 4 & $1.49 \pm 0.10$ & $ \phm{-}6.47 \pm 1.22$ \\
  263508  & $1.52122    \pm 0.00094    $ & 2 & $1.47 \pm 0.09$ & $ \phm{-}5.40 \pm 1.20$ \\
  269286  & $1.6593\phn \pm 0.00058    $ & 3 & $1.38 \pm 0.06$ & $ \phm{-}4.86 \pm 1.19$ \\
  240892  & $1.54939    \pm 0.00092    $ & 2 & $1.55 \pm 0.10$ & $ \phm{-}9.03 \pm 0.84$ \\
  205612  & $1.6751\phn \pm 0.0045\phn $ & 2 & $1.52 \pm 0.09$ & $       -3.92 \pm 1.76$ \\
  251833  & $1.42578    \pm 0.00057    $ & 2 & $1.71 \pm 0.18$ & $       -1.80 \pm 1.69$ \\
  228121  & $1.8084\phn \pm 0.0015\phn $ & 2 & $1.38 \pm 0.12$ & $ \phm{-}2.84 \pm 1.52$ \\
  321998  & $1.52263    \pm 0.00087    $ & 3 & $2.13 \pm 0.14$ & $       -1.25 \pm 1.59$ \\
  299038  & $1.81957    \pm 0.00098    $ & 3 & $2.06 \pm 0.12$ & $       -4.18 \pm 1.77$ \\
  209501  & $1.5401\phn \pm 0.0078\phn $ & 1 & $2.02 \pm 0.33$ & $ \phm{-}2.07 \pm 1.72$ \\
  253431  & $1.5423\phn \pm 0.0037\phn $ & 1 & $1.61 \pm 0.21$ & $       -1.47 \pm 1.72$ \\
  275414  & $1.8302\phn \pm 0.0147\phn $ & 1 & $2.01 \pm 0.24$ & $       -1.59 \pm 1.63$ \\
  \tableline
  Stacked & \nodata                      &   & $1.54 \pm 0.03$ & $ \phm{-}4.26 \pm 0.40$
  \enddata
  \tablenotetext{a}{
    Class 4: a spectrum  with a clear detection of both absorption lines and the 4000 \AA{} break. 
    Class 3: a spectrum  with a clear detection of the 4000 \AA{} break. 
    Class 2: a spectrum  with a relatively high-S/N continuum
    detection whose overall shape allows redshift measurement. 
    Class 1: a spectrum  with a low-S/N continuum detection for which the derived redshift is less reliable.
 }
\end{deluxetable}

\subsection{Comparison between Spectroscopic and Photometric Redshifts}
\label{sec:compz}

\begin{figure}[htbp]
  \begin{center}
    \includegraphics[width=0.95\linewidth]{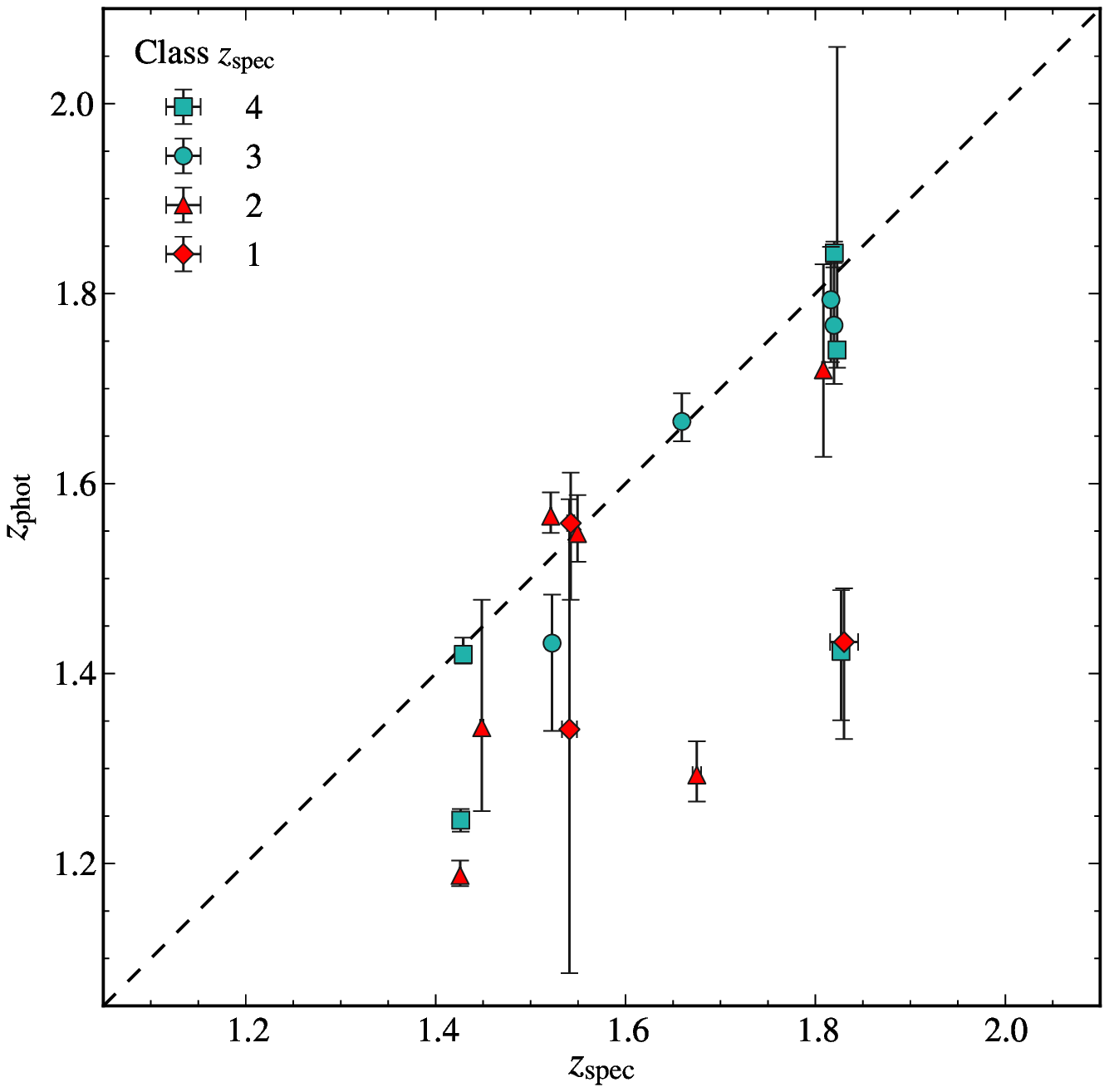}
  \end{center}
  \caption{
    Comparison between the spectroscopic redshifts of the present
    study and the photometric redshifts from
    \citet{ilbert:2009}. Different symbols refer to the quality Class
    of the derived spectroscopic redshifts as defined in the text and
    shown on the top-left corner of the figure.
  }
  \label{fig:zcomp}
\end{figure}

Figure \ref{fig:zcomp} shows the comparison between the spectroscopic
redshifts derived above and photometric redshifts from
\citet{ilbert:2009}.  This is virtually the first attempt to
systematically test the COSMOS 30-band photometric redshift against
measured spectroscopic redshift of PEGs at $z>1.4$.  A majority of the
objects show fairly good agreement between photometric and
spectroscopic redshifts, but several outliers also exist.  Of course,
such outliers can be ascribed to either of the two methods to measure
redshifts, and here we briefly try to identify the (main) culprit.

\begin{figure}[htbp]
  \begin{center}
    \includegraphics[width=0.95\linewidth]{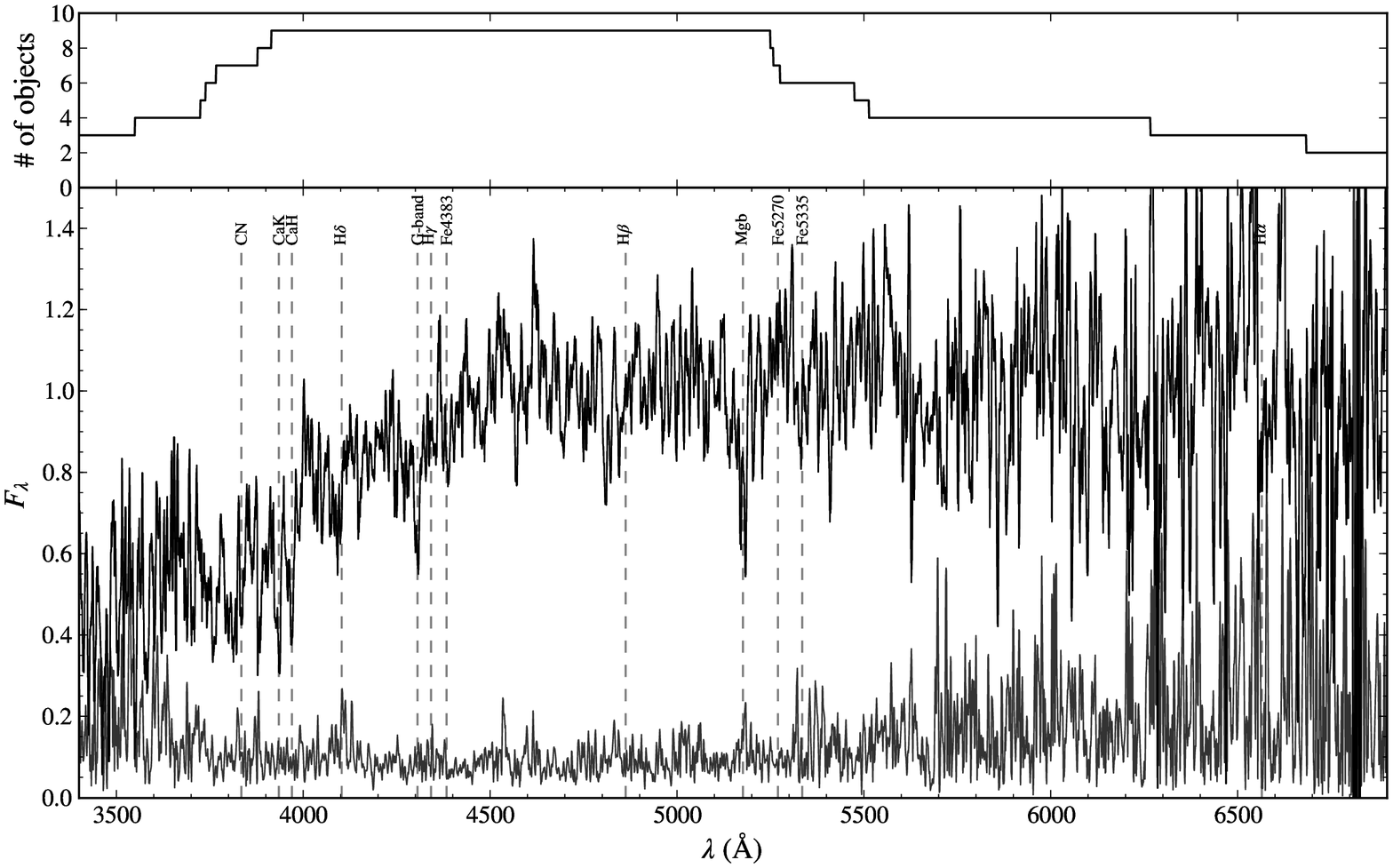}
  \end{center}
  \caption{
    The composite spectrum of up to 9 \pbzks whose redshifts  have
    Class 1 and 2  (i.e., lower quality). 
    The bottom panel shows the composite spectrum and its $1\sigma$ noise.
    The locations of major absorption lines are labeled and indicated by dashed-lines. 
    The top panel shows the number of objects used for stacking at each wavelength. 
  }
  \label{fig:stack12}
\end{figure}

First notice that two outliers have quite secure spectroscopic
redshift, having assigned Class 4 (namely, objects 217431 and 250093).
For these objects the photometric redshifts are clearly in error.  In
Figure \ref{fig:stack12} we show the stacked spectrum (see Section
\ref{sec:stacking}) of the 9 objects with low confidence class (1 and
2): several strong features are clearly present in this spectrum,
namely Ca II H\&K, the G band and Mg$b$, and these features are still
recognizable in the stacked spectrum including only the five Class 1
and 2 outliers seen in Figure \ref{fig:zcomp}.  We conclude that the
spectroscopic redshifts are correct also for the majority of these two
classes of objects and therefore in the present sample $\sim30$\%{} of
the PEG photometric redshifts are systematically underestimated.

More quantitatively, the average offsets are $\langle
z_\text{spec}-z_\text{phot}\rangle = 0.12$ or $\langle
(z_\text{spec}-z_\text{phot})/(1+z_\text{spec})\rangle=0.04$, and the
standard deviations of $z_\text{spec}-z_\text{phot}$ and
$(z_\text{spec}-z_\text{phot})/(1+z_\text{spec})$ are 0.14 and 0.053,
respectively.  The normalized median absolute deviation
\citep[NMAD;][]{hoaglin:1983}, $\sigma_\text{NMAD}=1.48\times
\text{median}(\left|z_\text{spec}-z_\text{phot}\right|/(1+z_\text{spec}))$,
which is used to estimate the accuracy of the COSMOS photometric
redshift by \citet{ilbert:2009}, is calculated as 0.050.  This is
similar to $\sigma_\text{NMAD}=0.054$ for the $1.5<z<3$ zCOSMOS-Deep
sample with $i^+\simeq24$ \citep{lilly:2007}.  However, they found
that $\sim 20$\%{} of the 147 zCOSMOS-Deep galaxies at $1.5<z<3$ show
catastrophic failure defined as
$\left|z_\text{spec}-z_\text{phot}\right|/(1+z_\text{spec})>0.15$,
whereas no such large catastrophic failures are found in the \pbzk
sample presented here.  Still, these results indicate that the
photometric redshifts from \citet{ilbert:2009} would lead to a
systematic underestimate of the volume density of high redshift PEGs,
though the present statistics is too limited to precisely quantify the
effect.

Notice that all our \pbzk selected objects with detected continuum
have $z_\text{spec}>1.4$,  including those with $z_\text{phot}<1.4$,
indicating that the passive \textit{BzK}-selection is indeed quite
effective to identify \textit{bona fide} 
PEGs at $z>1.4$ (see also Section \ref{sec:photoz}), 
especially when disregarding mid- and/or far-IR detected objects.

We also note that the clustering of the spectroscopic redshifts around
a few redshift spikes (including Class 1 and 2 redshifts, see Section
\ref{sec:overdensity}) lends support to the correct identification of
such redshifts.  No such clustering would indeed be expected from
random errors in the spectroscopic redshifts, unless spurious breaks
were introduced by the flux calibration, which does not appear to be
the case.

\subsubsection{Improving photometric redshifts for high redshift PEGs}
\label{sec:photoz}

  To explore the possibility of improving the performance of
  photometric redshifts, specifically for high-redshift PEGs, we have
  calculated new photometric redshifts using the 18 spectroscopic
  redshifts in this work as a training sample.  We used EAZY
  \citep{brammer:2008} to fit the 14-passband SEDs
  (\textit{UBgVrizJHKs} and the 4 IRAC bands, see Section
  \ref{sec:sedfitting}) of the 18 PEGs, and different sets of
  templates, namely the EAZY default templates
  \citep[cf.][]{brammer:2008,brammer:2011,whitaker:2011}, as well as
  passive synthetic populations built with \citet[][hereafter
  M05]{maraston:2005} and BC03 models, with a suitable range of ages
  and metallicities.

  Following \citet{ilbert:2009}, we estimated systematic photometric
  offsets relative to the default EAZY templates by running EAZY on
  the 18 PEGs having fixed the redshift to their spectroscopic value.
  This was done by measuring, for each band, the
  median offset between the observed flux of the galaxies and the flux
  of the best-fitting templates, iteratively correcting all passbands
  until convergence. We note that, besides
correcting for possible systematic errors in the photometric zero-points,
aperture corrections, and/or in the filter transmission
curves, such systematic offsets are specific to the adopted set of
  templates and reflect their possible limitations. Nonetheless, we
  show below that adopting  the systematic offsets determined
with the EAZY templates we obtain consistent photometric redshift
estimates also when using  the passive BC03 and M05 template sets.

\begin{figure*}[htbp]
  \begin{center}
    \includegraphics[width=0.95\linewidth]{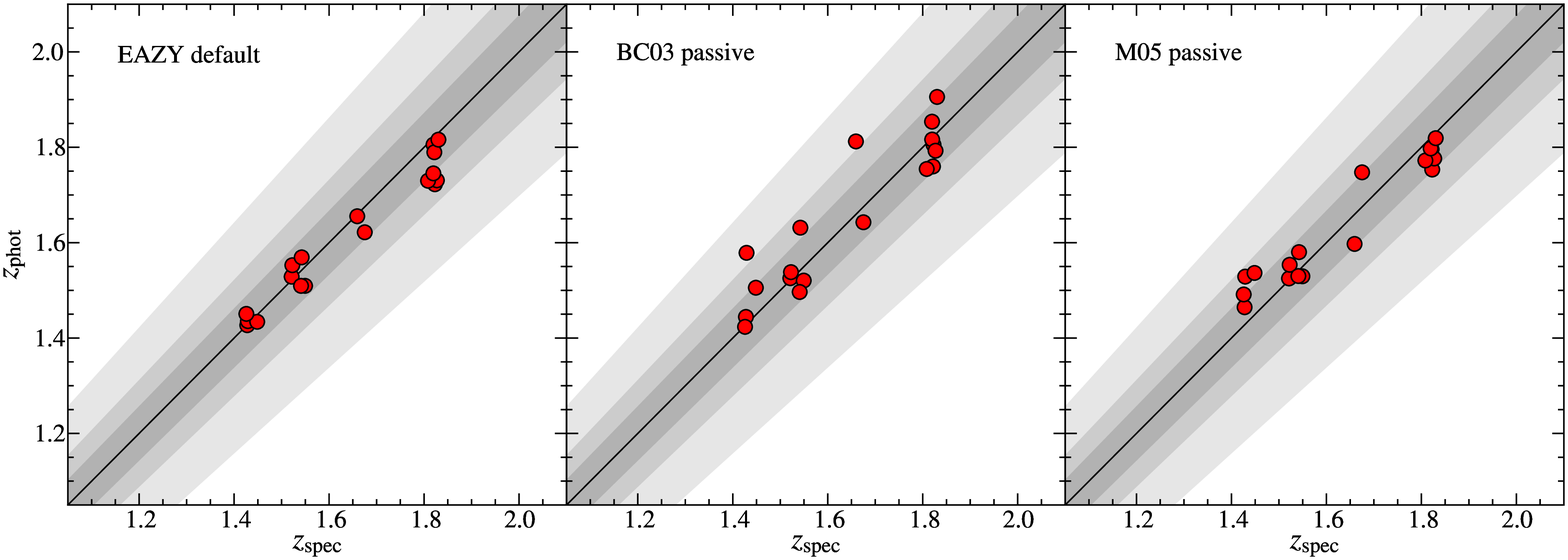}
  \end{center}
  \caption{
    Comparison between the spectroscopic redshifts from the present
    study and the corresponding photometric redshifts from different determinations: 
    \textit{Left}: $z_\text{phot}$ from this work using standard EAZY templates; 
    \textit{Middle and Right}: $z_\text{phot}$ from this work using EAZY with passive BC03 and M05 models, respectively 
    (see Section \ref{sec:photoz}). 
    All EAZY photometric redshifts were obtained using the 18 PEG
    spectroscopic redshifts in the present study as a training sample. 
    The solid line traces the bisector whereas the filled areas show deviations of $0.025$, $0.05$, and $0.1\times(1+z)$. 
    Readers can refer to Figure \ref{fig:zcomp} for a comparison with the photometric redshift by \citet{ilbert:2009}. 
  }
  \label{fig:zcomp2}
\end{figure*}

In Figure \ref{fig:zcomp2} the resulting $z_\text{phot}$ values are
compared to the corresponding $z_\text{spec}$ for the sample of our 18
PEGs, separately for each of the three template sets. The improvement
with respect to the original photometric redshifts is immediately
evident.  Using EAZY default templates the median of the discrepancy
$(z_\text{spec}-z_\text{phot})/(1+z_\text{spec})$ is 0.006, with a
scatter $\sigma_\text{NMAD} = 0.017$.  When using M05 and BC03
templates the median of the discrepancy is 0.004 and 0.0009,
respectively and $\sigma_\text{NMAD} \simeq 0.02$ in both cases.

Of course, we are comparing $z_\text{spec}$ and $z_\text{phot}$ for
the same 18 galaxies used as training sample to determine the
systematics offsets. However, within the limits due to the poor
statistics, similarly small median discrepancies and scatters are
found when using a random half of the sample for training and the
other half for checking $z_\text{spec}$ vs.  $z_\text{phot}$.  This
suggests that the quoted accuracy of the new photometric redshifts is
in fact reliable, at least for sources with similar SED, magnitudes
and redshift as the PEGs considered here. 
We note that the Ilbert et al.\ photometric redshifts were optimized
for $z\lesssim 1.3$, using a set of medium bands at optical wavelengths.
While they reached better than $\sim 1\%$ accuracy up to $z\sim 1.3$ for bright sources, 
the performance for fainter sources ($\mathit{i}>24$) and particularly 
for galaxies at $z\gtrsim 1.3$ is significantly degraded \citep{ilbert:2009}. 
The larger systematic offset
and scatter of the original photometric redshifts \citep{ilbert:2009} 
illustrated in Figure \ref{fig:zcomp} may at least partly be 
ascribed to the lack of PEGs in the zCOSMOS-Deep spectroscopic sample \citep{lilly:2007} 
that was used as a training sample, as it was designed to contain only star-forming galaxies.

The 18 PEGs we used for tuning photometric redshifts and determining
their accuracy are clearly on the bright tail of the general PEG
population at these redshifts. Nonetheless, assuming that photometric
redshifts derived as above remain reasonably accurate also for fainter
PEGs, we  measured the photometric redshifts
for all the \pbzk   MIPS-undetected  galaxies in the COSMOS sample
\citep{mccracken:2010} and compared their distribution to that
obtained with the original photometric redshifts from \citet{ilbert:2009}. 

\begin{figure}[htbp]
  \begin{center}
    \includegraphics[width=0.95\linewidth]{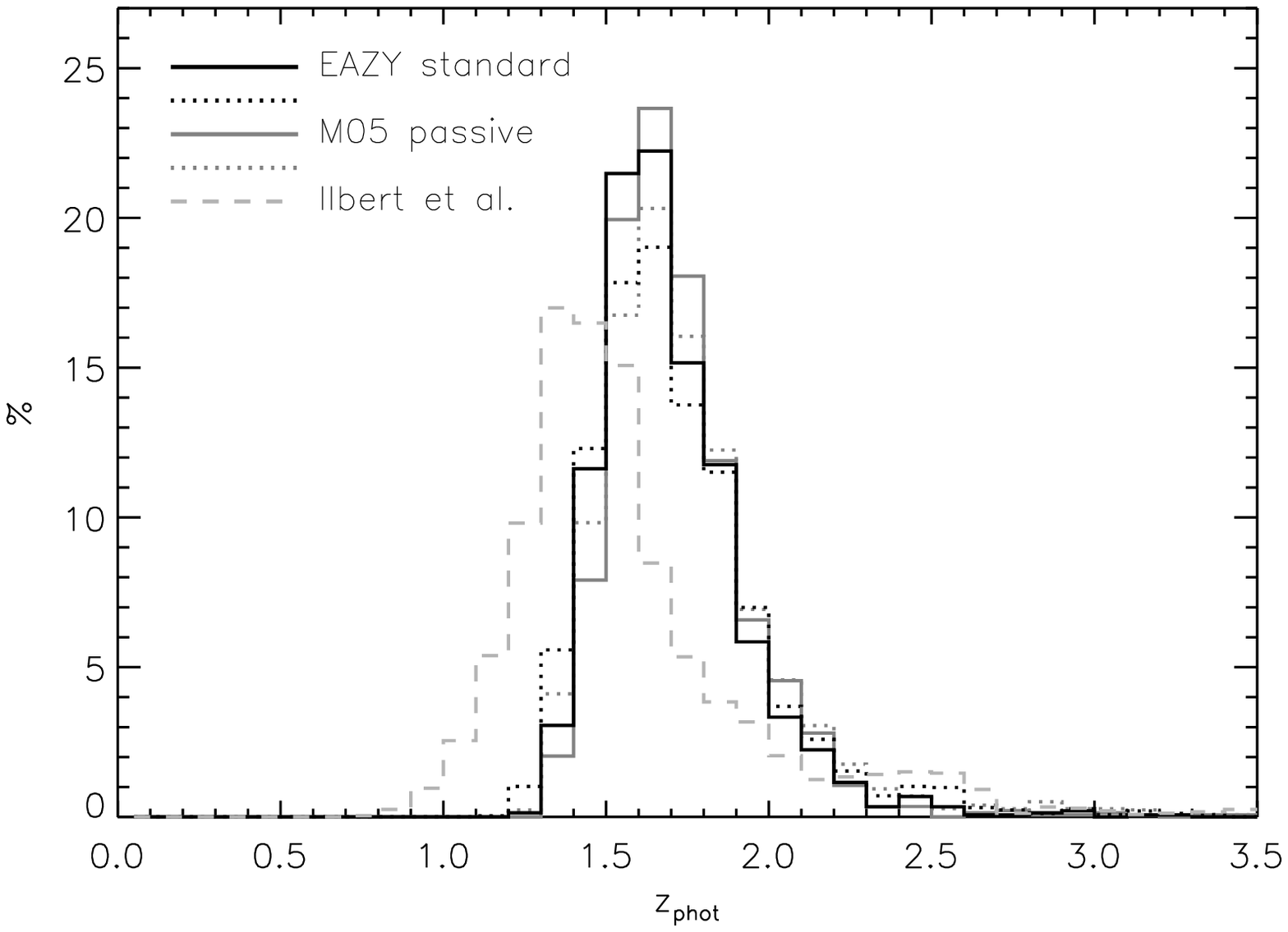}
  \end{center}
  \caption{ The distribution of photometric redshifts for
    \pbzk-selected PEGs.  The solid/dotted lines show the redshift
    distribution of 50\%/90\%{} of the full PEG sample over the COSMOS
    field using the photometric redshifts measured in this work using
    different sets of templates (standard EAZY and passive M05
    templates in black and dark gray, respectively). The dashed
    light-gray line shows the redshift distribution of 85\%{} of the
    full PEG sample using \citet{ilbert:2009} photometric redshifts.
    All EAZY photometric redshifts were obtained using the 18 PEG
    spectroscopic redshifts in this paper as a training sample.  }
 \label{fig:Nz}
\end{figure}

Figure \ref{fig:Nz} shows the $z_\text{phot}$ distribution of 90\%{}
of the full sample, for which a reliable $z_\text{phot}$ could be
obtained as judged from the $\chi^2$ of the best-fit and from the
$z_\text{phot}$ probability distribution (dotted black and dark-gray
lines, for $z_\text{phot}$ estimated with standard EAZY and passive
M05 templates, respectively). The solid lines show the distribution of
the 50\%{} of the sample with best constrained $z_\text{phot}$ (i.e.,
with the best $\chi^2$ values and tightest probability
distribution). The dashed light-gray line shows the distribution of
photometric redshifts from \citet{ilbert:2009}, for 85\%{} of their
full sample (after removing 2\% of the sources unmatched in the
\citeauthor{ilbert:2009} catalog or identified as X-ray sources, and a
further 13\%{} of sources located in masked areas). Apart from a small
shift in the peak of the distribution (from $z\sim1.5$ to $z\sim1.7$),
it is worth emphasizing that about one third of the sources are
located at $z_{\rm phot}<1.4$ according to the
\citeauthor{ilbert:2009}, compared to only a few percent with the new
determination trained on the present 18 spectroscopic PEGs.
This exercise shows that using spectroscopic high-$z$ PEGs as a training sample 
can substantially improve the estimation of photometric redshifts for this kind of galaxies, 
as illustrated by a comparison of Figure \ref{fig:zcomp} and Figure \ref{fig:zcomp2}.

\subsection{Note on a MIPS 24 \micron{} source}
\label{sec:mips}
\begin{figure}[htbp]
  \begin{center}
    \includegraphics[width=0.95\linewidth]{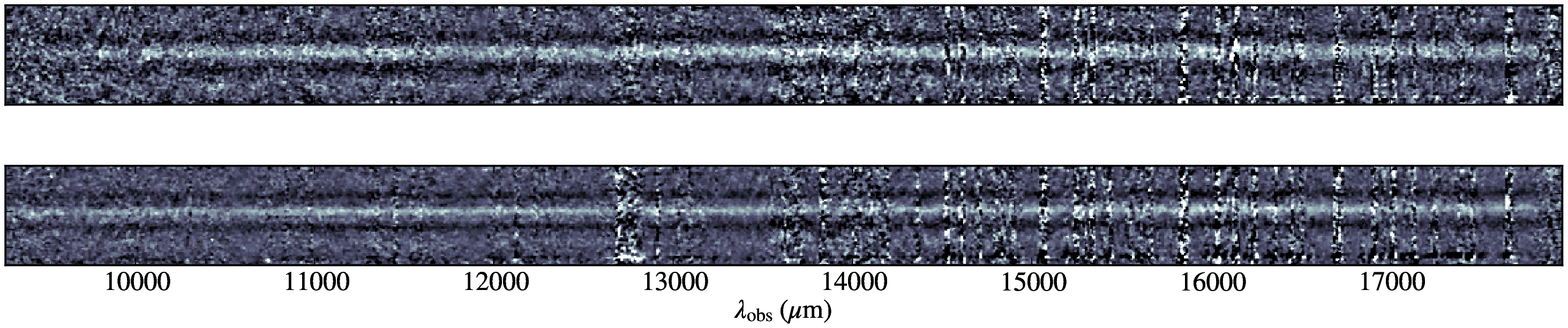}
  \end{center}
  \caption{
    The two dimensional spectra of the MIPS $24\,\mu\text{m}$ detected
    source, 313880, taken on February 6,  2010 (\textit{Top}) and on
    April 1,  2010 (\textit{Bottom}). 
    The  spectra are 1.9 arcsec wide in the spatial direction. 
  }
  \label{fig:2d313880}
\end{figure}

One of our galaxies pre-selected as a \pbzk, i.e., 313880 at $z=1.45$,
is detected at \spitzer/MIPS 24 \micron{} with a flux of 130
$\mu\text{Jy}$ which corresponds to a SFR of $\sim 90
M_\odot\text{yr}^{-1}$, adopting a conversion based on infrared SED
templates of \citet{chary:2001} and the Chabrier initial mass function
\citep[IMF;][]{chabrier:2003}. The \hst/ACS $i$-band (F814W filter)
morphology in Figure \ref{fig:pbzk_hst_spec_sed} suggests that this
object has a clumpy disk structure which could be the site of intense
star formation \citep{forsterschreiber:2011:clump2,genzel:2011}.
Surface profile fitting (see Section \ref{sec:size}) also shows this
object is best fit by an exponential profile.  Looking at the spectra
in Figure \ref{fig:pbzk_hst_spec_sed}, there seems to be an excess of
flux at $\sim 1.6$ \micron{}, indicating the presence of \hasp
emission as expected from the relatively high 24 \micron{} flux and
infrared-based SFR.  However, in the 2D spectra taken during two
nights separately as shown in Figure \ref{fig:2d313880}, there is no
evidence of emission lines and the excess is apparently caused by the
residual of the OH sky line subtraction.  Other emission lines such as
[\ion{O}{3}]$\lambda\lambda 4959,5007$ and \hbsp are also expected in
the \textit{J}-band where the sky line contamination is less severe
than in the \textit{H}-band, but none of them are
detected.  The other explanation for non-detection of emission lines
is that the object has large amount of dust which attenuates the
emission lines below the detection limit, ($3$--$10$)$\times10^{-17}$ erg
s$^{-1}$ cm$^{-2}$ ($3\sigma$).  The SED fit gives
$A_V=0.55^{+0.34}_{-0.45}$ mag which is one of the highest among the
objects in the present sample, but it does not seem enough to
attenuate emission lines below detection limits.  Given the degeneracy
of parameters in SED fitting, the derived $A_V$ could possibly be
underestimated in this particular case.  Here we conclude that this
object is probably a star-forming galaxy suffering from heavy dust
extinction, and therefore, we excluded it from the stacking analysis
(see Section \ref{sec:stacking}).

\section{Overdensities of Passively Evolving Galaxies at $\lowercase{z}>1.4$}
\label{sec:overdensity}

\begin{figure}[htbp]
  \begin{center}
    \includegraphics[width=0.95\linewidth]{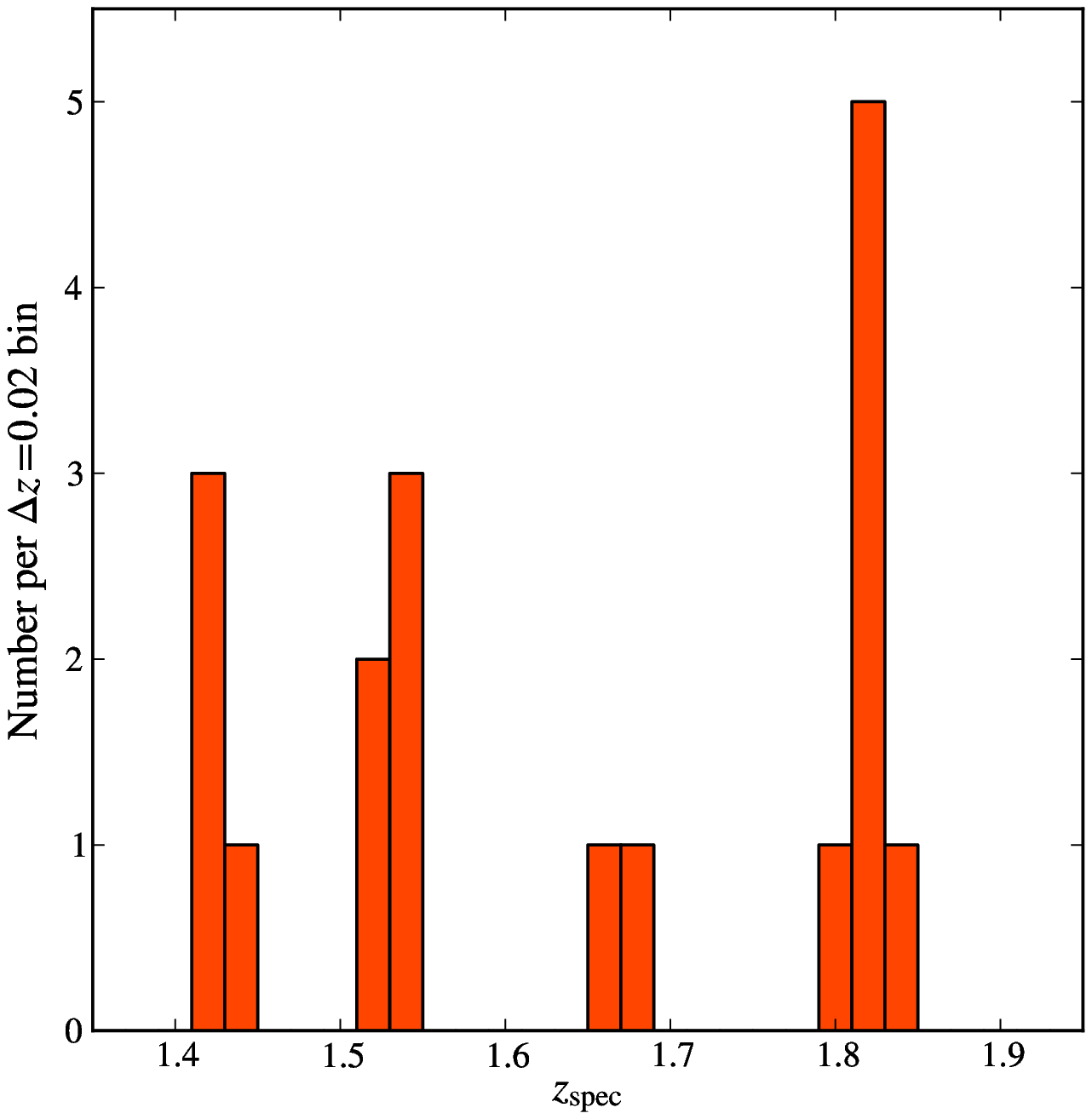}
  \end{center}
  \caption{
    The redshift distribution of spectroscopically identified \pbzks in the present sample. 
  }
  \label{fig:zhist}
\end{figure}

A cursory inspection of Figure \ref{fig:zcomp} is sufficient to
realize that spectroscopic redshifts are not randomly distributed, but
instead tend to cluster around a few spikes. Indeed, Figure
\ref{fig:zhist} shows the redshift distribution of the 18
spectroscopically identified \pbzks in our 3 masks.  There is a clear
redshift spike of 7 \pbzks at $z\simeq 1.82$ with $1\sigma$ scatter of
0.006.  The projected spatial extent of the overdensity is $\sim 6$
Mpc in diameter and $z=1.82\pm0.006$ corresponds to $\sim 19$ comoving
Mpc along the line of sight.  To our knowledge this is the first
spectroscopically confirmed overdensity of PEGs at such high redshift.

Another possible redshift-spike at $z\simeq 1.43$ includes 4 \pbzks.
Although the significance does not seem high, this $z=1.43$ spike is
also noticeable because two of its galaxies are the \textit{K}-band brightest
\pbzks over the whole COSMOS field \citep{mancini:2010}.

A third spike appears to be present at $z\simeq 1.53$, including 5
\pbzks, and the last two galaxies have also close redshifts, around
$z\simeq 1.67$. Unfortunately, this area lies outside the zCOSMOS-Deep
field, and therefore it is not possible to use its redshifts of
star-forming galaxies to check for the presence of spikes at the same
redshifts of those reported here. However, a fair number of \sbzks was
included in our MOIRCS masks (cf. Table \ref{tab:obslog}) and a
comparison of the redshift distributions of \pbzks and \sbzks will be
done in a future paper.

\begin{figure}[htbp]
  \begin{center}
    \includegraphics[width=0.95\linewidth]{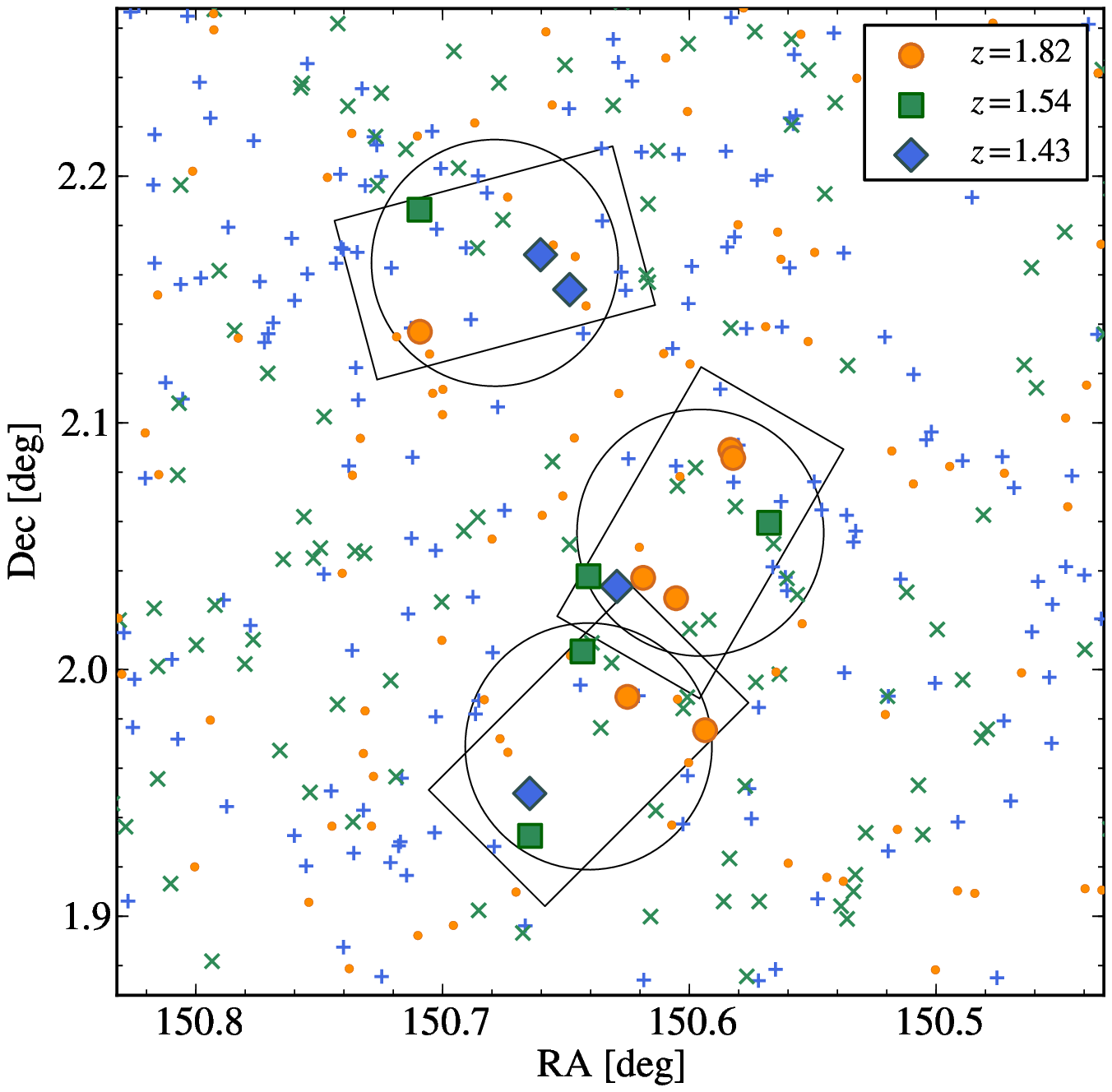}
  \end{center}
  \caption{ The spatial distribution of the \pbzks in the three major
    redshift spikes. The symbols are redshift-coded as indicated in
    the insert.  Small blue pluses, green crosses, and orange circles
    show objects with $K<23$ and photometric redshift within $\Delta z
    = \pm 0.02$ from the mean redshifts of the spikes at $z=1.43$,
    $1.54$, and $1.82$, respectively.  Solid boxes and circles
    represent the position of MOIRCS detectors and masks,
    respectively.  }
  \label{fig:map_spike}
\end{figure}

Figure \ref{fig:map_spike} shows the location on the sky of the
galaxies in these three redshift spikes at $z=1.43$, $1.53$, and
$1.82$, together with that of other photo-$z$ selected galaxies with
$\Delta z = \pm 0.02$ and $K<23$ from the mean redshift of each spike.
There may well be large spikes at these redshifts extending spatially
well beyond the area explored by the three MOIRCS FoVs, but our data
do not allow us to claim the presence of bound clusters.

The \textit{XMM-Newton} data \citep{hasinger:2007,finoguenov:2007}
indicate that there is no extended X-ray emission associated to these
redshift spikes down to ($6$--$8$)$\times10^{-15}$ erg\ s$^{-1}$\
cm$^{-2}$ ($2\sigma$) in $0.5$--$2$ keV band, which corresponds to the
$2\sigma$ limit of $L_\text{X}\simeq(1$--$2)\times10^{44}$ erg\ s$^{-1}$
and $M_{200}\simeq1.5\times10^{14}\,M_\odot$ at the redshifts of the
spikes (A.~Finoguenov, private communication).

\section{The Composite Spectrum of \pbzks}
\label{sec:stacking}

A composite spectrum has been constructed by stacking 17 out of the 18
spectroscopically identified \pbzks.  We excluded the object 313880
because of the MIPS $24\,\mu\text{m}$ detection (see Section
\ref{sec:mips}), which could cause a contamination of a star-forming
galaxy into a passively evolving population.  We did not exclude
objects with Class 1 and 2 since the spectroscopic identifications
appear to be correct as investigated in Section \ref{sec:compz} and
low-S/N does not affect the result after weighting as described below.

After being de-redshifted to the rest-frame, each individual spectrum
was first normalized by the mean flux at rest-frame $4500<\lambda
(\text{\AA})<5200$.  Then the spectra were linearly interpolated into
a 1 \AA{} linearly spaced wavelength grid.  The associated noise
spectra were normalized by the same factor as that for the object
spectrum and interpolated to the rest-frame in quadrature.  The
spectra were then coadded with weights proportional to the inverse
square of the noise spectra at each rest-frame wavelength.

To estimate a realistic error and any biases related to the
stacking, we applied the jackknife method to the spectra used for
stacking.  We made 17 composite spectra in the same way, but removing
one object at a time from the stacking. Then the standard deviation of the flux
at each wavelength pixel is estimated as
\begin{equation}
  \sigma_\text{Jack}^2 = \frac{N-1}{N}\sum_{i=1}^{N}\left(f - f_{(i)}\right)^2, 
\end{equation}
where $N$ is a number of objects in the sample (i.e., $N=17$ here), $f$ is a flux
of composite spectrum of $N$ spectra, and $f_{(i)}$ is the flux of the
composite spectrum made of $(N-1)$ spectra by removing $i$-th spectrum.  We
use $\sigma_\text{Jack}$ as the $1\sigma$ error of the
composite spectrum.  The composite spectrum was also corrected for a
sampling bias derived from the jackknife estimator as
\begin{equation}
  f' = f - \left(N-1\right)\left(\langle f_{(i)}\rangle - f\right),
\end{equation}
where $\langle f_{(i)}\rangle$ is the average of $f_{(i)}$. The typical
correction factor due to the bias estimate is $<2$\%.  We adopt
the bias corrected spectrum $f'$ as the final composite spectrum,
which is inevitably dominated by the objects with highest
S/N ratios as well as by observed-frame wavelength less contaminated
by the sky residuals.

\begin{figure*}[htbp]
  \begin{center}
    \includegraphics[width=0.8\linewidth]{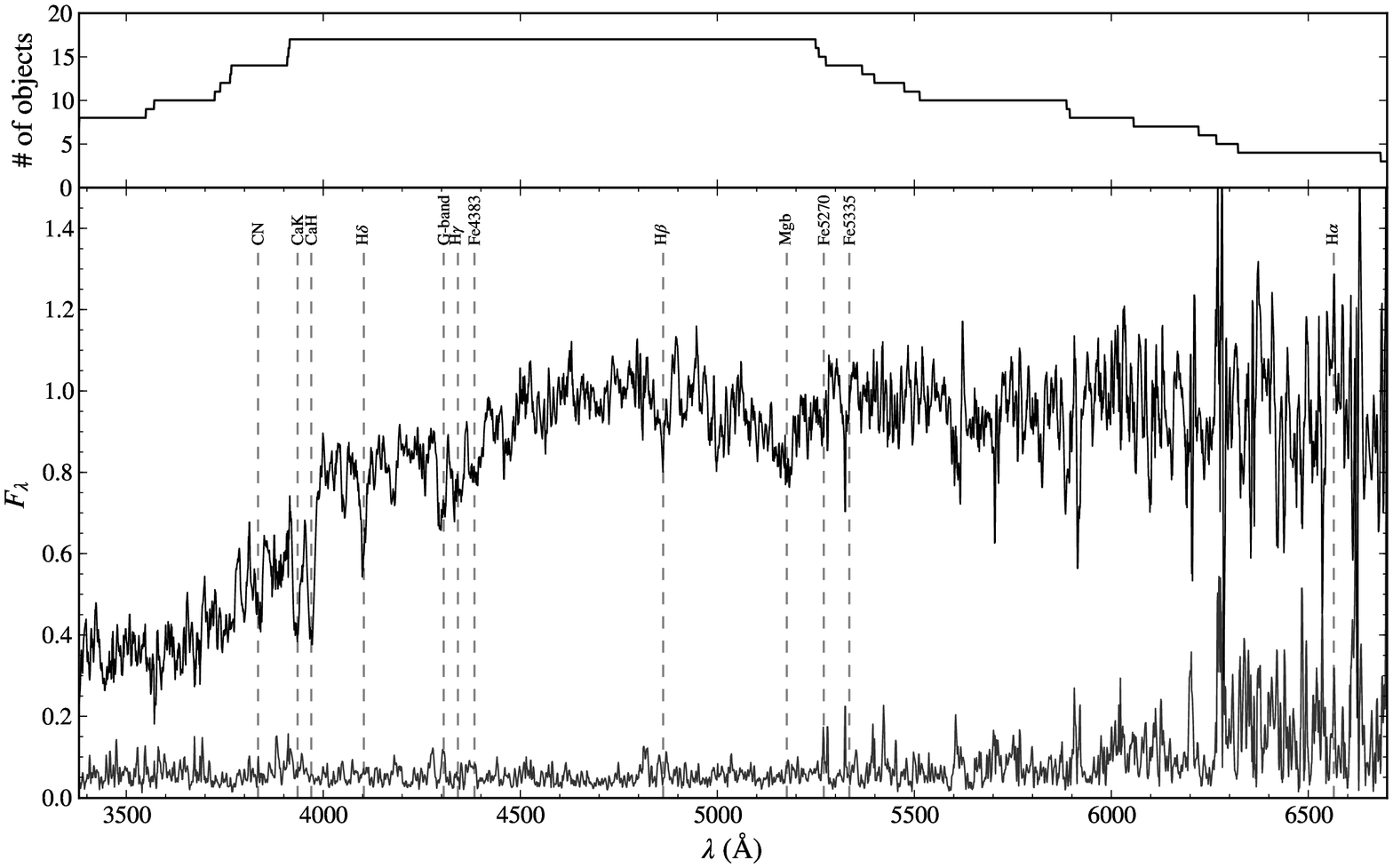}
  \end{center}
  \caption{
    The same as Figure \ref{fig:stack12}, but now co-adding the
    spectra of all 17 MIPS-undetected \pbzks in the present sample. 
  }
  \label{fig:composite}
\end{figure*}

The composite spectrum obtained in this way (equivalent to a MOIRCS
integration time of about 140 hours) is shown in Figure
\ref{fig:composite}.  At least 11 spectral lines/features are clearly
visible in this spectrum, and identified in Figure
\ref{fig:composite}; specifically, CN, \ion{Ca}{2} H\&K, H$\delta$,
the G band, H$\gamma$, Fe4383, Fe5270, Fe5335, H$\beta$, and
Mg$b$. Although the number of stacked objects are small at longer wavelength 
where contamination from the residual OH-airglow subtraction is severe, 
there could be an indication of excess flux at the location of \ha. 
We have measured the equivalent width (EW) of the potential \hasp emission line 
from the stacked spectrum divided by the best-fit model (see Section \ref{sec:ana_comp}) 
accounting the underlying absorption, and we found $\text{EW}(\text{\ha})=3.5$--$5$ \AA{} 
depending on the width of the integration window to derive the EW (1--2.5 times of FWHM of the instrument).  
In case of $EW=5$ \AA, it can be translated into $\text{SFR}=5.6\,M_\odot\,\text{yr}^{-1}$ 
and $1.9\,M_\odot\,\text{yr}^{-1}$ for a object at $z=1.69$ with 
the brightest and median H-band magnitudes, respectively.  
In addition to \ha, we have also carried out a similar measurement 
on the position of [\ion{O}{2}]$\lambda3727$, and found corresponding SFR of $4\,M_\odot\,\text{yr}^{-1}$ 
and $1.6\,M_\odot\,\text{yr}^{-1}$ for the brightest and median cases 
with an error of about 20-40\%.  Looking at the individual spectra extending to \ha, 
307881 is the brightest and could mainly contribute to this feature in the composite spectrum.  
However, the feature around \hasp{} in the spectra of 307881 appears to be rather broad, 
mostly contaminated by adjacent OH residuals.  Indeed, when we remove this object from the stacking, 
the feature disappeared, while the estimated SFR at the position of [\ion{O}{2}] still remains the same.  
Therefore, the residual star formation, if any, could be at most $\sim5\,M_\odot\,\text{yr}^{-1}$ 
and most likely $<1$--$2\,M_\odot\,\text{yr}^{-1}$, which is $>100$--$200$ times 
smaller than those of galaxies with $M_\star=10^{11}\,M_\odot$ 
on the star formation main sequence at $z=2$ \citep[e.g.,][]{daddi:2007:sfr,pannella:2009,rodighiero:2011}.
The composite spectrum is analyzed is Section \ref{sec:ana_comp}.

\section{Stellar Population Properties of the Program Galaxies}
\label{sec:stellarpopulation}

\subsection{Broad-band SED Fitting}
\label{sec:sedfitting}
Having fixed the redshifts to their spectroscopic values, we have
derived the physical properties of the galaxies from their broad-band SEDs. 
Besides the Subaru/Suprime-Cam \textit{B} and \textit{z'} band and CFHT/WIRCam $K$-band photometry 
to apply the \textit{BzK} selection technique, 
we used the  multi-band photometry including the  CFHT/MegaCam \textit{u} band,  the
Subaru/Suprime-Cam \textit{g'}, \textit{V}, \textit{r'}, and \textit{i'} bands, 
and the CFHT/WIRCam \textit{J} and \textit{H} bands as well as the four \spitzer/IRAC bands.  
All magnitudes from \textit{u} to $K$  are converted to total magnitudes by applying 
aperture corrections both for point sources and for extended objects
\citep[][]{mccracken:2010,mancini:2011}. 
The IRAC band magnitudes are corrected to the total magnitude following \citet{ilbert:2009}. 

The SED fitting was carried out by using FAST \citep[Fitting and
Assessment of Synthetic Templates;][]{kriek:2009}.  FAST includes
internal dust reddening and provides various physical quantities of
the galaxies such as stellar mass, age, and metallicity.  For the
templates, we used the composite stellar population models generated
from the SSPs of Charlot \& Bruzual (2007, in preparation; hereafter
CB07) with the Chabrier IMF.
It has been argued that the IMF may not be universal among local early-type galaxies, 
but could depend on the galaxy stellar mass \citep{cappellari:2012}. 
If so, the masses derived here would be underestimated by up to a factor 2--3 for the most massive objects. 
We assume a universal IMF in this paper, as it was done in all other similar studies to which we compare our results.

The adopted star formation histories (SFHs) assume an exponentially
declining SFR with various $e$-folding times ($\tau$) ranging from
$\log(\tau/\text{yr})=7.0$ to $10$ with an interval of
$\Delta\log\tau=0.1$ and ages ranging between
$\log\text{(age/yr)}=7.0$ to $10$ with an interval of
$\Delta\log\text{age}=0.05$.  The metallicity $Z$ is fixed for each
model (no attempt is made at mimicking chemical evolution) and is
chosen from the set $Z=0.004$, $0.008$, $0.02$, and $0.05$.  Dust
extinction was also applied following the recipe by
\citet{calzetti:2000} with $A_V=0$--$3$ mag with an intervals of
$0.05$ mag.  Ages are required to be less than the age of the Universe
at the observed redshift.  The time-like free parameters in the fit
are therefore the age and $\tau$.

Then the best values and 68\%{} confidence intervals of the stellar
mass, timescale of the star-formation ($\tau$), age from the onset of
star formation, reddening, and metallicity for each galaxy are derived
based on the likelihood distribution of each template.  The
SFR-weighted ages are also computed by using the best-fit $\tau$ and
age.  Derived physical parameters are listed in Table
\ref{tab:propsed} and the observed galaxy SEDs and the best-fit
templates are shown in the right panels of Figure
\ref{fig:pbzk_hst_spec_sed}.

All but a few of the \pbzks have SEDs consistent with $A_V<0.4$ or
$E(B-V)<0.1$, i.e., as expected the derived dust extinction is very
small for most of the objects.  When considering that reddening and
age are partly degenerate one can conclude that none of our objects
suffers major dust obscuration, as expected for passive galaxies.
Stellar population age of the \pbzks are typically $\gtrsim 2$
Gyr.  
All best-fit SEDs have a short \textit{e}-folding time
$\tau\lesssim500\,\text{Myr}$ which is typically much shorter than the
derived age, i.e., the present star formation is already well
suppressed. Stellar masses range from $\sim 4\times 10^{10}\,M_\odot$
to $\sim 4\times 10^{11}\,M_\odot$, whereas best fit metallicities are
near-solar or slightly sub-solar with large uncertainties that cover
well above the solar metallicity.

It has been recently argued that an increasing SFR is more appropriate
than a declining one in the case of high-redshift star-forming
galaxies \citep[e.g.,][]{renzini:2009,maraston:2010,papovich:2011}.
In particular, an exponentially increasing SFR has been suggested,
with $\text{SFR}\propto \tau^{-1}\exp(+t/\tau)$ where
$\tau\simeq500$~Myr.  In the case of \pbzks such SFR needs to be
truncated at some point, in order mimic the quenching of their star
formation.  Most galaxies in the present sample show a strong 4000
\AA{} break and some of them strong Balmer absorption lines (see
Section \ref{sec:lidx}), indicating that the age of the dominant
stellar populations must be $\gtrsim 1$ Gyr, hence star formation has
to be quenched $\gtrsim 1$ Gyr before the observed epoch.  As shown in
Table \ref{tab:propsed}, assuming instead a declining SFR the best fit
star formation timescale turns out to be typically about 10 times
shorter than the age of the galaxies which is well in excess of 1 Gyr.
Therefore, for these quenched galaxies both exponentially declining
and increasing star formation scenarios would result in almost
identical stellar population properties being selected by the best fit
procedure.

%
%
\begin{deluxetable*}{ccccccc}
  \tablewidth{0pt}
  \tablecolumns{7}
  \tablecaption{Physical Parameters from Broad-band SED Fitting\label{tab:propsed}}
  \tablehead{
    \colhead{ID} &
    \colhead{$\log M_\star$} &
    \colhead{$\log \tau$} &
    \colhead{$\log t_0$\tablenotemark{a}} &
    \colhead{$\log t_\text{sf}$\tablenotemark{b}} &
    \colhead{$A_V$} &
    \colhead{$Z$} 
    \\
    \colhead{} &
    \colhead{($M_\odot$)} &
    \colhead{(yr)} &
    \colhead{(yr)} &
    \colhead{(yr)} &
    \colhead{(mag)} &
    \colhead{}
  }
  \startdata
 254025 & $11.39_{-0.03}^{+0.15}$ & $8.20_{-0.15}^{+0.20}$ & $9.10_{-0.11}^{+0.20}$ & $9.04$ & $0.20_{-0.20}^{+0.09}$ & $0.008_{-0.004}^{+0.018}$  \\
 217431 & $11.57_{-0.15}^{+0.03}$ & $8.70_{-0.25}^{+0.01}$ & $9.60_{-0.23}^{+0.00}$ & $9.54$ & $0.25_{-0.10}^{+0.25}$ & $0.008_{-0.004}^{+0.011}$  \\
 307881 & $11.50_{-0.11}^{+0.03}$ & $8.60_{-0.24}^{+0.04}$ & $9.55_{-0.17}^{+0.04}$ & $9.50$ & $0.20_{-0.11}^{+0.15}$ & $0.008_{-0.003}^{+0.008}$  \\
 233838 & $11.39_{-0.25}^{+0.16}$ & $8.30_{-1.30}^{+0.22}$ & $9.30_{-0.31}^{+0.20}$ & $9.25$ & $0.20_{-0.15}^{+0.34}$ & $0.008_{-0.004}^{+0.013}$  \\
 277491 & $11.31_{-0.12}^{+0.11}$ & $8.20_{-0.11}^{+0.35}$ & $9.05_{-0.06}^{+0.31}$ & $8.98$ & $0.60_{-0.59}^{+0.05}$ & $0.020_{-0.013}^{+0.030}$  \\
 313880 & $10.85_{-0.27}^{+0.19}$ & $8.10_{-1.10}^{+0.48}$ & $9.05_{-0.13}^{+0.34}$ & $9.00$ & $0.55_{-0.45}^{+0.34}$ & $0.008_{-0.004}^{+0.016}$  \\
 250093 & $11.00_{-0.08}^{+0.13}$ & $8.10_{-0.14}^{+0.35}$ & $9.05_{-0.12}^{+0.26}$ & $9.00$ & $0.20_{-0.12}^{+0.41}$ & $0.020_{-0.014}^{+0.013}$  \\
 263508 & $10.86_{-0.21}^{+0.10}$ & $7.90_{-0.90}^{+0.59}$ & $9.00_{-0.11}^{+0.38}$ & $8.96$ & $0.35_{-0.35}^{+0.22}$ & $0.020_{-0.015}^{+0.015}$  \\
 269286 & $11.01_{-0.28}^{+0.02}$ & $8.40_{-1.40}^{+0.12}$ & $9.40_{-0.43}^{+0.06}$ & $9.35$ & $0.00_{-0.00}^{+0.16}$ & $0.004_{-0.000}^{+0.017}$  \\
 240892 & $11.04_{-0.13}^{+0.19}$ & $8.40_{-1.40}^{+0.22}$ & $9.40_{-0.17}^{+0.20}$ & $9.35$ & $0.00_{-0.00}^{+0.24}$ & $0.020_{-0.013}^{+0.005}$  \\
 205612 & $10.90_{-0.06}^{+0.13}$ & $8.30_{-0.11}^{+0.15}$ & $9.15_{-0.07}^{+0.11}$ & $9.08$ & $0.20_{-0.20}^{+0.10}$ & $0.020_{-0.004}^{+0.014}$  \\
 251833 & $10.74_{-0.05}^{+0.27}$ & $8.20_{-0.35}^{+0.43}$ & $9.10_{-0.22}^{+0.35}$ & $9.04$ & $0.30_{-0.20}^{+0.57}$ & $0.020_{-0.014}^{+0.019}$  \\
 228121 & $11.14_{-0.03}^{+0.14}$ & $8.40_{-0.20}^{+0.21}$ & $9.35_{-0.10}^{+0.20}$ & $9.30$ & $0.00_{-0.00}^{+0.30}$ & $0.020_{-0.014}^{+0.010}$  \\
 321998 & $11.03_{-0.09}^{+0.12}$ & $8.50_{-0.14}^{+0.21}$ & $9.45_{-0.12}^{+0.15}$ & $9.40$ & $0.05_{-0.05}^{+0.28}$ & $0.020_{-0.012}^{+0.016}$  \\
 299038 & $11.05_{-0.15}^{+0.08}$ & $8.40_{-1.40}^{+0.12}$ & $9.40_{-0.42}^{+0.10}$ & $9.35$ & $0.10_{-0.10}^{+0.34}$ & $0.008_{-0.002}^{+0.017}$  \\
 209501 & $10.62_{-0.13}^{+0.19}$ & $8.40_{-1.40}^{+0.30}$ & $9.35_{-0.35}^{+0.25}$ & $9.30$ & $0.00_{-0.00}^{+0.41}$ & $0.020_{-0.013}^{+0.012}$  \\
 253431 & $10.73_{-0.30}^{+0.23}$ & $8.40_{-1.40}^{+0.32}$ & $9.35_{-0.43}^{+0.25}$ & $9.30$ & $0.35_{-0.35}^{+0.36}$ & $0.008_{-0.004}^{+0.018}$  \\
 275414 & $11.05_{-0.24}^{+0.16}$ & $8.30_{-1.30}^{+0.23}$ & $9.30_{-0.35}^{+0.18}$ & $9.25$ & $0.20_{-0.20}^{+0.37}$ & $0.008_{-0.004}^{+0.018}$
  \enddata
  \tablenotetext{a}{Age from the onset of star-formation.}
  \tablenotetext{b}{Age weighted by SFR derived as $\int_0^{t_0} (t_0-t)\psi(t)dt/\int_0^{t_0}\psi(t)dt$, where $\psi(t)$ is the SFR and $t_0$ and $\tau$ are the best-fit age and $\tau$, respectively.}
\end{deluxetable*}

\subsection{Rest-frame \textit{UVJ} colors}
\label{sec:uvj}

\begin{figure}[htbp]
  \begin{center}
    \includegraphics[width=0.95\linewidth]{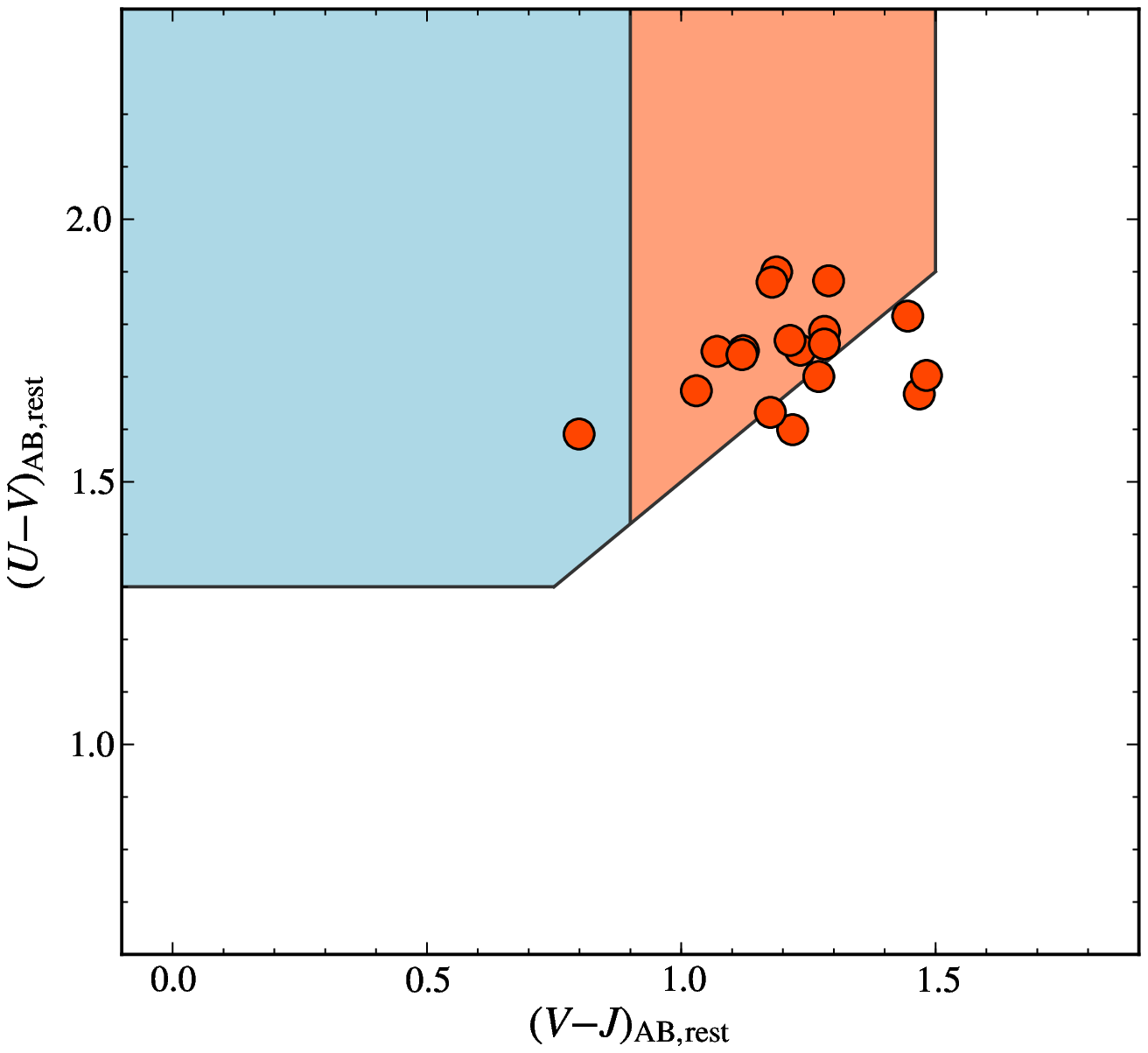}
  \end{center}
  \caption{ The rest-frame
    $(\mathit{U}-\mathit{V})$--$(\mathit{V}-\mathit{J})$ diagram for
    the 18 \pbzks in the sample (circles).  The colored areas
    represent the selection box for young (blue) and old (red)
    passively evolving galaxies, respectively, following the
    definition by \citet{whitaker:2012}.  }
  \label{fig:uvj}
\end{figure}

Passively evolving galaxies are known to lie in a distinct region on
rest-frame $(\mathit{U}-\mathit{V})$ vs. $(\mathit{V}-\mathit{J})$
two-color diagram \citep[e.g.,][]{wuyts:2007,whitaker:2011}.  This can
be used to check the passive nature of \pbzks in the present sample.
Figure \ref{fig:uvj} shows the \textit{UVJ} diagram where we have
adopted the identical filter combination as that used in
\citet{whitaker:2011}, namely the Johnson \textit{UBV} system
\citep{maizapellaniz:2006} and the 2MASS \textit{J} band.  The
rest-frame \textit{UVJ} colors were computed by convolving the
best-fit templates with the filter response curves without applying
atmospheric and instrumental throughputs.  The selection box for
passive galaxies is defined as
$(\mathit{U}-\mathit{V})>0.8\times(\mathit{V}-\mathit{J})+0.7$,
$(\mathit{U}-\mathit{V})>1.3$, and $(\mathit{V}-\mathit{J})<1.5$, and
young and old galaxies are further separated at
$(\mathit{V}-\mathit{J})=0.9$ following \citet{whitaker:2012}.

As shown in Figure \ref{fig:uvj}, most of objects have rest-frame
\textit{UVJ} colors consistent with passive evolution.  The colors of
some outliers are still within $\sim 0.2$ mag from the selection box,
and have an offset consistent with possible systematics in the SED
fitting and photometric errors.  In fact, galaxies with stellar ages
of $\gtrsim 1$ Gyr can still be found around
$(\mathit{U}-\mathit{V})\simeq1.7$ and
$(\mathit{V}-\mathit{J})\simeq1.5$ \citep{whitaker:2012}.  The MIPS 24
\micron source, 313880, is well within the selection box for passive
galaxies, which suggests that the object may be a contaminant to the
\textit{UVJ} selection, or is affected by systematics in the SED
fitting.  From this \textit{UVJ} color test we conclude that in all 18
objects star formation has been already quenched, or is very low.

\subsection{Rest-frame Optical Absorption Line Indices and the Dn4000 Index}
\label{sec:lidx}

To further characterize the stellar population content of the \pbzks
in this sample some of the Lick indices and the strength of the 4000
\AA\ break were measured on our rest-frame optical spectra.  To this
end we used the \texttt{Lick\_EW} program as a part of the
\texttt{EZ\_Ages} IDL code
package\footnote{\url{http://astro.berkeley.edu/\~{}graves/ez\_ages.html}}
\citep{graves:2008,schiavon:2007}, following the definition of the
Lick indices by \citet{worthey:1994} and \citet{worthey:1997}.  For
this purpose one needs to specify a stellar velocity dispersion and
the instrument dispersion: we adopted 27 \AA{} in the observed frame
for the FWHM of the instrument configuration, and a stellar velocity
dispersion of $\sigma_*=270$ \kms{} for 254025 (see Section
\ref{sec:veldisp}).  For all other objects we used the stellar
mass--stellar velocity dispersion relation of local galaxies at
$0.05<z<0.07$, determined by using stellar velocity dispersions from
the NYU Value-Added Galaxy Catalog \citep{blanton:2005} and stellar
masses from the MPA-JHU catalog \citep{kauffmann:2003,salim:2007}.
The stellar masses of SDSS galaxies are converted to a Chabrier IMF
from a Kroupa IMF \citet{kroupa:2001}, albeit the correction factor
for the velocity dispersions is small ($\sim10$\%{} and $3.5$\%{}
increases for the stellar mass and stellar velocity dispersion,
respectively).  It is quite possible that high redshift PEGs are more
compact than local ones (cf. Section \ref{sec:size} below), hence may
have higher velocity dispersion compared to the local $M_*-\sigma$
relation. However, we use here the local relation because with the
exception of the value reported by \citet{vandokkum:2009}, all other
objects at $z>1.4$ with measured $\sigma$ still follow such relation
\citep{cappellari:2009:gmass,cenarro:2009,onodera:2010:pbzk,vandesande:2011}.

Given the low-S/N of our spectra (typically $\text{S/N}<5$ in the 60
\kms{} interval) and the pixel to pixel fluctuations in S/N due to sky
emission lines, we modified the \texttt{Lick\_EW} code to use median
flux values to compute the pseudo-continua for the indices.  Moreover,
absorption lines are not clearly visible in most of the spectra, and
therefore we restricted our analysis only to objects in which lines
are clearly detected and allow a fairly accurate measurement of the
relative index. Measurements have been carried out for all objects. In
particular, here we focus on the \hdf{} index, which is less affected
by OH lines and by emission line filling, if any.  Ideally,
$\text{H}\beta$ would be a more useful index because it is relatively
isolated from other metal absorption lines, hence independent of
metallicity of the stellar population. However, for many of our
objects $\text{H}\beta$ is strongly contaminated by sky emission
lines.

Besides the \hdf{} index, we measured the strength of 4000 \AA{} break
as quantified by the Dn4000 index, for which we follow the definition
of \citet{balogh:1999} which uses relatively narrow wavelength windows
($\Delta\lambda=100$\,\AA{}) for red and blue continua.  The Dn4000
index is more robustly derived because the wavelength intervals are
wider than those used for absorption line indices.  Errors on Dn4000
were estimated through Monte Carlo simulations by artificially adding
random noise assuming the normal distribution based on the error
spectra at each wavelength pixel.  Then 68-percentile intervals are
derived from 10,000 realizations per object.

\begin{figure*}[htbp]
  \begin{center}
    \includegraphics[width=0.9\linewidth]{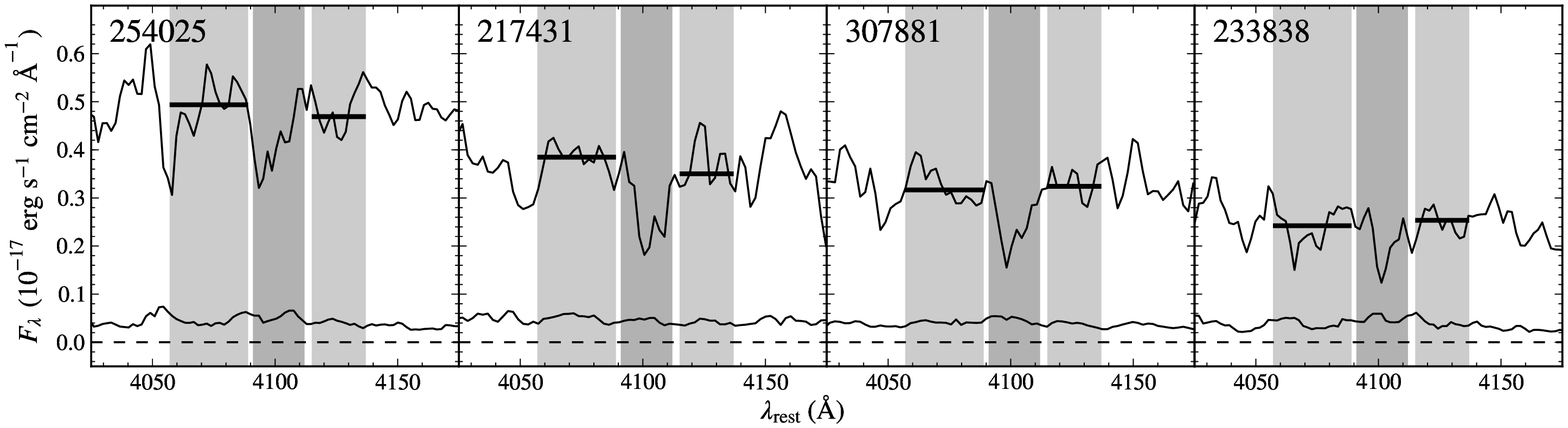}
  \end{center}
  \caption{ Zoom-in of the spectra around H$\delta$ for the four
    objects with the highest S/N ratio.  The solid line in each panel
    shows the object spectrum and the corresponding $1\sigma$ noise.
    The left and right gray-shaded regions indicate respectively the
    blue and red continuum windows used used in the definition of the
    $\text{H}\delta_F$ index, whereas the central ones show the window
    including the H$\delta$ absorption line.  The horizontal thick
    lines indicate the level of blue and red continua for each
    object. The noise spectrum is shown at the bottom of each panel.
  }
  \label{fig:zoom_hdelta}
\end{figure*}

\begin{figure*}[htbp]
  \begin{center}
    \includegraphics[width=0.95\linewidth]{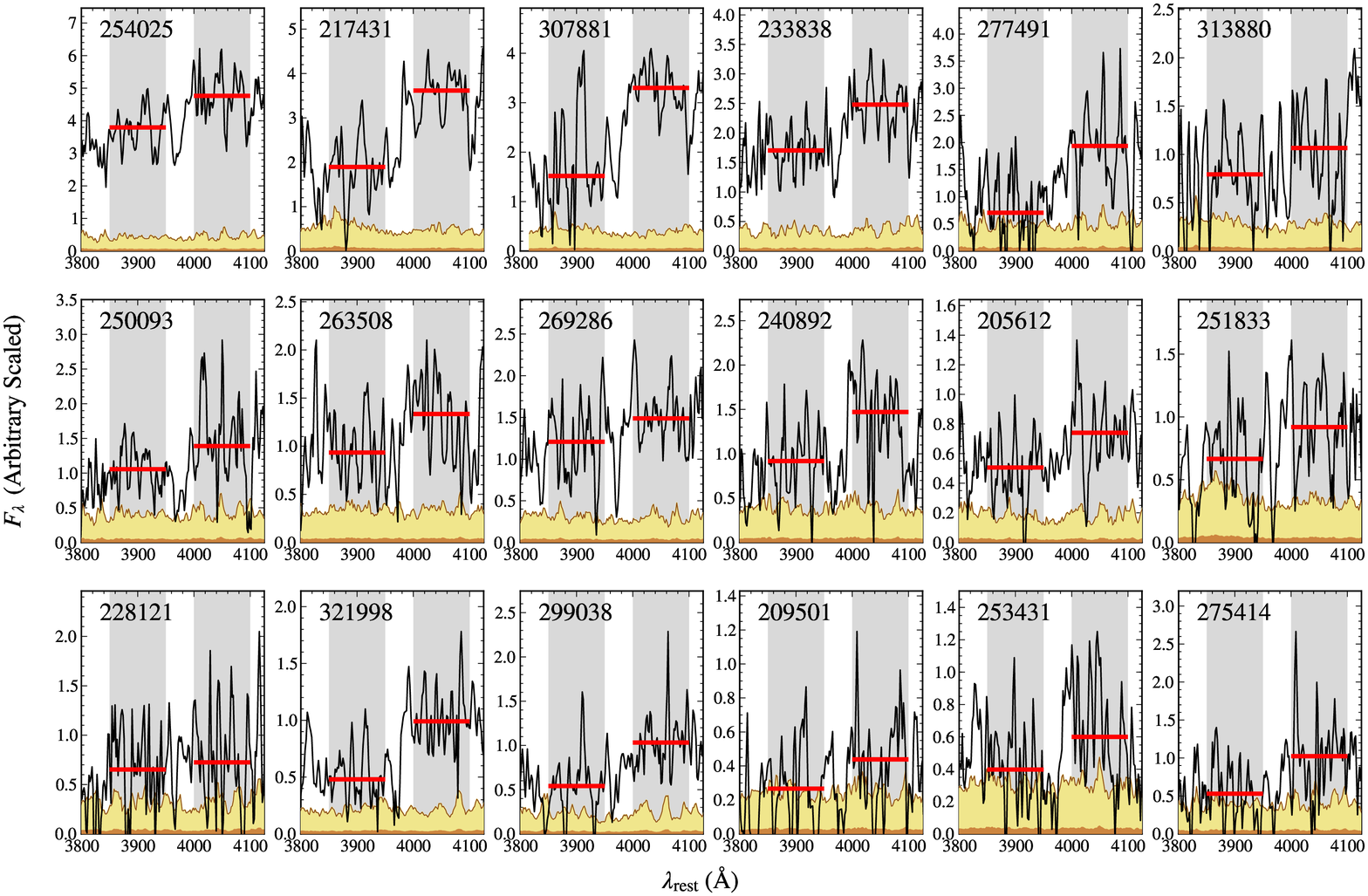}
  \end{center}
  \caption{ Zoom-in of the spectra around the 4000\AA\ break. The
    solid line in each panel shows the spectrum and their $1\sigma$
    noise is shown by the light-yellow shaded regions. The gray shaded
    regions indicate the blue and red continuum windows used to
    measure the Dn4000 index, and the horizontal thick lines indicate
    the adopted continuum level on each side of the break.  The brown
    shaded region at the bottom of each panel represents the $1\sigma$
    noise divided by $\sqrt{100}$ (the size of each continuum window),
    which shows the level of the effective noise associated to the
    flux in these continuum windows.  }
  \label{fig:zoom_d4000}
\end{figure*}

Figure \ref{fig:zoom_hdelta} shows the zoom-in of the spectra around
H$\delta$ for the objects with high-S/N ($\text{S/N}>5$) for which the
absorption line is relatively well detected. On the other hand, Figure
\ref{fig:zoom_d4000} shows the spectra around the 4000 \AA{} break for
all objects with spectroscopic redshift.  For almost all the objects,
Dn4000 appears to be measured reasonably well, while for the objects
not shown in Figure \ref{fig:zoom_hdelta} the region around H$\delta$
is dominated by the noise.

\begin{figure*}[htbp]
  \begin{center}
    \includegraphics[width=0.47\linewidth]{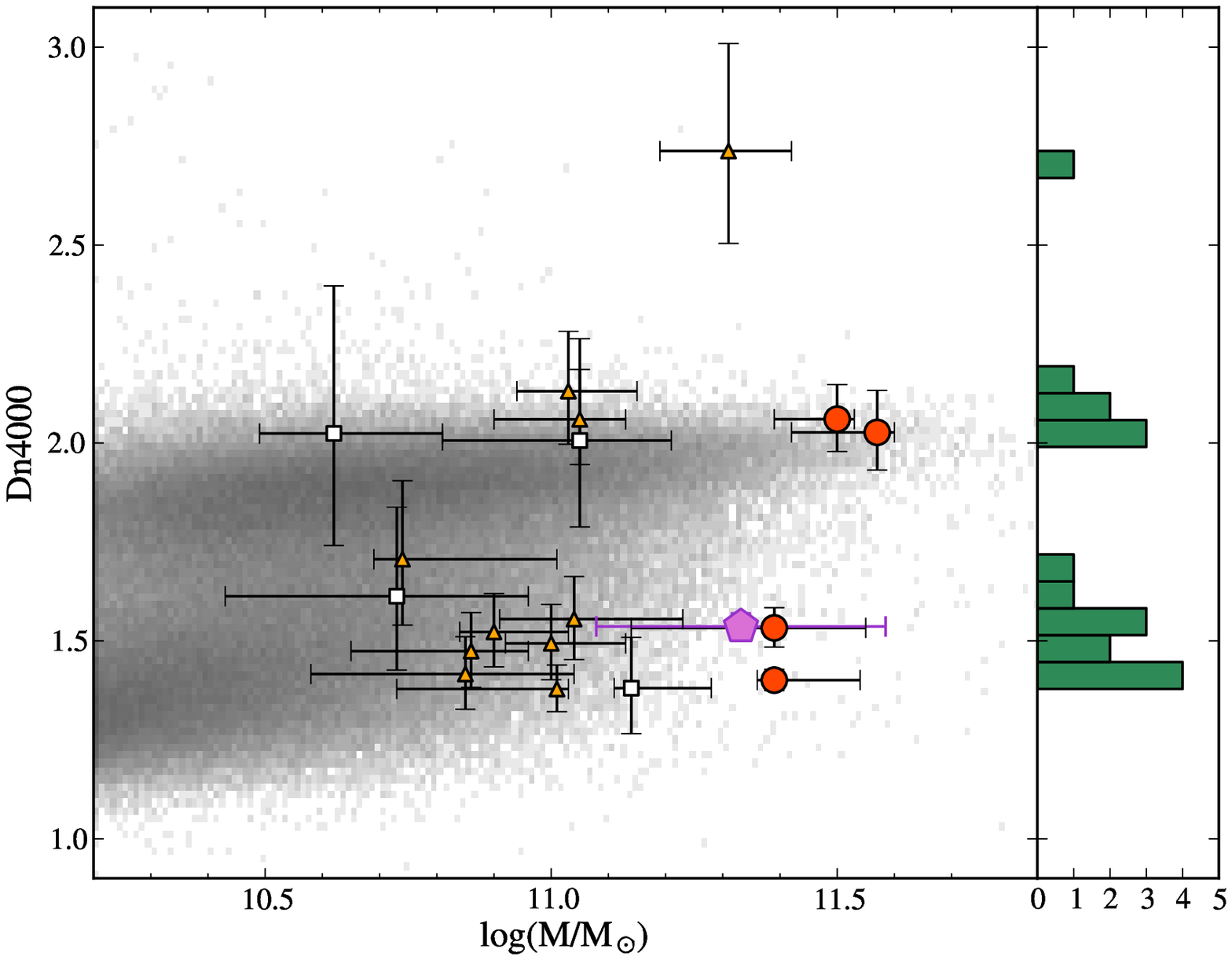}
    \includegraphics[width=0.47\linewidth]{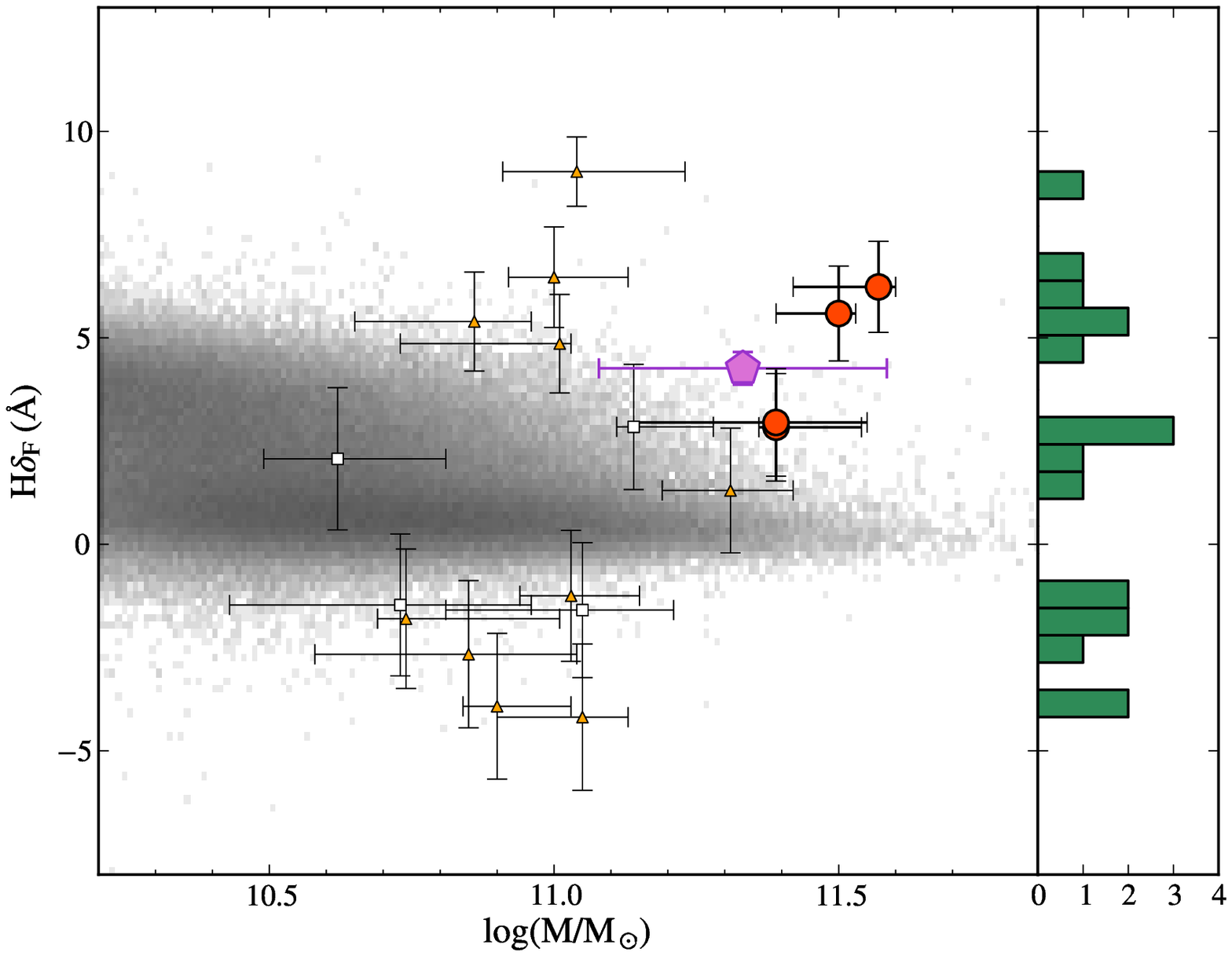}
  \end{center}
  \caption{\textit{Left}: the Dn4000 index as a function of stellar
    mass for the 18 \pbzks with spectroscopic redshifts.
    \textit{Right}: the H$\delta_F$ index as a function of stellar
    mass.  The number of objects in each bin of Dn4000 and \hdf{} are
    shown as histograms on the right panel of each plot.  Individual
    objects are shown with red circles, yellow triangles, and white
    squares corresponding to $\text{S/N}>5$, $3<\text{S/N}<5$, and
    $\text{S/N}<3$ in 60 \kms{} bins, respectively. The pink pentagon
    refers to the composite spectrum where the stellar mass is a
    S/N-weighted average among those used for the stacking.  The
    underlying gray shaded area shows the distribution of the local
    SDSS galaxies including both passive and star-forming objects,
    after correction for emission line filling in the case of emission
    line galaxies.  }
  \label{fig:mass_line}
\end{figure*}

Figure \ref{fig:mass_line} shows the Dn4000 and \hdf{} as a function
of the stellar mass, along with the corresponding relation for local
galaxies at $z\simeq0.06$ extracted from the MPA-JHU SDSS database
\citep{kauffmann:2003}, corrected for emission line filling. Local
galaxies fall in two sequences, one for passive galaxies with higher
Dn4000 and lower \hdf{} at a given stellar mass, and another for
star-forming galaxies with lower Dn4000 and higher \hdf.

\begin{figure}[htbp]
  \begin{center}
    \includegraphics[width=0.95\linewidth]{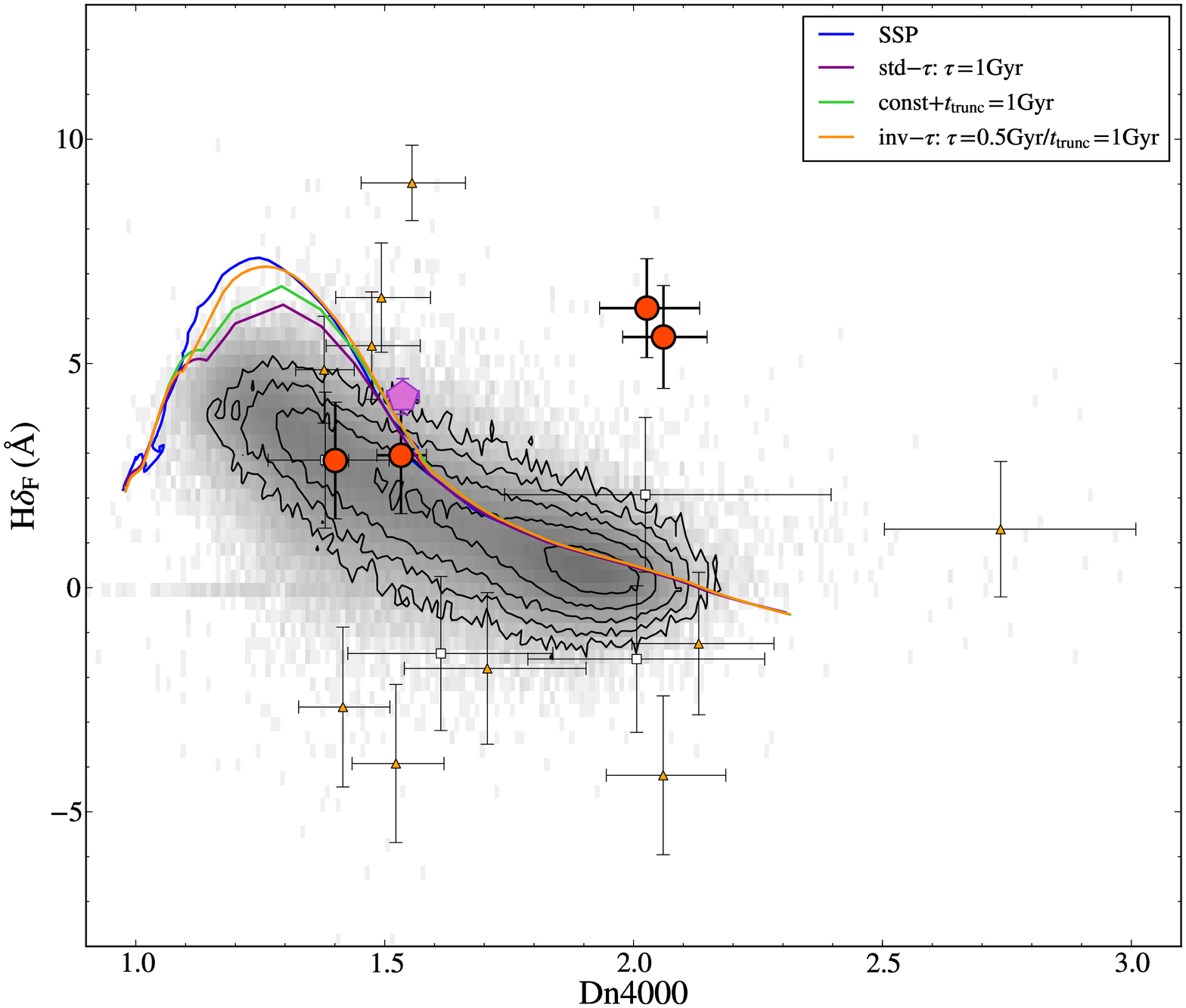}
  \end{center}
  \caption{ The $\text{H}\delta_F$ vs. Dn4000 plot; symbols correspond
    to those in Figure \ref{fig:mass_line}.  The contours show the
    distribution of the local SDSS passive galaxies whereas the gray
    shaded area refers to both passive and star-forming local SDSS
    galaxies.  The solid lines show the evolution of stellar
    populations with various star formation histories (SFH): an
    instantaneous burst (i.e., a SSP, blue), exponentially declining
    SFHs with $e$-folding time of 1 Gyr (purple), constant star
    formation rate over 1 Gyr followed by instantaneous quenching to
    $\text{SFR}=0$ (green), and exponentially increasing star
    formation rate with $\tau=0.5$ Gyrs with a truncation at 1 Gyr
    (orange), as also indicated in the insert. The CB07 evolutionary
    synthesis models have been used.  }
  \label{fig:d4k_hdelt}
\end{figure}

The objects with S/N $> 5$ are also the most massive \pbzks in the
sample. As illustrated in Figure \ref{fig:d4k_hdelt} which shows
\hdf{} as a function of Dn4000, two among them (217431 and 307881)
show simultaneously high Dn4000 ($\sim 2$, comparable to those of the
local early-type galaxies) and high \hdf{} ($\sim 5$ \AA), much
stronger than typical of local PEGs of similar stellar mass.  A
combination of high Dn4000 ($> 1.5$) and high \hdf{} ($> 3$ \AA)
appears to be typical of local and moderate redshift post-starburst
galaxies \citep[e.g.,][]{vergani:2010}, although in this particular
case the combination (Dn4000, \hdf) $\simeq (2, 5)$ appears to be
rather extreme \citep[cf. e.g.,][]{leborgne:2006}.  This can be also
seen in Figure \ref{fig:d4k_hdelt} in which the location of these two
objects cannot be explained by any of the overplotted star formation
histories, namely instantaneous burst, exponentially declining SFR,
constant and exponentially increasing SFR with truncation.  We cannot
exclude that Dn4000 may have been somewhat overestimated, given the
relatively narrow wavelength range available shortwards of the break
(see the corresponding spectra in Figure \ref{fig:pbzk_hst_spec_sed}),
although the sensitivity curves used for the flux calibration are
still smooth and do not suffer from a sharp drop at the edge.  By
taking these indices at face value, we suggest that these two galaxies
may have been quenched a relatively long time ago compared to the star
formation timescale, which makes Dn4000 larger, and experienced a
recent episode of star formation which has enhanced the H$\delta$
absorption, although it appears to be difficult to enhance \hdf{}
without changing significantly Dn4000.

The spectra of the two other objects with high S/N and the composite
spectrum show a Dn4000 and \hdf{} combination more akin that of local
star-forming galaxies.  However, there are virtually no local star
forming galaxies with comparable stellar mass and Dn4000 as at this
stellar mass almost all local galaxies are already passive since many
Gyr. This is also the case for \hdf{} as there is almost no local
counterpart with similar stellar mass and \hdf.  The location of these
two galaxies in Figure \ref{fig:d4k_hdelt}, i.e.,
$\text{Dn4000}\simeq1.5$ and $\text{\hdf}\simeq3$ \AA, are typical of
the early stages of a post-starburst, i.e., of a very recently
quenched galaxy \citep{balogh:1999}.

\subsection{Stellar Populations and Star Formation History from the Composite Spectrum}
\label{sec:ana_comp}

The high S/N of the composite spectrum allows us to study the
``average'' stellar population content of our PEGs at $z>1.4$ by using
both the detailed shape of the continuum and the absorption lines.
For this purpose, we used
\starlight\footnote{\url{http://www.starlight.ufsc.br}}
\citep{cidfernandes:2005,cidfernandes:2009}.  \starlight is a program
to fit an observed spectrum with a combination of template spectra
which usually consist of population synthesis models.  These template
spectra are selected from the CB07 SSP library spanning 29 ages
logarithmically spaced between 10 Myrs and 5 Gyrs (the age of the
Universe at $z=1.4$ is $\sim 4.5$ Gyrs) and 4 metallicities from
$Z=0.004$ to $Z=0.05$.  No intrinsic reddening was assumed as there
are no emission lines in the composite spectrum indicative of a star
formation activity.  We leave the total velocity dispersion and
velocity offset as free parameters.

\begin{figure}[htbp]
  \begin{center}
    \includegraphics[width=0.95\linewidth]{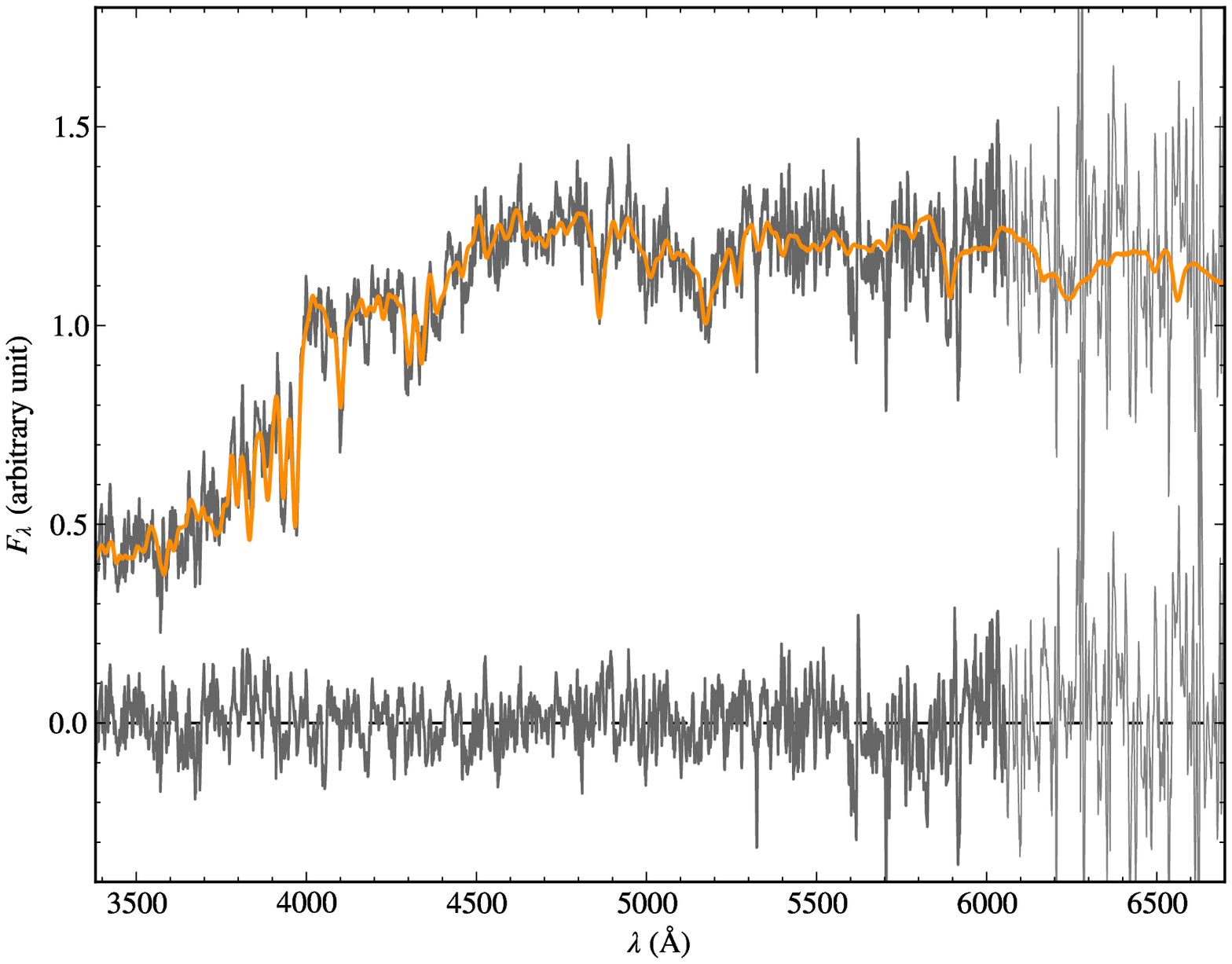}
  \end{center}
  \caption{ The \starlight fit to the composite spectrum.  The gray
    line is the composite spectrum being thicker in the wavelength
    range used for the fit. The orange line shows the \starlight best
    fit model having allowed age and metallicity to span the
    $0.01<\text{Age(Gyr)}<5$ and $0.004<Z<0.05$ ranges.  The gray line
    around zero gives the difference between the model and composite
    spectrum.}
  \label{fig:starlight_spec}
\end{figure}

\begin{figure*}[htbp]
  \begin{center}
    \includegraphics[width=0.95\linewidth]{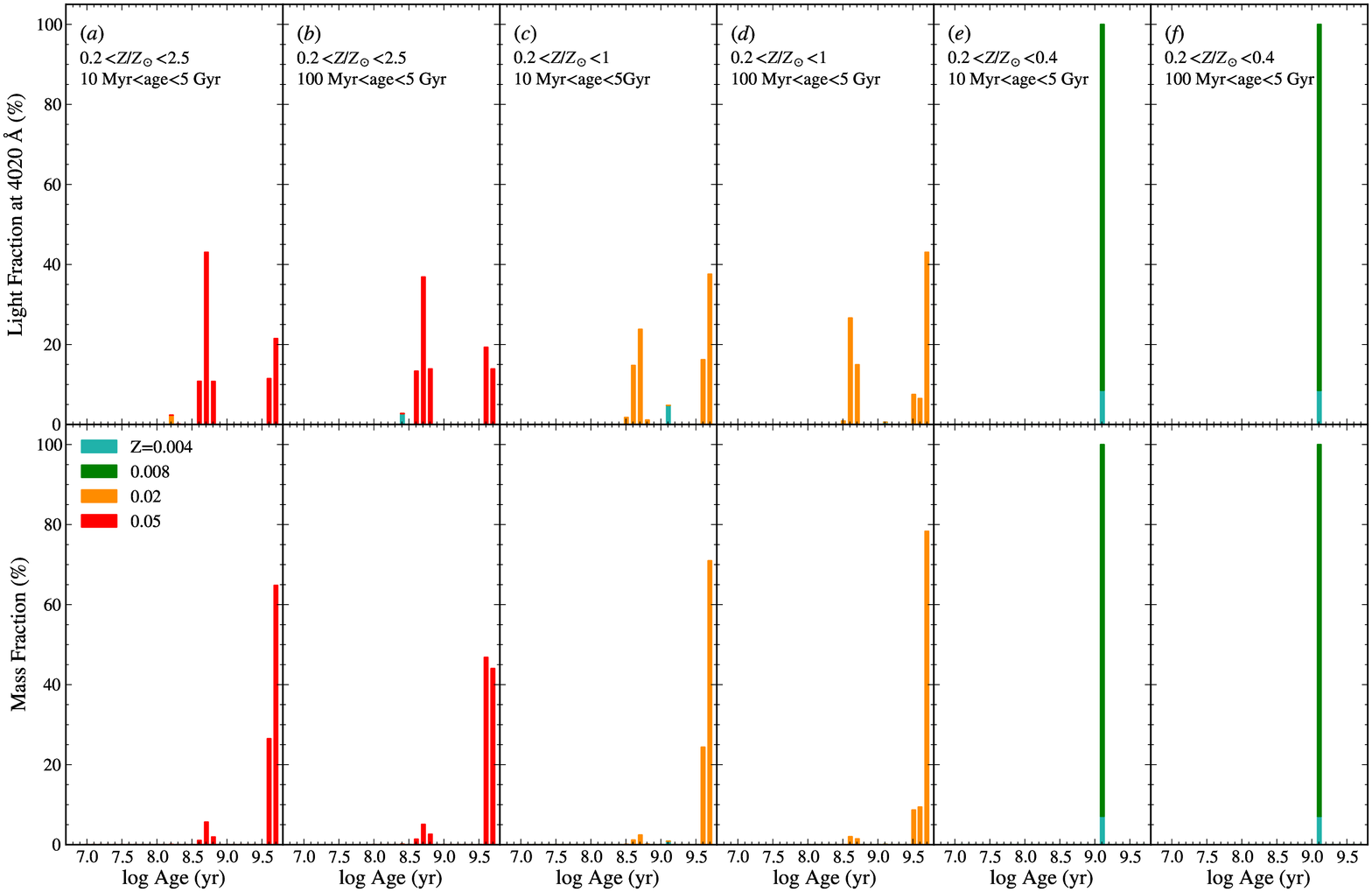}
  \end{center}
  \caption{ The age and metallicity distributions of the stellar
    populations resulting from the \starlight best fit to the
    composite spectrum. The distributions refer to the fractional
    contributions to the light at 4020 \AA{} (\textit{top}) and mass
    (\textit{bottom}).  On each top panel the age and metallicity
    ranges allowed during the fit are indicated, whereas metallicities
    are color-coded on the bottom-left panel.  }
  \label{fig:starlight_sfh}
\end{figure*}

The results of the fitting by \starlight are shown in Figure
\ref{fig:starlight_spec}.  The composite spectrum is well reproduced
by the model with the rms of the residual being $\lesssim8\%{}$ of the
observed flux and with a reduced-$\chi^2=1.02$.  The composition of
stellar populations giving the best fitting spectra is shown in the
leftmost panels of Figure \ref{fig:starlight_sfh}.  The light at 4020
\AA{} is dominated by $400$--$800$ Myr stellar populations with $Z =
2.5Z_\odot$, whereas the mass is dominated by $\sim 3$ Gyr old stellar
populations, again with $Z = 2.5Z_\odot$.  The luminosity and mass
weighted ages are $\langle \log t/\text{yr} \rangle_L = 9.01$ and
$\langle \log t/\text{yr} \rangle_M = 9.59$, respectively.

To check the dependence on the choice of template spectra, the
composite spectrum was fitted by restricting the metallicity within
the range $Z=0.008$--$0.02$ and/or dropping the templates with age
less than $100$ Myrs.  The quality of the fit is essentially
indistinguishable from the previous one, with a
reduced-$\chi^2=1.02$--$1.14$ and the resulting SFHs are also shown in
Figure \ref{fig:starlight_sfh}.  There is no indication for the
presence of young stellar population with age of $\lesssim 400$ Myr in
all fitting results, i.e., there is no detectable B-type star
contribution, consistent with the strong Balmer absorption lines
(mainly produced by A-type stars) seen in the composite spectrum.  The
lack of B-type stars is also consistent with no emission line being
detected in the residual spectrum in Figure \ref{fig:starlight_spec}.
When the metallicity range is limited to sub-solar (two rightmost
panels, using only templates with $Z=0.004$ and $0.008$), the
composite spectrum is reproduced by a $\sim1$ Gyr old stellar
population formed in an almost instantaneous burst.  On the other
hand, if solar or super-solar metallicities are allowed in the
fitting, the mass results to be dominated by $\gtrsim3$ Gyr old
stellar populations with a small amount of younger, $\sim 400$--$800$
Myr old stellar populations, very similar to SFR-weighted ages derived
from SED fitting.

\begin{figure}[htbp]
  \begin{center}
    \includegraphics[width=0.95\linewidth]{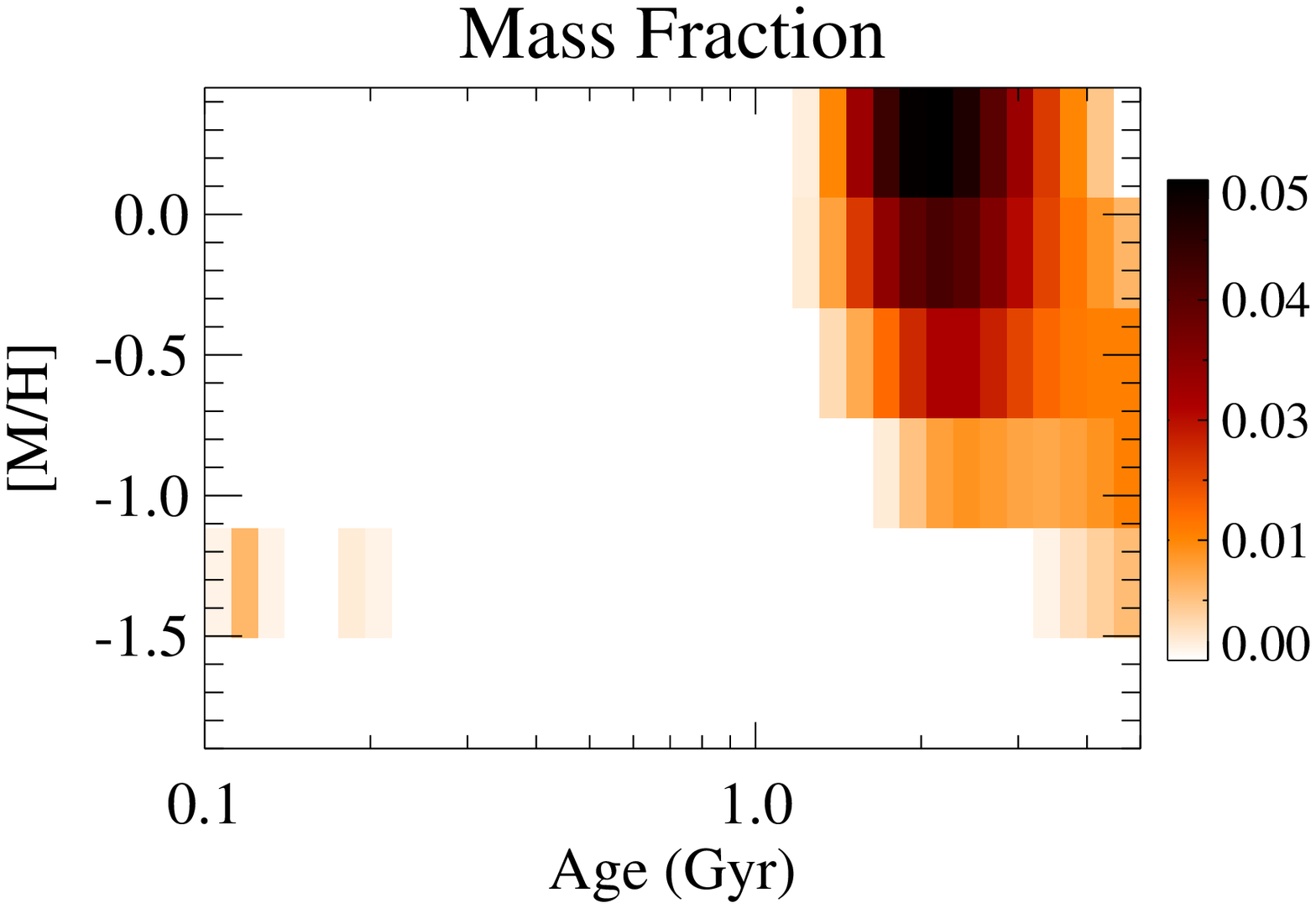}
  \end{center}
  \caption{ The age and metallicity distribution obtained from the
    spectral fitting to the composite spectrum with pPXF
    \citep{cappellari:2004:ppxf}.  The colors indicate the mass
    fraction in each age and metallicity bin.  }
  \label{fig:ppxf_sfh}
\end{figure}

An independent check has been carried out by doing the spectral
fitting using the \ppxf routine \citep{cappellari:2004:ppxf}.  The fit
used as templates a set of 210 MILES SSP
\citep{vazdekis:2010,falconbarroso:2011} with a Kroupa IMF
\citet{kroupa:2001} arranged in a regular grid of 35 different ages,
logarithmically spaced between 0.1 and 5 Gyr, and 6 metallicities,
namely $\text{[M/H]}=-1.71$, $-1.31$, $-0.71$, $-0.40$, $0.00$, and
$0.20$.  To reduce the noise in the recovery of the SFH and
metallicity distribution we employed linear regularization of the
weights during the fitting \citep{press:1992}, as implemented in the
current version of pPXF.  The regularization level was adjusted to
increase the $\chi^2$ from the best un-regularized fit by
$\Delta\chi^2=\sqrt{2\times N_{\rm pix}}$, with $N_{\rm pix}$ the
number of spectral pixels.  In this way the solution represents the
smoothest one that is still consistent with the observed spectrum.
The obtained result is shown in Figure \ref{fig:ppxf_sfh} as a mass
fraction within each age and metallicity interval.  Since the best-fit
spectrum is indistinguishable from that from STARLIGHT, we do not show
it here.  The inferred distribution is dominated by a $\sim 2.5$ Gyr
stellar populations with solar to super-solar metallicities, with at
most a small ($<1$\%) contribution from young (100-200 Myr)
populations.  The mass-weighted age is $\langle \log t/{\rm
  yr}\rangle_M=9.40$, which is very close to that found using
\starlight.

These results show once more the effects of the well known
age--metallicity degeneracy, in particular, from the fit with \ppxf
which produces nearly constant contours along a line with decreasing
age and increasing metallicity.  Combining the two completely
independent experiments, derived by using different codes and spectral
libraries, any significant contribution from young populations can be
excluded and we can conclude that the composite spectrum is well
fitted by old ($\gtrsim 1$ Gyr), metal-rich stellar populations.  The
very small contributions by $\lesssim 400$ Myr old populations found
by \starlight and by $\sim 200$ Myr old populations found by \ppxf are
not significant, and should rather be interpreted as upper limits.

\section{Structural and Kinematical Parameters}
\label{sec:size_sigma}

A puzzling property of high-redshift PEGs has emerged in recent years
and is currently much debated in the literature. Indeed, several of
them appear to have effective radii $\sim 2$--$5$ smaller compared to
local ellipticals of the same stellar mass
\citep[e.g.,][]{daddi:2005:pbzk, trujillo:2007, cimatti:2008,
  vandokkum:2008, saracco:2009, cassata:2011}, which implies densities
within their respective effective radius that are $\sim 10-100$ times
higher than that to their counterparts at $z=0$. No general consensus
has yet emerged about the physical mechanisms that would cause such
observed size evolution, although accretion of an extended envelope
via minor mergers is one widely-proposed scenario.  On the other hand,
$z>1.4$ PEGs with size comparable to that of local ellipticals do also
exist \citep{mancini:2010,saracco:2010}, indicating that galaxies with
a diversity of structural properties co-exists at these high
redshifts.  Possible biases that could lead to size underestimates
have also been discussed
\citep[e.g.,][]{daddi:2005:pbzk,hopkins:2009,mancini:2010}, but it is
understood that observational biases could account for only part of
the effect.  An independent way to check this issues is by measuring
the stellar velocity dispersion ($\sigma_*$) from absorption lines.
If they are truly compact and highly dense, their $\sigma_*$ should be
higher than that of local ellipticals.  So far $\sigma_*$ has been
measured for only 5 or 6 individual PEGs at $z>1.4$
\citep{cappellari:2009:gmass,cenarro:2009,vandokkum:2009,newman:2010,onodera:2010:pbzk,vandesande:2011},
and with one exception \citep{vandokkum:2009} they appear structurally
and dynamically similar to local ellipticals.

The structural parameters of the galaxies in the present sample are
then presented and discussed in this section.  We attempted the
extraction of the velocity dispersion with \ppxf for all galaxies.
However, due to (i) the low S/N, combined with (ii) the presence of
systematic sky residuals, and (iii) the low spectral resolution, a
reliable results could be obtained only for the galaxy with the
highest S/N.

\subsection{Surface Brightness Fits and Measurement of Galaxy Sizes}
\label{sec:size}

  We have measured the structural properties of these galaxies by
  fitting the 2D \sersic profile with GALFIT version 3.0
  \citep{peng:2002:galfit,peng:2010:galfit} on the HST/ACS
  \textit{i}-band (F814W) mosaic version 2.0 over the COSMOS field
  which has a reduced pixel scale of 0.03 arcsec pix$^{-1}$
  \citep[][]{koekemoer:2007,massey:2010}, obtained from the original
  images with the MultiDrizzle routine \citep[][]{koekemoer:2002}.
  The structural properties of three of these 18 galaxies (namely
  254025, 307881, and 217431) have already been measured by
  \citet{mancini:2010} on the previous COSMOS HST/ACS release which
  used the original ACS pixel scale of 0.05 arcsec pix$^{-1}$.  This
  difference has slightly affected the GALFIT results (see the next
  section).  Here we have paid special attention to the treatment of
  neighboring objects, such as in object 254025 which has one bright
  and two faint neighbors. Therefore, in this and similar cases we
  used the segmentation map generated by SExtractor
  \citep{bertin:1996} to mask out the faint neighbors and fit the
  brightest ones together with the main object.

For each galaxy, we constructed the PSF by combining unsaturated stars
in the field as close as possible to each  target, 
though using the single nearest star does not change the result at all.
Indeed, the \sersic index $n$ and effective radius \re{} vary on average 
within 2\%{} of  the values  reported in Table \ref{tab:size}.
Some of the objects are not bright enough in the \textit{i}-band 
(e.g., $\mathit{i}_\text{AB}>24$) to  allow a robust estimate of the structural parameters, 
though we have carried out the measurement for the whole sample. 

The level of the sky background is the most crucial parameter which
affects the estimates of \re{} and the \sersic index 
whereas the estimate of the $i$-band magnitude, position angle, and
axial ratio are less affected by the adopted sky level. 
To account for the effect of the sky background, 
we ran GALFIT with different assumptions for the sky value: 
(i) sky as free parameter in the fit, (ii) sky fixed to the so-called ``pedestal'' GALFIT estimate, 
and (iii) sky manually measured from  empty regions near the main object. 
By combining the results from such  GALFIT runs, we derived \re, $n$,
\textit{i}-band magnitude, and axial ratio as the mid point between
their maximum and minimum values  and quoted half such range as the
corresponding error. This method provides a more reliable estimate of
the actual uncertainties of these measurements compared to the small formal
GALFIT errors from the $\chi^2$ test.

Four of the 18 galaxies (namely, 250093, 275414, 277491, and 313880)
are better fit with a pure exponential profile rather than a
\sersic profile with $n>1$.  For these objects the normal GALFIT fit
did not converge when leaving $n$ as a free parameter, whereas an
acceptable fit can be achieved when $n$ is fixed to unity. 
The resulting structural parameters are reported  in Table
\ref{tab:size}, where effective radii are \textit{circularized}
(i.e., $\re=a_\text{e}\sqrt{b/a}$ where $a_\text{e}$ is the effective semi-major axis 
as given by GALFIT and $b/a$ is the axis ratio). 
The disk scale lengths for the objects best-fit with the exponential profile 
are translated into effective radii by multiplying by 1.678. 

Concerning the three galaxies included in the \citet{mancini:2010} sample, 
we note that the \re{} values reported in their Table~1 and 
Figure~4 do not refer to the \textit{circularized} effective radii, but 
still to the effective semi-major axes $a_\text{e}$. 
Like for the other PEGs, here we use properly circularized effective radii for objects 307881, 254025, and 217431, 
using $b/a=0.58$, $0.65$, and $0.50$, respectively (Mancini et al.\ Erratum in preparation). 
We conclude that the radius of 307881 is in good agreement (within 10\% and $1\sigma$) 
with the value reported in \citet{mancini:2010}.  Instead, the effective radii of 254025, and 217431 
are $\sim 30\%$, and $\sim 45\%$ smaller, respectively, than derived in \citet{mancini:2010}, 
though still within $\sim 1.5\sigma$ of the previous value given the large errors on \re. 
The \sersic indices are all in agreement within $\sim 20\%$ (and within $\sim 1$--$1.5\sigma$), 
and the other free parameters of the fit (i.e., magnitude, position angle, and axial ratio) 
are fully consistent (within $2$--$6$\%, and $1\sigma$) within the published values.
All these differences can be traced back to the different pixel size (hence PSF sampling) 
of the new data and to the different methods of estimating the sky background.

%
%
\begin{deluxetable}{ccccc}
  \tablewidth{0pt}
  \tablecolumns{5}
  \tablecaption{Structural Properties\label{tab:size}}
  \tablehead{
    \colhead{ID} &
    \multicolumn{2}{c}{\re} &
    \colhead{\sersic $n$} &
    \colhead{$i_{814}$}
    \\
    \colhead{} &
    \colhead{(arcsec)} &
    \colhead{(kpc)} &
    \colhead{} &
    \colhead{(mag)}
    \\
    \colhead{(1)} &
    \colhead{(2)} &
    \colhead{(3)} &
    \colhead{(4)} &
    \colhead{(5)} 
  }
  \startdata
  254025  &  $ \phn  0.375 \pm \phn  0.072 $  &  $  3.16 \pm  0.61 $  &  $ 3.36 \pm 0.42 $  &  $ 22.83 \pm  0.13 $  \\
  217431  &  $ \phn  0.851 \pm \phn  0.231 $  &  $  7.19 \pm  1.95 $  &  $ 3.83 \pm 0.56 $  &  $ 22.38 \pm  0.18 $  \\
  307881  &  $ \phn  0.318 \pm \phn  0.014 $  &  $  2.68 \pm  0.12 $  &  $ 2.29 \pm 0.10 $  &  $ 22.92 \pm  0.04 $  \\
  233838  &  $ \phn  0.266 \pm \phn  0.037 $  &  $  2.25 \pm  0.31 $  &  $ 3.07 \pm 0.34 $  &  $ 23.74 \pm  0.09 $  \\
  277491  &  $ \phn  0.292 \pm \phn  0.013 $  &  $  2.46 \pm  0.11 $  &  $ 1.00 $\tablenotemark{a} &  $ 24.31 \pm  0.05 $  \\
  313880  &  $ \phn  0.460 \pm \phn  0.063 $  &  $  3.89 \pm  0.53 $  &  $ 1.00 $\tablenotemark{a} &  $ 23.53 \pm  0.19 $  \\
  250093  &  $ \phn  0.356 \pm \phn  0.018 $  &  $  3.00 \pm  0.15 $  &  $ 1.00 $\tablenotemark{a} &  $ 24.33 \pm  0.05 $  \\
  263508  &  $ \phn  0.101 \pm \phn  0.004 $  &  $  0.86 \pm  0.03 $  &  $ 3.23 \pm 0.21 $  &  $ 23.76 \pm  0.03 $  \\
  269286  &  $ \phn  0.122 \pm \phn  0.014 $  &  $  1.03 \pm  0.12 $  &  $ 4.96 \pm 0.72 $  &  $ 23.82 \pm  0.06 $  \\
  240892  &  $ \phn  0.153 \pm \phn  0.016 $  &  $  1.29 \pm  0.13 $  &  $ 3.01 \pm 0.34 $  &  $ 24.24 \pm  0.07 $  \\
  205612  &  $ \phn  0.245 \pm \phn  0.033 $  &  $  2.08 \pm  0.28 $  &  $ 2.36 \pm 0.26 $  &  $ 24.17 \pm  0.11 $  \\
  251833  &  $ \phn  0.190 \pm \phn  0.011 $  &  $  1.61 \pm  0.09 $  &  $ 2.13 \pm 0.13 $  &  $ 24.01 \pm  0.04 $  \\
  228121  &  $ \phn  0.329 \pm \phn  0.102 $  &  $  2.78 \pm  0.86 $  &  $ 4.07 \pm 0.89 $  &  $ 24.40 \pm  0.20 $  \\
  321998  &  $ \phn  0.216 \pm \phn  0.040 $  &  $  1.83 \pm  0.34 $  &  $ 4.40 \pm 0.65 $  &  $ 24.15 \pm  0.10 $  \\
  299038  &  $ \phn  0.114 \pm \phn  0.002 $  &  $  0.96 \pm  0.02 $  &  $ 1.89 \pm 0.05 $  &  $ 24.94 \pm  0.02 $  \\
  209501  &  $ \phn  0.077 \pm \phn  0.000 $  &  $  0.65 \pm  0.00 $  &  $ 1.63 \pm 0.01 $  &  $ 24.97 \pm  0.00 $  \\
  253431  &  $ \phn  0.317 \pm \phn  0.022 $  &  $  2.69 \pm  0.18 $  &  $ 4.23 \pm 0.19 $  &  $ 24.46 \pm  0.04 $  \\
  275414  &  $ \phn  0.269 \pm \phn  0.015 $  &  $  2.27 \pm  0.13 $  &  $ 1.00$\tablenotemark{a} &  $ 25.07 \pm  0.05 $
  \enddata
  \tablenotetext{a}{
    The fitting has been carried out by fixing $n=1$ for these objects. 
  }
  \tablecomments{
    (1) ID; (2) effective radius in arcseconds; (3) effective radius
    in kpc; (4) \sersic index; (5) \hst/{\it ACS} $i$-band magnitude
    recovered by 
   GALFIT. The effective radii are circularized and the disk scale
   lengths for the objects best fitted  with the exponential profile
   are converted into effective radii. 
    }
\end{deluxetable}

\subsection{Stellar mass--size relation}

\begin{figure}[htbp]
  \begin{center}
    \includegraphics[width=0.95\linewidth]{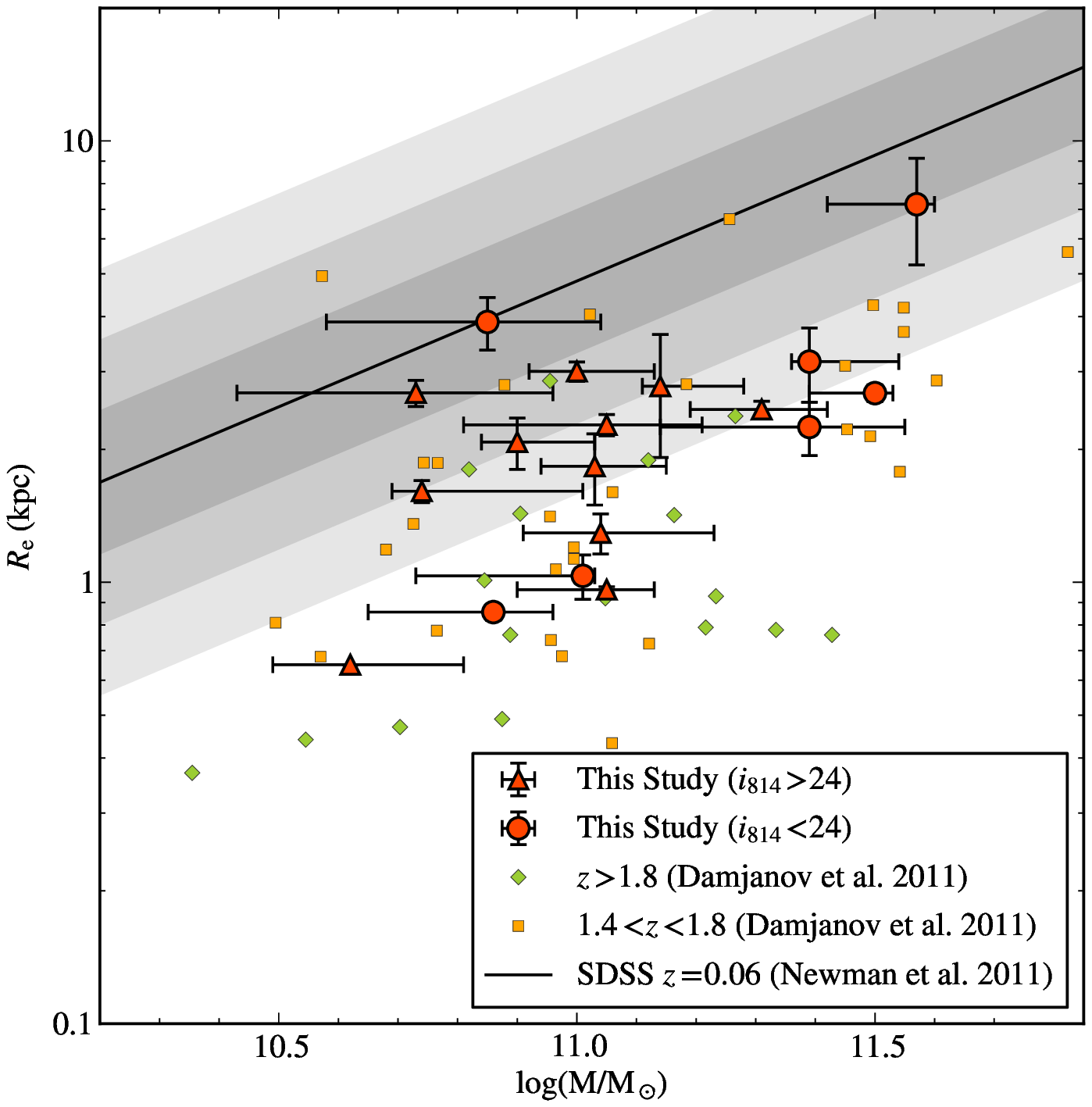}
  \end{center}
  \caption{ The stellar mass--size relation of our spectroscopically
    identified \pbzks at $1.4\lesssim z\lesssim 1.8$ with
    $\mathit{i}_{814}<24$ (red circles) and $\mathit{i}_{814}>24$ (red
    triangles), where $\mathit{i}_{814}$ is the \hst/{\it ACS} F814W
    magnitude recovered by GALFIT.  The orange squares and green
    diamonds represent passively evolving galaxies from the
    compilation of \citet{damjanov:2011} with spectroscopic redshifts
    respectively in the range $1.4<z<1.8$ (the same as in our study)
    and at $z>1.8$. The solid line refers to the mean mass--size
    relation for $z=0.06$ passive galaxies in SDSS from
    \citet{newman:2012} and the gray shaded bands show the $1\sigma$,
    $2\sigma$ and $3\sigma$ deviations from such relation.  }
  \label{fig:masssize}
\end{figure}

Figure \ref{fig:masssize} shows the resulting effective radii of the
program galaxies as a function of the stellar mass.  For comparison,
the local relation derived by \citet{newman:2012} is also shown,
together with the subset of $z>1.4$ PEGs from the compilation of 465
PEGs with spectroscopic redshift by \citet{damjanov:2011}.  For both
data sets we have converted the stellar masses into those appropriate
for the Chabrier IMF by using correction factors shown in Table 2 of
\citet{bernardi:2010}. 
  The majority of these galaxies appear to be more compact compared to
  local PEGs of the same mass, although about a third of them lies on,
  or close to, the local relation (within $\sim 2\sigma$), whereas 7
  of them (i.e., $\sim 40$\%) are more than $3\sigma$ from the local
  relation.  When only the brighter objects with $\mathit{i}<24$ are
  considered, then $\sim 50$\%{} are classified as compact.

Note that the sizes of high-$n$ galaxies, in particular the fainter
ones, may have been systematically underestimated as shown by
simulations \citep{mancini:2010}, an effect that may not be confined
only to high redshift galaxies. For example, the measured effective
radius of M87 can vary by up to a factor of two, depending on the
depth of the adopted photometry \citep{kormendy:2009}. These
uncertainties may well affect the derived fractions of normal and
compact objects in our sample.

\begin{figure}[htbp]
  \begin{center}
    \includegraphics[width=0.95\linewidth]{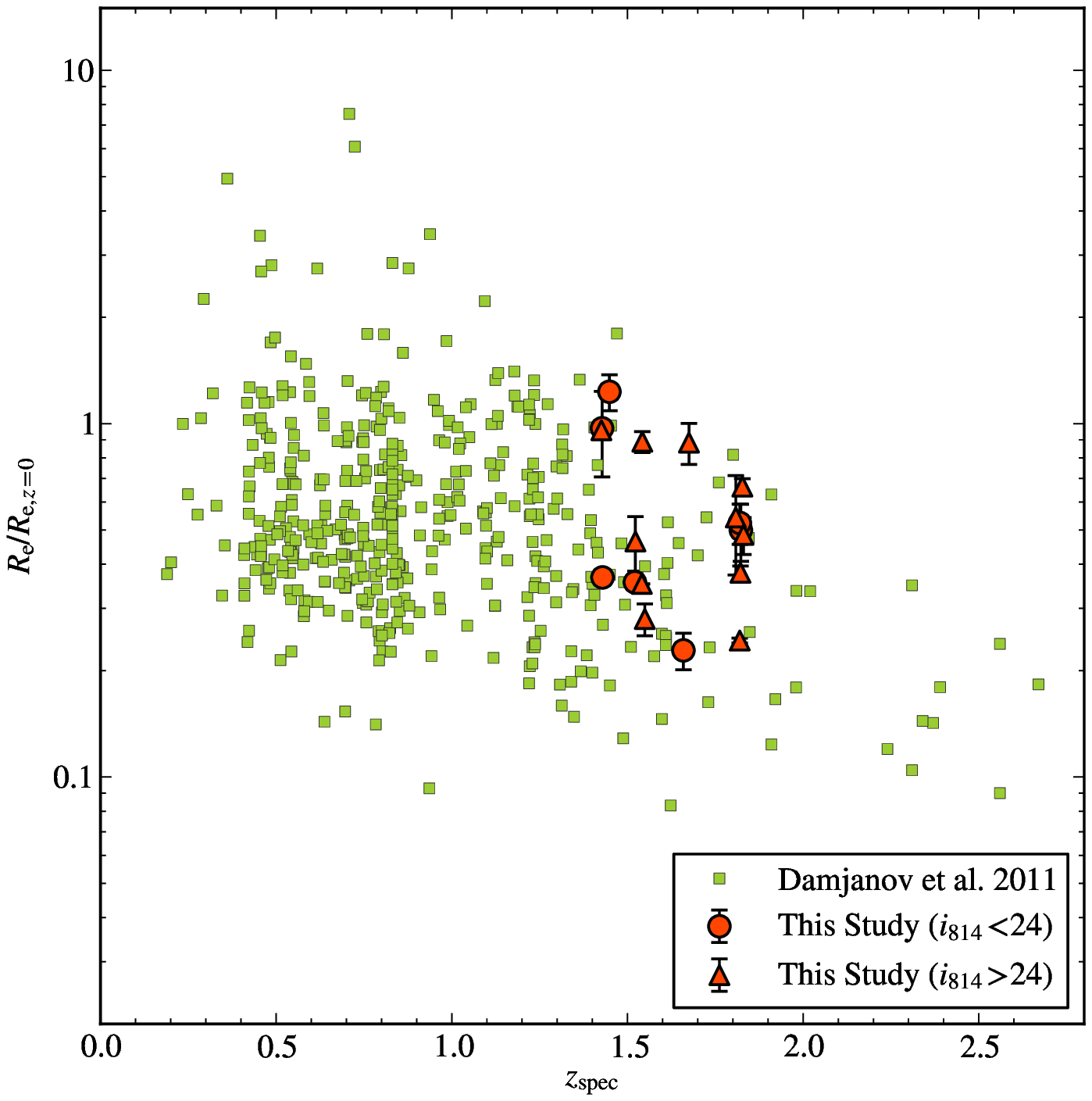}
  \end{center}
  \caption{
    The effective radii of the \pbzks presented here (circles) and
    those of the passive galaxies with spectroscopic redshift
    from \citet{damjanov:2011}, normalized to the mass--size relation at $z=0.06$ from \citet{newman:2012}
    at the same stellar mass. 
  }
  \label{fig:zdsize}
\end{figure}

  Figure \ref{fig:zdsize} shows the effective radius of the 18 galaxies
  (normalized to the local relation) vs.\ their spectroscopic redshift, 
  along with the objects in the compilation of \citet{damjanov:2011}.  
  About 40\%{} of our sample show $\re/r_{\text{e},z=0}<0.3$ while previous studies have found 
  a slightly stronger evolution with $\gtrsim 50\%$ of PEGs at $z\simeq1.6$ with $M\gtrsim10^{11}\,M_\odot$ 
  having $\re/r_{\text{e},z=0}\lesssim0.3$ \citep[e.g.,][]{trujillo:2007,buitrago:2008,damjanov:2011,newman:2012}. 
  \citet{cooper:2012} have shown that PEGs at $0.4 < z < 1.2$ in high
  density regions tend to be bigger by up to $\sim 25\%$ than those in
  low density regions.  If this trend were to continue towards higher
  redshifts then the present PEG sample may be slightly biased towards
  larger sizes compared to those in the general field, as our galaxies
  are preferentially located in overdense regions  (see Section 4).

\subsection{Relation between Size and Age Indicators}
\label{sec:agesize}

One effect contributing to the evolution of the mass--size relation of
PEGs comes from the progressive quenching of larger star-forming
galaxies \citep{vanderwel:2009:quench,valentinuzzi:2010,cassata:2011}.
In this scenario, one expect that younger, i.e., recently quenched,
PEGs have size close to the local relation, while older ones, i.e.,
quenched at an earlier time, are more compact.  If so, one would
expect some correlation between the SED-derived age and a departure
from the local mass-size relation.  Based on the SED ages of 62
spectroscopically identified PEGs at $0.9<z<2$, \citet{saracco:2011}
claimed that the normal size high-$z$ elliptical galaxies have
relatively younger stellar population ages compared to the compact
ones that would show a wider range of formation redshifts with a large
fraction of them having formed at $z>5$.  In contrast,
\citet{whitaker:2012} found that younger PEGs at $1.5<z<2.0$ are more
centrally concentrated and may have slightly smaller sizes.  To check
for this effect in our sample, we have compared the size and age
indicators, namely ages from broad-band SED fitting and Dn4000
indices.

\begin{figure*}[htbp]
  \begin{center}
    \includegraphics[width=0.45\linewidth]{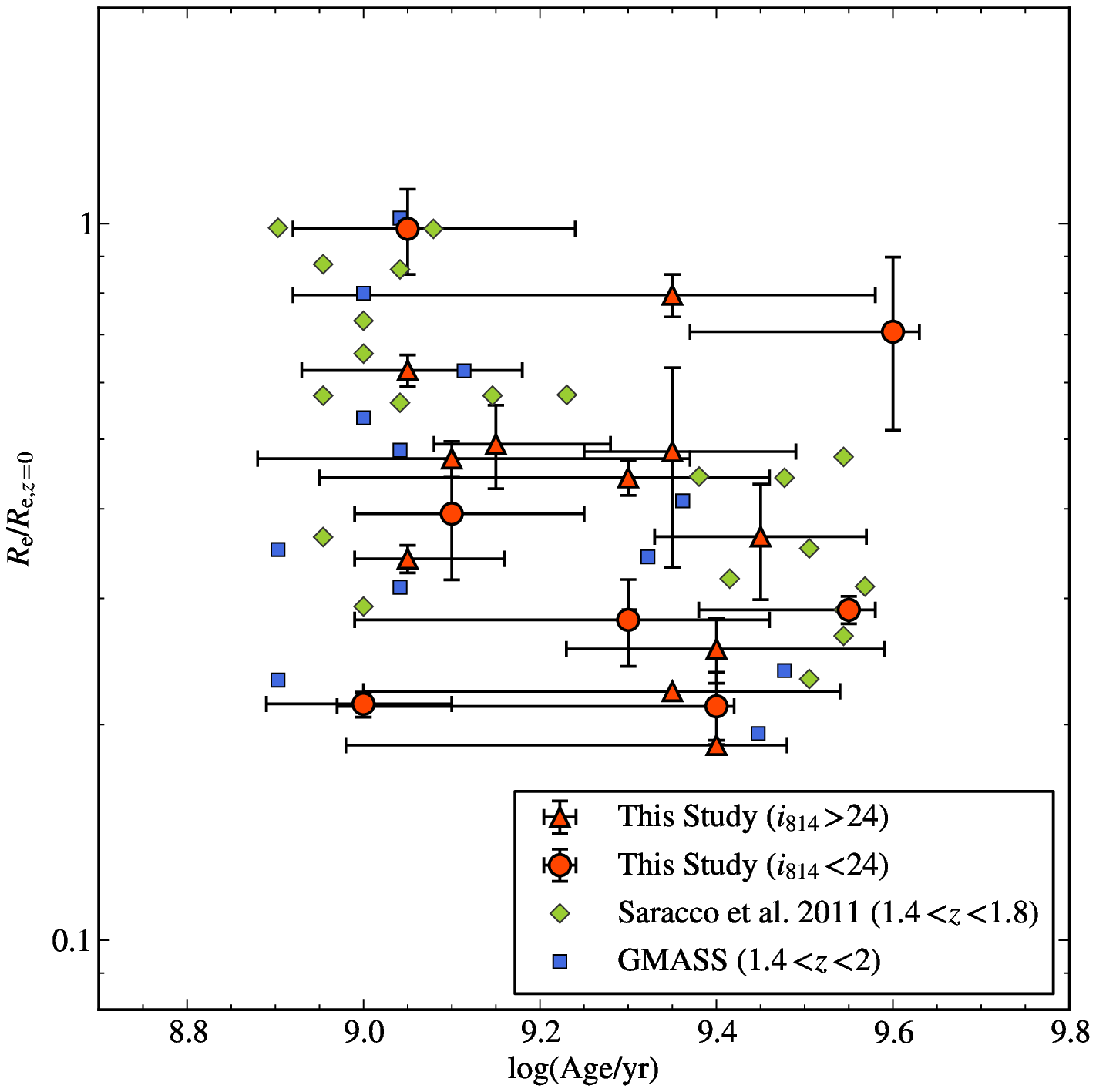}
    \includegraphics[width=0.45\linewidth]{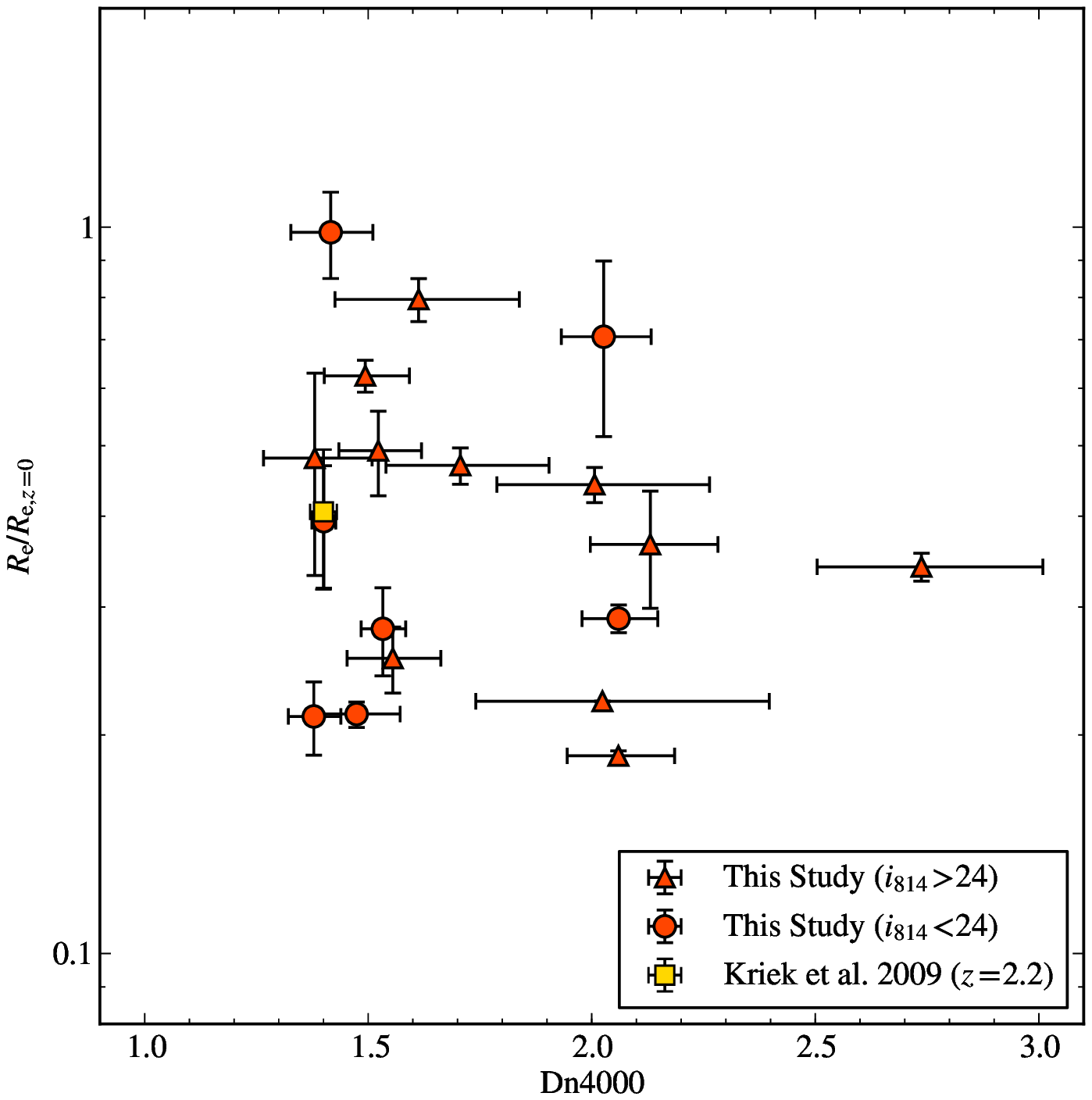}
  \end{center}
  \caption{ \textit{Left}: The effective radius of passively evolving
    galaxies with spectroscopic redshifts at $z>1.4$ as a function of
    their stellar population age from SED fitting.  Red circles and
    triangles, green diamonds, and blue squares represent the \pbzk
    sample in our study (depending on their F814W magnitude as in
    Figure \ref{fig:masssize}), the passively evolving galaxies
    studied by \citet{saracco:2011} at $1.4<z<1.8$, and those from the
    GMASS survey \citep{cimatti:2008} at $1.4<z<2$.  \textit{Right}:
    Same as the left panel, but expressed as a function of Dn4000. Red
    symbols are the same as in the left panel, and a yellow square
    represents a galaxy at $z=2.2$ from \citet{kriek:2009}.  }
  \label{fig:agesize}
\end{figure*}

  The left panel of Figure \ref{fig:agesize} compares the size
  normalized to the local mass-size relation with the age from the SED
  fitting. Besides the present \pbzk sample, the SED ages of
  spectroscopically identified PEGs at similar redshifts from
  \citet{cimatti:2008} and \citet{saracco:2011} are also shown for
  comparison.  As pointed out by \citet{saracco:2011}, in their data
  there appears to be a weak trend with compact galaxies showing a
  wider range of SED ages compared to normal-size ones that tend to
  show younger ages.  Splitting our sample at $\log
  (r_\text{e}/r_{\text{e},z=0})=-0.48$ ($3\sigma$ deviation from the
  local mass--size relation), the median of $\log (\text{Age/yr})$ are
  9.15 and 9.40 for the normal and compact galaxies, respectively,
  i.e., the compact galaxies appear to be $\sim 70\%$ (or about 1 Gyr)
  older than the normal ones.  When only the objects with
  $\mathit{i}_{814}<24$ are considered, the median $\log
  (\text{Age/yr})$ are 9.10 and 9.35 for normal and compact objects,
  respectively.  Note that the uncertainties in the SED ages are
  typically comparable to these differences.

  The right panel of Figure \ref{fig:agesize} shows the
  relation between the normalized size and Dn4000, which can be taken
  as a proxy for stellar population age.  For comparison, a compact
  quenched galaxy at $z=2.2$ \citep{kriek:2009} is also shown.
  Excluding the object with exceptionally large Dn4000 ($\sim 2.75$),
  there seems to be essentially no trend between size and Dn4000.  The
  median Dn4000 of the normal and compact objects are 1.61 and 1.55,
  respectively, a difference comparable to the typical $1\sigma$ error
  in Dn4000 measurement.  Restricting the sample to the brightest
  objects, the median Dn4000 of normal and compact objects are 1.42
  and 1.50, respectively.

  We conclude that in our sample there is no strong correlation
  between the size of high redshift PEGs and stellar population age.
  A Similar conclusion is also found for the brightest objects for
  which the size measurements are more reliable.  To draw any firm
  conclusions it is essential to construct a larger sample with
  robustly measured size and Dn4000.

\subsection{Stellar Velocity Dispersion of the \pbzk Galaxy 254025}
\label{sec:veldisp}

For the brightest object in the sample (254025) a $1\sigma$ upper
limit to its stellar velocity dispersion ($\sigma_*<304$ \kms) was
placed by \citet{onodera:2010:pbzk} based on our MOIRCS 2009 data.  By
adding the data taken in 2010 we have been able to upgrade this upper
limit to a true measurement of $\sigma_*$.  We used a FWHM of 27 \AA{}
at the observed wavelength for the instrumental profile and assumed
the resolution to be constant in linear wavelength scale across the
wavelength range of interest.  We have restricted the analysis to the
$3550$--$4450$ \AA{} rest frame wavelength range 
with $\text{S/N}=11.5$ in 60 km s$^{-1}$ spectral interval, 
which encompasses features such as CN, \ion{Ca}{2} H\&K, H$\delta$, 
the G band, H$\gamma$ and Fe 4383.  We determined a dispersion
$\sigma_*=270\pm105$ \kms{} from 300 realizations of \ppxf fits that
consider not only the effect of random noise, but also explore the
effect of sky residuals and different input parameters in \ppxf, as
described in Section \ref{sec:specz}.  Here for $\sigma_* $ we take
the median of the distribution and the $1\sigma$ error is determined
as $(\sigma_{*,84}-\sigma_{*,16})/2$ where $\sigma_{*,84}$ and
$\sigma_{*,16}$ correspond respectively to the 84- and 16-percentiles
of the probability distribution.  We emphasize that the quoted error
is very conservative, as a formal random error of only 32 \kms{} is
indicated by single \ppxf fits. We quote instead an error of 105
\kms{} derived above as a more realistic value that includes an
estimate of the systematic errors.

\begin{figure*}[htbp]
  \begin{center}
    \includegraphics[width=0.95\linewidth]{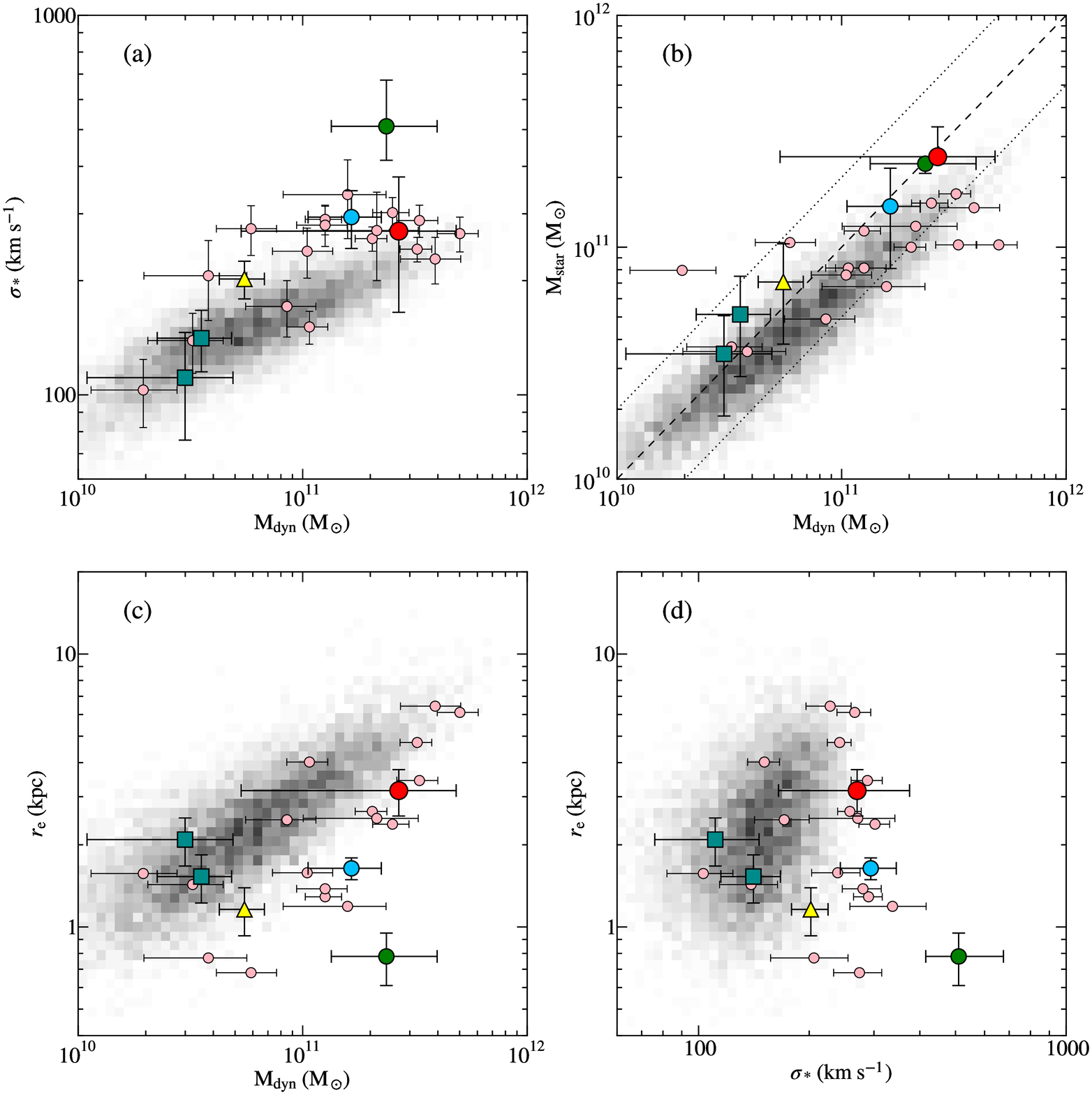}
  \end{center}
  \caption{ A comparison of the structural/dynamical properties of
    high-redshift early-type galaxies with measured velocity
    dispersions (symbols with error bars) with those from SDSS
    elliptical galaxies at $z\simeq0.06$ (shown as gray scale).
    (\textit{a}): The stellar velocity dispersions vs.\ the dynamical
    mass.  (\textit{b}): Comparison between the dynamical and the
    stellar masses; the diagonal dashed line refers to equal mass and
    the dotted lines show the range around it by a factor of 2.
    (\textit{c}): The effective radii vs.\ the dynamical masses.
    (\textit{d}): The effective radii vs.\ the stellar velocity
    dispersions.  In all panels, the symbols refer to: the galaxy
    254025 from the present study (red filled circle); an object at
    $z=2.19$ \citep[green filled circle;][]{vandokkum:2009}; two GMASS
    galaxies at $z=1.41$ and the stacked properties of the GMASS
    galaxies at $1.6<z<2.0$ \citep[blue squares and yellow triangle,
    respectively;][]{cappellari:2009:gmass}; an object at $z=1.80$
    \citep[cyan circle;][]{vandesande:2011}; and objects at
    $1.05<z<1.60$ \citep[small pink circles;][]{newman:2010}.  }
  \label{fig:dynprop}
\end{figure*}

Figure \ref{fig:dynprop} compares the structural and dynamical
properties of the galaxy 254025 with those of local early type
galaxies and with those from the small sample of $z>1$ PEGs for which
a stellar velocity dispersion has been measured. 
Dynamical masses are derived as $M_\text{dyn}=5\re\sigma_*^2/G$ 
where $G$ is the gravitational constant, 
and we obtained $\log(M_\text{dyn}/M_\odot)=11.43\pm0.35$ for the galaxy 254025.
Figure \ref{fig:dynprop} is an update of
the similar figure presented in \citet{onodera:2010:pbzk}. We can thus
confirm that this galaxy closely follows all the local relations
involving the dynamical mass, the effective radius and the
stellar velocity dispersion within the error bars.  Again, this is a proof of the
coexistence at high redshifts of PEGs that appear structurally and
dynamically similar to their local counterparts with other galaxies
that are more compact and are correspondingly characterized by a
higher velocity dispersion.

\section{Number and Stellar Mass Densities}
\label{sec:number}

As our spectroscopic identification of the PEGs with $K<21$ or
$M_*\gtrsim 10^{11}\,M_\odot$ is nearly complete over the surveyed
area, we can estimate the abundance of massive PEGs at $z>1.4$, to
within cosmic variance errors.  The redshift range covered by the
present sample $1.43\lesssim z\lesssim1.83$ over the survey area of
66.2 arcmin$^2$ contains a comoving volume of $\sim 82,000$ Mpc$^3$.
Thus, from the 16 objects with $K<21$ we estimate a comoving number
density of $1.9\times10^{-4}$ Mpc$^{-3}$, and the corresponding
stellar mass density is $3.0\times10^{7}\,M_\odot$ Mpc$^{-3}$, with an
uncertainty of $\sim 25\%{}$ from Poisson statistics.  These
should be regarded as upper limits since the field was chosen for
its overdensity of PEGs.  However they are in agreement with
estimates derived from a complete sample of galaxies in the Hubble
Ultra Deep Field of $3.4\times10^{-4}$ Mpc$^{-3}$ and
$1.9$--$4.6\times10^{7}\,M_\odot$ Mpc$^{-3}$ (after converting to
Chabrier IMF), as measured by \citet{daddi:2005:pbzk} for
spectroscopically identified PEGs at $1.39<z<2.00$ down to the limit
$K=23$, although their explored comoving volume is $\sim 4$ times
smaller.

\citet{brammer:2011} have derived number and mass densities of  PEGs at $0.4<z_{\rm phot}<2.2$ from the NEWFIRM Medium Band Survey (NMBS), 
covering an area of $\sim 0.4$ deg$^{2}$, and for the mean redshift of our galaxies ($z=1.69$) found  
$(1.0\pm0.3)\times10^{-4}$ Mpc$^{-3}$ and $(1.6\pm0.4)\times10^{7}\,M_\odot$ Mpc$^{-3}$, respectively.  
Our estimates are somewhat higher, as expected given the mentioned overdensity bias. 

Comparing to the corresponding local values, respectively $\sim4\times10^{-4}$ Mpc$^{-3}$ and 
$\sim8\times10^{7}\,M_\odot$ Mpc$^{-3}$ from \citet{baldry:2004}, 
taken at face value our result suggests that PEGs at $1.4\lesssim z \lesssim 1.8$ 
account for $\sim50\%$ of the number density and $\sim35$--$40\%$ of the stellar mass density of local PEGs with $M_*>10^{11}\,M_\odot$.

\section{Summary and Conclusions}
\label{sec:summary}

Mapping the 3D space distribution of passively evolving galaxies at
high redshift is an endeavour of critical importance in the broader
context of galaxy evolution and observational cosmology. However, it
is perhaps also the most challenging observationally, as it requires
many hundreds of hours of 8m-class telescope time with current
instrumentation. Emission lines are absent (or very weak) in the
spectra of PEGs, and therefore measuring redshifts and velocity
dispersions must rely uniquely on absorption lines. With spectrometers
working at optical wavelengths, for such objects at $z\gtrsim 1.4$
this is possible using absorption lines over an intrinsically very
faint UV continuum, which requires extremely long integration times.
Using near-IR, multiobject spectrometers offers a potentially
attractive alternative, as the strongest spectral features over a
strong continuum become accessible. However, ground based, infrared
spectroscopy of faint, high redshift objects is affected by a higher
sky and thermal background which largely limits this advantage.  Using
the MOIRCS instrument at the Subaru telescope we have started a pilot
experiment aimed at studying a sample of bright PEGs at $z\gtrsim 1.4$
while assessing the feasibility of a more ambitious project.

With this objective, we have first identified the strongest
concentrations of high-$z$ candidate PEGs over the COSMOS field, and
focused our observational effort over one such concentration and its
immediate vicinity. A sample of 34 objects has been observed, with
integration times ranging from 280 to 550 minutes, and the continuum
has been detected in 18 of them. The redshift of these detected
galaxies are concentrated in three spikes at $z=1.43$, 1.53 and 1.82,
and the last two galaxies have quite similar redshifts around
1.67. This spectroscopically confirms the strong clustering of high
redshift PEGs, previously hinted only from their 2-point correlation
function \citep[e.g.,][]{kong:2006,mccracken:2010}.  A comparison with
COSMOS photometric redshifts shows that $\sim 2/3$ of them agree
closely with the spectroscopic redshifts, whereas the other $\sim 1/3$
of them are systematically underestimated by up to $\sim 25\%$.  Thus,
these photometric redshifts can lead to a sizable underestimate of the
volume density of PEGs at $z\gtrsim 1.4$.   However, we show that
photometric redshift training with these spectroscopic redshifts can
significantly improve their performance for high-$z$ PEGs.

Having fixed the redshifts to their spectroscopic values, we have
explored the stellar population content of these galaxies, measuring
their masses, luminosity-weighted ages and metallicities 
using various  SED-fitting techniques. Stellar masses range from
$\sim 4\times 10^{10}$ to $\sim 4\times 10^{11}\, M_\odot$, ages from
$\sim 1$ to $\sim 3$ Gyr, whereas near-solar metallicity is found
for most galaxies. 

The Lick \hdf{} and the Dn4000 indices allowed us to investigate
further the stellar population content and evolutionary stage of the
sample galaxies.  We find that about 40\%{} of the \pbzks show large
Dn4000 (around 2) which indicates that these galaxies are already
quenched long ago compared to their timescale of star formation.  On
the other hand, about 60\%{} of them have Dn4000 values close to those
of young star-forming or post-starburst galaxies, which indicates that
these galaxies have been quenched more recently, given the lack of the
star-forming activity (except for one object with a mid-IR detection).
This is also confirmed for the four most massive and brightest objects
and the composite spectrum with measurable H$\delta$ absorption line.
Two of them (as well as the composite spectrum, see below) show
moderate Dn4000 and \hdf{}, consistent with the early stages of a
post-starburst galaxy where star formation has been quenched very
recently.  The other two bright galaxies show both large Dn4000 and
\hdf{} and appear to be already passive, possibly with a recent
episode of star-formation, though the combination of two indices seems
rather extreme.

The composite spectrum of the 17 galaxies exhibits several
well-detected spectral features, notably CN, \ion{Ca}{2} H\&K, the G
band, H$\beta, \gamma, \delta$, Mg$b$ and various iron lines. A fit to
this composite spectrum indicates that it is dominated by old,
metal-rich stellar populations with a mass-weighted age of $\sim 3.5$
Gyr.  The fit is somewhat dependent on the allowed age and metallicity
ranges, but the existence of stellar population younger than $\sim
400$ Myr is always excluded by the fits, which is confirmed by the use
of another fitting code adopting a different input SSP library.

The structural parameters, (circularized) effective radius and \sersic
index, have been measured for these galaxies over the \hst/ACS
\textit{i}-band image of the COSMOS field. A majority of the 18
galaxies have effective radii somewhat lower than equally massive
galaxies in the local Universe, with some up to three times
lower. However, almost about a third of the galaxies appear to be on, 
or close to, the local $\re-M_*$ relation, confirming that compact and
apparently normal PEGs coexist at these redshifts.

Finally, the stellar velocity dispersion of the brightest galaxy in
the sample was measured as $\sigma_* = 270\pm 105$
\kms{}. Combined with its effective radius and stellar mass, this
value of $\sigma_*$ confirms that this $z=1.82$ galaxy closely follows 
all the local structural and dynamical relations for PEGs.

  The nearly complete identification of PEGs at the top end of
  their stellar mass function at the observed redshift range allows us
  to estimate their number and stellar mass densities, still with the
  proviso that they refer to an overdensity region.
  Thus, the PEGs with $M*\gtrsim 10^{11}\,M_\odot$ at $1.4\gtrsim z
  \gtrsim 1.8$ we found in the studied field correspond to about
  50\%{} of the number and about 35--40\%{} of the stellar mass
  densities of local PEGs with similar stellar mass. 

Without counting the time lost to technical problems and bad weather,
this project has used 4--5 nights of Subaru telescope time.  Over the
whole 2 square degree COSMOS field one can roughly count $\sim 20$
concentrations of candidate high-$z$ PEGs similar to the one explored
in the present study.  Thus, some 80--100 nights would be required to
map all such concentrations in a similar fashion with Subaru/MOIRCS.
  Obviously, even more would be required to map the distribution of
  all massive PEGs over the entire COSMOS field.  A coordinated effort
  with the upcoming near-IR multi-object spectrographs at 8--10 m
  class telescopes (such as Gemini/FLAMINGOS-II, VLT/KMOS, and
  Keck/MOSFIRE) would be required to make such large project feasible
  in the near future.

\acknowledgments

We thank the staffs of the Subaru telescope, especially Ichi Tanaka,
for supporting the observations.  We also thank Mariska Kriek for a
permission to use FAST, Stephan Charlot for providing the stellar
population synthesis models before publication, Alexis Finoguenov for
computing the upper limits from XMM-Newton data for the overdense
regions, Masayuki Tanaka for checking the existence of red sequence
cluster around redshift spikes, and Masafumi Yagi for fruitful
discussion.  We are grateful to the anonymous referee for her/his
comments.  This work has been partly supported by the ASI grant
``COFIS-Analisi Dati'' and by the INAF grant ``PRIN-2008'' and is
supported by a Grant-in-Aid for Science Research (No.~23224005) by the
Japanese Ministry of Education, Culture, Sports, Science and
Technology.

{\textit{Facilities:}} \facility{Subaru Telescope (NAOJ)}


\begin{thebibliography}{103}
\expandafter\ifx\csname natexlab\endcsname\relax\def\natexlab#1{#1}\fi

\bibitem[{{Baldry} {et~al.}(2004){Baldry}, {Glazebrook}, {Brinkmann},
  {Ivezi{\'c}}, {Lupton}, {Nichol}, \& {Szalay}}]{baldry:2004}
{Baldry}, I.~K., {Glazebrook}, K., {Brinkmann}, J., {Ivezi{\'c}}, {\v Z}.,
  {Lupton}, R.~H., {Nichol}, R.~C., \& {Szalay}, A.~S. 2004, \apj, 600, 681

\bibitem[{{Balogh} {et~al.}(1999){Balogh}, {Morris}, {Yee}, {Carlberg}, \&
  {Ellingson}}]{balogh:1999}
{Balogh}, M.~L., {Morris}, S.~L., {Yee}, H.~K.~C., {Carlberg}, R.~G., \&
  {Ellingson}, E. 1999, \apj, 527, 54

\bibitem[{{Bernardi} {et~al.}(2010){Bernardi}, {Shankar}, {Hyde}, {Mei},
  {Marulli}, \& {Sheth}}]{bernardi:2010}
{Bernardi}, M., {Shankar}, F., {Hyde}, J.~B., {Mei}, S., {Marulli}, F., \&
  {Sheth}, R.~K. 2010, \mnras, 404, 2087

\bibitem[{{Bertin} \& {Arnouts}(1996)}]{bertin:1996}
{Bertin}, E., \& {Arnouts}, S. 1996, \aaps, 117, 393

\bibitem[{{Blanton} {et~al.}(2005){Blanton}, {Schlegel}, {Strauss},
  {Brinkmann}, {Finkbeiner}, {Fukugita}, {Gunn}, {Hogg}, {Ivezi{\'c}}, {Knapp},
  {Lupton}, {Munn}, {Schneider}, {Tegmark}, \& {Zehavi}}]{blanton:2005}
{Blanton}, M.~R., {et~al.} 2005, \aj, 129, 2562

\bibitem[{{Bolzonella} {et~al.}(2010){Bolzonella}, {Kova{\v c}}, {Pozzetti},
  {Zucca}, {Cucciati}, {Lilly}, {Peng}, {Iovino}, {Zamorani}, {Vergani},
  {Tasca}, {Lamareille}, {Oesch}, {Caputi}, {Kampczyk}, {Bardelli}, {Maier},
  {Abbas}, {Knobel}, {Scodeggio}, {Carollo}, {Contini}, {Kneib}, {Le
  F{\`e}vre}, {Mainieri}, {Renzini}, {Bongiorno}, {Coppa}, {de la Torre}, {de
  Ravel}, {Franzetti}, {Garilli}, {Le Borgne}, {Le Brun}, {Mignoli},
  {Pell{\'o}}, {Perez-Montero}, {Ricciardelli}, {Silverman}, {Tanaka},
  {Tresse}, {Bottini}, {Cappi}, {Cassata}, {Cimatti}, {Guzzo}, {Koekemoer},
  {Leauthaud}, {Maccagni}, {Marinoni}, {McCracken}, {Memeo}, {Meneux},
  {Porciani}, {Scaramella}, {Aussel}, {Capak}, {Halliday}, {Ilbert},
  {Kartaltepe}, {Salvato}, {Sanders}, {Scarlata}, {Scoville}, {Taniguchi}, \&
  {Thompson}}]{bolzonella:2010}
{Bolzonella}, M., {et~al.} 2010, \aap, 524, A76

\bibitem[{{Brammer} {et~al.}(2008){Brammer}, {van Dokkum}, \&
  {Coppi}}]{brammer:2008}
{Brammer}, G.~B., {van Dokkum}, P.~G., \& {Coppi}, P. 2008, \apj, 686, 1503

\bibitem[{{Brammer} {et~al.}(2009){Brammer}, {Whitaker}, {van Dokkum},
  {Marchesini}, {Labb{\'e}}, {Franx}, {Kriek}, {Quadri}, {Illingworth}, {Lee},
  {Muzzin}, \& {Rudnick}}]{brammer:2009}
{Brammer}, G.~B., {et~al.} 2009, \apjl, 706, L173

\bibitem[{{Brammer} {et~al.}(2011){Brammer}, {Whitaker}, {van Dokkum},
  {Marchesini}, {Franx}, {Kriek}, {Labb{\'e}}, {Lee}, {Muzzin}, {Quadri},
  {Rudnick}, \& {Williams}}]{brammer:2011}
---. 2011, \apj, 739, 24

\bibitem[{{Bruzual} \& {Charlot}(2003)}]{bruzual:2003}
{Bruzual}, G., \& {Charlot}, S. 2003, \mnras, 344, 1000

\bibitem[{{Buitrago} {et~al.}(2008){Buitrago}, {Trujillo}, {Conselice},
  {Bouwens}, {Dickinson}, \& {Yan}}]{buitrago:2008}
{Buitrago}, F., {Trujillo}, I., {Conselice}, C.~J., {Bouwens}, R.~J.,
  {Dickinson}, M., \& {Yan}, H. 2008, \apjl, 687, L61

\bibitem[{{Calzetti} {et~al.}(2000){Calzetti}, {Armus}, {Bohlin}, {Kinney},
  {Koornneef}, \& {Storchi-Bergmann}}]{calzetti:2000}
{Calzetti}, D., {Armus}, L., {Bohlin}, R.~C., {Kinney}, A.~L., {Koornneef}, J.,
  \& {Storchi-Bergmann}, T. 2000, \apj, 533, 682

\bibitem[{{Cappellari} \& {Emsellem}(2004)}]{cappellari:2004:ppxf}
{Cappellari}, M., \& {Emsellem}, E. 2004, \pasp, 116, 138

\bibitem[{{Cappellari} {et~al.}(2009){Cappellari}, {di Serego Alighieri},
  {Cimatti}, {Daddi}, {Renzini}, {Kurk}, {Cassata}, {Dickinson},
  {Franceschini}, {Mignoli}, {Pozzetti}, {Rodighiero}, {Rosati}, \&
  {Zamorani}}]{cappellari:2009:gmass}
{Cappellari}, M., {et~al.} 2009, \apjl, 704, L34

\bibitem[{{Cappellari} {et~al.}(2012){Cappellari}, {McDermid}, {Alatalo},
  {Blitz}, {Bois}, {Bournaud}, {Bureau}, {Crocker}, {Davies}, {Davis}, {de
  Zeeuw}, {Duc}, {Emsellem}, {Khochfar}, {Krajnovi{\'c}}, {Kuntschner},
  {Lablanche}, {Morganti}, {Naab}, {Oosterloo}, {Sarzi}, {Scott}, {Serra},
  {Weijmans}, \& {Young}}]{cappellari:2012}
---. 2012, \nat, 484, 485

\bibitem[{{Cassata} {et~al.}(2011){Cassata}, {Giavalisco}, {Guo}, {Renzini},
  {Ferguson}, {Koekemoer}, {Salimbeni}, {Scarlata}, {Grogin}, {Conselice},
  {Dahlen}, {Lotz}, {Dickinson}, \& {Lin}}]{cassata:2011}
{Cassata}, P., {et~al.} 2011, \apj, 743, 96

\bibitem[{{Cenarro} \& {Trujillo}(2009)}]{cenarro:2009}
{Cenarro}, A.~J., \& {Trujillo}, I. 2009, \apjl, 696, L43

\bibitem[{{Chabrier}(2003)}]{chabrier:2003}
{Chabrier}, G. 2003, \pasp, 115, 763

\bibitem[{{Chary} \& {Elbaz}(2001)}]{chary:2001}
{Chary}, R., \& {Elbaz}, D. 2001, \apj, 556, 562

\bibitem[{{Cid Fernandes} {et~al.}(2005){Cid Fernandes}, {Mateus}, {Sodr{\'e}},
  {Stasi{\'n}ska}, \& {Gomes}}]{cidfernandes:2005}
{Cid Fernandes}, R., {Mateus}, A., {Sodr{\'e}}, L., {Stasi{\'n}ska}, G., \&
  {Gomes}, J.~M. 2005, \mnras, 358, 363

\bibitem[{{Cid Fernandes} {et~al.}(2009){Cid Fernandes}, {Schoenell}, {Gomes},
  {Asari}, {Schlickmann}, {Mateus}, {Stasinska}, {Sodr{\'e}}, {Torres-Papaqui},
  \& {Seagal Collaboration}}]{cidfernandes:2009}
{Cid Fernandes}, R., {et~al.} 2009, in Revista Mexicana de Astronomia y
  Astrofisica Conference Series, Vol.~35, Revista Mexicana de Astronomia y
  Astrofisica Conference Series, 127--132

\bibitem[{{Cimatti} {et~al.}(2006){Cimatti}, {Daddi}, \&
  {Renzini}}]{cimatti:2006}
{Cimatti}, A., {Daddi}, E., \& {Renzini}, A. 2006, \aap, 453, L29

\bibitem[{{Cimatti} {et~al.}(2004){Cimatti}, {Daddi}, {Renzini}, {Cassata},
  {Vanzella}, {Pozzetti}, {Cristiani}, {Fontana}, {Rodighiero}, {Mignoli}, \&
  {Zamorani}}]{cimatti:2004}
{Cimatti}, A., {et~al.} 2004, \nat, 430, 184

\bibitem[{{Cimatti} {et~al.}(2008){Cimatti}, {Cassata}, {Pozzetti}, {Kurk},
  {Mignoli}, {Renzini}, {Daddi}, {Bolzonella}, {Brusa}, {Rodighiero},
  {Dickinson}, {Franceschini}, {Zamorani}, {Berta}, {Rosati}, \&
  {Halliday}}]{cimatti:2008}
---. 2008, \aap, 482, 21

\bibitem[{{Cooper} {et~al.}(2012){Cooper}, {Griffith}, {Newman}, {Coil},
  {Davis}, {Dutton}, {Faber}, {Guhathakurta}, {Koo}, {Lotz}, {Weiner},
  {Willmer}, \& {Yan}}]{cooper:2012}
{Cooper}, M.~C., {et~al.} 2012, \mnras, 419, 3018

\bibitem[{{Daddi} {et~al.}(2004){Daddi}, {Cimatti}, {Renzini}, {Fontana},
  {Mignoli}, {Pozzetti}, {Tozzi}, \& {Zamorani}}]{daddi:2004:bzk}
{Daddi}, E., {Cimatti}, A., {Renzini}, A., {Fontana}, A., {Mignoli}, M.,
  {Pozzetti}, L., {Tozzi}, P., \& {Zamorani}, G. 2004, \apj, 617, 746

\bibitem[{{Daddi} {et~al.}(2005){Daddi}, {Renzini}, {Pirzkal}, {Cimatti},
  {Malhotra}, {Stiavelli}, {Xu}, {Pasquali}, {Rhoads}, {Brusa}, {di Serego
  Alighieri}, {Ferguson}, {Koekemoer}, {Moustakas}, {Panagia}, \&
  {Windhorst}}]{daddi:2005:pbzk}
{Daddi}, E., {et~al.} 2005, \apj, 626, 680

\bibitem[{{Daddi} {et~al.}(2007){Daddi}, {Dickinson}, {Morrison}, {Chary},
  {Cimatti}, {Elbaz}, {Frayer}, {Renzini}, {Pope}, {Alexander}, {Bauer},
  {Giavalisco}, {Huynh}, {Kurk}, \& {Mignoli}}]{daddi:2007:sfr}
---. 2007, \apj, 670, 156

\bibitem[{{Damjanov} {et~al.}(2011){Damjanov}, {Abraham}, {Glazebrook},
  {McCarthy}, {Caris}, {Carlberg}, {Chen}, {Crampton}, {Green}, {J{\o}rgensen},
  {Juneau}, {Le Borgne}, {Marzke}, {Mentuch}, {Murowinski}, {Roth}, {Savaglio},
  \& {Yan}}]{damjanov:2011}
{Damjanov}, I., {et~al.} 2011, \apjl, 739, L44

\bibitem[{{Falc{\'o}n-Barroso} {et~al.}(2011){Falc{\'o}n-Barroso},
  {S{\'a}nchez-Bl{\'a}zquez}, {Vazdekis}, {Ricciardelli}, {Cardiel}, {Cenarro},
  {Gorgas}, \& {Peletier}}]{falconbarroso:2011}
{Falc{\'o}n-Barroso}, J., {S{\'a}nchez-Bl{\'a}zquez}, P., {Vazdekis}, A.,
  {Ricciardelli}, E., {Cardiel}, N., {Cenarro}, A.~J., {Gorgas}, J., \&
  {Peletier}, R.~F. 2011, \aap, 532, A95

\bibitem[{{Finoguenov} {et~al.}(2007){Finoguenov}, {Guzzo}, {Hasinger},
  {Scoville}, {Aussel}, {B{\"o}hringer}, {Brusa}, {Capak}, {Cappelluti},
  {Comastri}, {Giodini}, {Griffiths}, {Impey}, {Koekemoer}, {Kneib},
  {Leauthaud}, {Le F{\`e}vre}, {Lilly}, {Mainieri}, {Massey}, {McCracken},
  {Mobasher}, {Murayama}, {Peacock}, {Sakelliou}, {Schinnerer}, {Silverman},
  {Smol{\v c}i{\'c}}, {Taniguchi}, {Tasca}, {Taylor}, {Trump}, \&
  {Zamorani}}]{finoguenov:2007}
{Finoguenov}, A., {et~al.} 2007, \apjs, 172, 182

\bibitem[{{F{\"o}rster Schreiber} {et~al.}(2011){F{\"o}rster Schreiber},
  {Shapley}, {Genzel}, {Bouch{\'e}}, {Cresci}, {Davies}, {Erb}, {Genel},
  {Lutz}, {Newman}, {Shapiro}, {Steidel}, {Sternberg}, \&
  {Tacconi}}]{forsterschreiber:2011:clump2}
{F{\"o}rster Schreiber}, N.~M., {et~al.} 2011, \apj, 739, 45

\bibitem[{{Genzel} {et~al.}(2011){Genzel}, {Newman}, {Jones}, {F{\"o}rster
  Schreiber}, {Shapiro}, {Genel}, {Lilly}, {Renzini}, {Tacconi}, {Bouch{\'e}},
  {Burkert}, {Cresci}, {Buschkamp}, {Carollo}, {Ceverino}, {Davies}, {Dekel},
  {Eisenhauer}, {Hicks}, {Kurk}, {Lutz}, {Mancini}, {Naab}, {Peng},
  {Sternberg}, {Vergani}, \& {Zamorani}}]{genzel:2011}
{Genzel}, R., {et~al.} 2011, \apj, 733, 101

\bibitem[{{Graves} \& {Schiavon}(2008)}]{graves:2008}
{Graves}, G.~J., \& {Schiavon}, R.~P. 2008, \apjs, 177, 446

\bibitem[{{Hasinger} {et~al.}(2007){Hasinger}, {Cappelluti}, {Brunner},
  {Brusa}, {Comastri}, {Elvis}, {Finoguenov}, {Fiore}, {Franceschini}, {Gilli},
  {Griffiths}, {Lehmann}, {Mainieri}, {Matt}, {Matute}, {Miyaji}, {Molendi},
  {Paltani}, {Sanders}, {Scoville}, {Tresse}, {Urry}, {Vettolani}, \&
  {Zamorani}}]{hasinger:2007}
{Hasinger}, G., {et~al.} 2007, \apjs, 172, 29

\bibitem[{{Hoaglin} {et~al.}(1983){Hoaglin}, {Mosteller}, \&
  {Tukey}}]{hoaglin:1983}
{Hoaglin}, D.~C., {Mosteller}, F., \& {Tukey}, J.~W. 1983, {Understanding
  robust and exploratory data anlysis}, ed. {Hoaglin, D.~C., Mosteller, F., \&
  Tukey, J.~W.}

\bibitem[{{Hopkins} {et~al.}(2009){Hopkins}, {Bundy}, {Murray}, {Quataert},
  {Lauer}, \& {Ma}}]{hopkins:2009}
{Hopkins}, P.~F., {Bundy}, K., {Murray}, N., {Quataert}, E., {Lauer}, T.~R., \&
  {Ma}, C.-P. 2009, \mnras, 398, 898

\bibitem[{{Ichikawa} {et~al.}(2006){Ichikawa}, {Suzuki}, {Tokoku}, {Uchimoto},
  {Konishi}, {Yoshikawa}, {Yamada}, {Tanaka}, {Omata}, \&
  {Nishimura}}]{ichikawa:2006:moircs}
{Ichikawa}, T., {et~al.} 2006, in Society of Photo-Optical Instrumentation
  Engineers (SPIE) Conference Series, Vol. 6269, Society of Photo-Optical
  Instrumentation Engineers (SPIE) Conference Series

\bibitem[{{Ilbert} {et~al.}(2009){Ilbert}, {Capak}, {Salvato}, {Aussel},
  {McCracken}, {Sanders}, {Scoville}, {Kartaltepe}, {Arnouts}, {Le Floc'h},
  {Mobasher}, {Taniguchi}, {Lamareille}, {Leauthaud}, {Sasaki}, {Thompson},
  {Zamojski}, {Zamorani}, {Bardelli}, {Bolzonella}, {Bongiorno}, {Brusa},
  {Caputi}, {Carollo}, {Contini}, {Cook}, {Coppa}, {Cucciati}, {de la Torre},
  {de Ravel}, {Franzetti}, {Garilli}, {Hasinger}, {Iovino}, {Kampczyk},
  {Kneib}, {Knobel}, {Kovac}, {Le Borgne}, {Le Brun}, {F{\`e}vre}, {Lilly},
  {Looper}, {Maier}, {Mainieri}, {Mellier}, {Mignoli}, {Murayama}, {Pell{\`o}},
  {Peng}, {P{\'e}rez-Montero}, {Renzini}, {Ricciardelli}, {Schiminovich},
  {Scodeggio}, {Shioya}, {Silverman}, {Surace}, {Tanaka}, {Tasca}, {Tresse},
  {Vergani}, \& {Zucca}}]{ilbert:2009}
{Ilbert}, O., {et~al.} 2009, \apj, 690, 1236

\bibitem[{{Ilbert} {et~al.}(2010){Ilbert}, {Salvato}, {Le Floc'h}, {Aussel},
  {Capak}, {McCracken}, {Mobasher}, {Kartaltepe}, {Scoville}, {Sanders},
  {Arnouts}, {Bundy}, {Cassata}, {Kneib}, {Koekemoer}, {Le F{\`e}vre}, {Lilly},
  {Surace}, {Taniguchi}, {Tasca}, {Thompson}, {Tresse}, {Zamojski}, {Zamorani},
  \& {Zucca}}]{ilbert:2010}
---. 2010, \apj, 709, 644

\bibitem[{{Kauffmann} {et~al.}(2003){Kauffmann}, {Heckman}, {White}, {Charlot},
  {Tremonti}, {Brinchmann}, {Bruzual}, {Peng}, {Seibert}, {Bernardi},
  {Blanton}, {Brinkmann}, {Castander}, {Cs{\'a}bai}, {Fukugita}, {Ivezic},
  {Munn}, {Nichol}, {Padmanabhan}, {Thakar}, {Weinberg}, \&
  {York}}]{kauffmann:2003}
{Kauffmann}, G., {et~al.} 2003, \mnras, 341, 33

\bibitem[{{Koekemoer} {et~al.}(2002){Koekemoer}, {Fruchter}, {Hook}, \&
  {Hack}}]{koekemoer:2002}
{Koekemoer}, A.~M., {Fruchter}, A.~S., {Hook}, R.~N., \& {Hack}, W. 2002, in
  The 2002 HST Calibration Workshop : Hubble after the Installation of the ACS
  and the NICMOS Cooling System, ed. S.~Arribas, A.~Koekemoer, \& B.~Whitmore
  (Baltimore, MD: STScI), 337, ed. S.~{Arribas}, A.~{Koekemoer}, \&
  B.~{Whitmore}, 337

\bibitem[{{Koekemoer} {et~al.}(2007){Koekemoer}, {Aussel}, {Calzetti}, {Capak},
  {Giavalisco}, {Kneib}, {Leauthaud}, {Le F{\`e}vre}, {McCracken}, {Massey},
  {Mobasher}, {Rhodes}, {Scoville}, \& {Shopbell}}]{koekemoer:2007}
{Koekemoer}, A.~M., {et~al.} 2007, \apjs, 172, 196

\bibitem[{{Kong} {et~al.}(2006){Kong}, {Daddi}, {Arimoto}, {Renzini},
  {Broadhurst}, {Cimatti}, {Ikuta}, {Ohta}, {da Costa}, {Olsen}, {Onodera}, \&
  {Tamura}}]{kong:2006}
{Kong}, X., {et~al.} 2006, \apj, 638, 72

\bibitem[{{Kormendy} {et~al.}(2009){Kormendy}, {Fisher}, {Cornell}, \&
  {Bender}}]{kormendy:2009}
{Kormendy}, J., {Fisher}, D.~B., {Cornell}, M.~E., \& {Bender}, R. 2009, \apjs,
  182, 216

\bibitem[{{Kriek} {et~al.}(2009){Kriek}, {van Dokkum}, {Labb{\'e}}, {Franx},
  {Illingworth}, {Marchesini}, \& {Quadri}}]{kriek:2009}
{Kriek}, M., {van Dokkum}, P.~G., {Labb{\'e}}, I., {Franx}, M., {Illingworth},
  G.~D., {Marchesini}, D., \& {Quadri}, R.~F. 2009, \apj, 700, 221

\bibitem[{{Kriek} {et~al.}(2006){Kriek}, {van Dokkum}, {Franx}, {Quadri},
  {Gawiser}, {Herrera}, {Illingworth}, {Labb{\'e}}, {Lira}, {Marchesini},
  {Rix}, {Rudnick}, {Taylor}, {Toft}, {Urry}, \& {Wuyts}}]{kriek:2006:passive}
{Kriek}, M., {et~al.} 2006, \apjl, 649, L71

\bibitem[{{Kriek} {et~al.}(2008){Kriek}, {van Dokkum}, {Franx}, {Illingworth},
  {Marchesini}, {Quadri}, {Rudnick}, {Taylor}, {F{\"o}rster Schreiber},
  {Gawiser}, {Labb{\'e}}, {Lira}, \& {Wuyts}}]{kriek:2008:survey}
---. 2008, \apj, 677, 219

\bibitem[{{Kroupa}(2001)}]{kroupa:2001}
{Kroupa}, P. 2001, \mnras, 322, 231

\bibitem[{{Le Borgne} {et~al.}(2006){Le Borgne}, {Abraham}, {Daniel},
  {McCarthy}, {Glazebrook}, {Savaglio}, {Crampton}, {Juneau}, {Carlberg},
  {Chen}, {Marzke}, {Roth}, {J{\o}rgensen}, \& {Murowinski}}]{leborgne:2006}
{Le Borgne}, D., {et~al.} 2006, \apj, 642, 48

\bibitem[{{Le Floc'h} {et~al.}(2009){Le Floc'h}, {Aussel}, {Ilbert},
  {Riguccini}, {Frayer}, {Salvato}, {Arnouts}, {Surace}, {Feruglio},
  {Rodighiero}, {Capak}, {Kartaltepe}, {Heinis}, {Sheth}, {Yan}, {McCracken},
  {Thompson}, {Sanders}, {Scoville}, \& {Koekemoer}}]{lefloch:2009}
{Le Floc'h}, E., {et~al.} 2009, \apj, 703, 222

\bibitem[{{Lilly} {et~al.}(2007){Lilly}, {Le F{\`e}vre}, {Renzini}, {Zamorani},
  {Scodeggio}, {Contini}, {Carollo}, {Hasinger}, {Kneib}, {Iovino}, {Le Brun},
  {Maier}, {Mainieri}, {Mignoli}, {Silverman}, {Tasca}, {Bolzonella},
  {Bongiorno}, {Bottini}, {Capak}, {Caputi}, {Cimatti}, {Cucciati}, {Daddi},
  {Feldmann}, {Franzetti}, {Garilli}, {Guzzo}, {Ilbert}, {Kampczyk}, {Kovac},
  {Lamareille}, {Leauthaud}, {Borgne}, {McCracken}, {Marinoni}, {Pello},
  {Ricciardelli}, {Scarlata}, {Vergani}, {Sanders}, {Schinnerer}, {Scoville},
  {Taniguchi}, {Arnouts}, {Aussel}, {Bardelli}, {Brusa}, {Cappi}, {Ciliegi},
  {Finoguenov}, {Foucaud}, {Franceschini}, {Halliday}, {Impey}, {Knobel},
  {Koekemoer}, {Kurk}, {Maccagni}, {Maddox}, {Marano}, {Marconi}, {Meneux},
  {Mobasher}, {Moreau}, {Peacock}, {Porciani}, {Pozzetti}, {Scaramella},
  {Schiminovich}, {Shopbell}, {Smail}, {Thompson}, {Tresse}, {Vettolani},
  {Zanichelli}, \& {Zucca}}]{lilly:2007}
{Lilly}, S.~J., {et~al.} 2007, \apjs, 172, 70

\bibitem[{{Lilly} {et~al.}(2009){Lilly}, {Le Brun}, {Maier}, {Mainieri},
  {Mignoli}, {Scodeggio}, {Zamorani}, {Carollo}, {Contini}, {Kneib}, {Le
  F{\`e}vre}, {Renzini}, {Bardelli}, {Bolzonella}, {Bongiorno}, {Caputi},
  {Coppa}, {Cucciati}, {de la Torre}, {de Ravel}, {Franzetti}, {Garilli},
  {Iovino}, {Kampczyk}, {Kovac}, {Knobel}, {Lamareille}, {Le Borgne}, {Pello},
  {Peng}, {P{\'e}rez-Montero}, {Ricciardelli}, {Silverman}, {Tanaka}, {Tasca},
  {Tresse}, {Vergani}, {Zucca}, {Ilbert}, {Salvato}, {Oesch}, {Abbas},
  {Bottini}, {Capak}, {Cappi}, {Cassata}, {Cimatti}, {Elvis}, {Fumana},
  {Guzzo}, {Hasinger}, {Koekemoer}, {Leauthaud}, {Maccagni}, {Marinoni},
  {McCracken}, {Memeo}, {Meneux}, {Porciani}, {Pozzetti}, {Sanders},
  {Scaramella}, {Scarlata}, {Scoville}, {Shopbell}, \&
  {Taniguchi}}]{lilly:2009}
---. 2009, \apjs, 184, 218

\bibitem[{{Ma{\'{\i}}z Apell{\'a}niz}(2006)}]{maizapellaniz:2006}
{Ma{\'{\i}}z Apell{\'a}niz}, J. 2006, \aj, 131, 1184

\bibitem[{{Mancini} {et~al.}(2010){Mancini}, {Daddi}, {Renzini}, {Salmi},
  {McCracken}, {Cimatti}, {Onodera}, {Salvato}, {Koekemoer}, {Aussel},
  {Floc'h}, {Willott}, \& {Capak}}]{mancini:2010}
{Mancini}, C., {et~al.} 2010, \mnras, 401, 933

\bibitem[{{Mancini} {et~al.}(2011){Mancini}, {F{\"o}rster Schreiber},
  {Renzini}, {Cresci}, {Hicks}, {Peng}, {Vergani}, {Lilly}, {Carollo},
  {Pozzetti}, {Zamorani}, {Daddi}, {Genzel}, {Maraston}, {McCracken},
  {Tacconi}, {Bouch{\'e}}, {Davies}, {Oesch}, {Shapiro}, {Mainieri}, {Lutz},
  {Mignoli}, \& {Sternberg}}]{mancini:2011}
---. 2011, \apj, 743, 86

\bibitem[{{Maraston}(2005)}]{maraston:2005}
{Maraston}, C. 2005, \mnras, 362, 799

\bibitem[{{Maraston} {et~al.}(2010){Maraston}, {Pforr}, {Renzini}, {Daddi},
  {Dickinson}, {Cimatti}, \& {Tonini}}]{maraston:2010}
{Maraston}, C., {Pforr}, J., {Renzini}, A., {Daddi}, E., {Dickinson}, M.,
  {Cimatti}, A., \& {Tonini}, C. 2010, \mnras, 407, 830

\bibitem[{{Massey} {et~al.}(2010){Massey}, {Stoughton}, {Leauthaud}, {Rhodes},
  {Koekemoer}, {Ellis}, \& {Shaghoulian}}]{massey:2010}
{Massey}, R., {Stoughton}, C., {Leauthaud}, A., {Rhodes}, J., {Koekemoer}, A.,
  {Ellis}, R., \& {Shaghoulian}, E. 2010, \mnras, 401, 371

\bibitem[{{McCarthy} {et~al.}(2004){McCarthy}, {Le Borgne}, {Crampton}, {Chen},
  {Abraham}, {Glazebrook}, {Savaglio}, {Carlberg}, {Marzke}, {Roth},
  {J{\o}rgensen}, {Hook}, {Murowinski}, \& {Juneau}}]{mccarthy:2004}
{McCarthy}, P.~J., {et~al.} 2004, \apjl, 614, L9

\bibitem[{{McCracken} {et~al.}(2010){McCracken}, {Capak}, {Salvato}, {Aussel},
  {Thompson}, {Daddi}, {Sanders}, {Kneib}, {Willott}, {Mancini}, {Renzini},
  {Cook}, {Le F{\`e}vre}, {Ilbert}, {Kartaltepe}, {Koekemoer}, {Mellier},
  {Murayama}, {Scoville}, {Shioya}, \& {Tanaguchi}}]{mccracken:2010}
{McCracken}, H.~J., {et~al.} 2010, \apj, 708, 202

\bibitem[{{Newman} {et~al.}(2012){Newman}, {Ellis}, {Bundy}, \&
  {Treu}}]{newman:2012}
{Newman}, A.~B., {Ellis}, R.~S., {Bundy}, K., \& {Treu}, T. 2012, \apj, 746,
  162

\bibitem[{{Newman} {et~al.}(2010){Newman}, {Ellis}, {Treu}, \&
  {Bundy}}]{newman:2010}
{Newman}, A.~B., {Ellis}, R.~S., {Treu}, T., \& {Bundy}, K. 2010, \apjl, 717,
  L103

\bibitem[{{Oke} \& {Gunn}(1983)}]{oke:1983}
{Oke}, J.~B., \& {Gunn}, J.~E. 1983, \apj, 266, 713

\bibitem[{{Onodera} {et~al.}(2010){Onodera}, {Daddi}, {Gobat}, {Cappellari},
  {Arimoto}, {Renzini}, {Yamada}, {McCracken}, {Mancini}, {Capak}, {Carollo},
  {Cimatti}, {Giavalisco}, {Ilbert}, {Kong}, {Lilly}, {Motohara}, {Ohta},
  {Sanders}, {Scoville}, {Tamura}, \& {Taniguchi}}]{onodera:2010:pbzk}
{Onodera}, M., {et~al.} 2010, \apjl, 715, L6

\bibitem[{{Pannella} {et~al.}(2009){Pannella}, {Carilli}, {Daddi}, {McCracken},
  {Owen}, {Renzini}, {Strazzullo}, {Civano}, {Koekemoer}, {Schinnerer},
  {Scoville}, {Smol{\v c}i{\'c}}, {Taniguchi}, {Aussel}, {Kneib}, {Ilbert},
  {Mellier}, {Salvato}, {Thompson}, \& {Willott}}]{pannella:2009}
{Pannella}, M., {et~al.} 2009, \apjl, 698, L116

\bibitem[{{Papovich} {et~al.}(2011){Papovich}, {Finkelstein}, {Ferguson},
  {Lotz}, \& {Giavalisco}}]{papovich:2011}
{Papovich}, C., {Finkelstein}, S.~L., {Ferguson}, H.~C., {Lotz}, J.~M., \&
  {Giavalisco}, M. 2011, \mnras, 412, 1123

\bibitem[{{Peng} {et~al.}(2002){Peng}, {Ho}, {Impey}, \&
  {Rix}}]{peng:2002:galfit}
{Peng}, C.~Y., {Ho}, L.~C., {Impey}, C.~D., \& {Rix}, H.-W. 2002, \aj, 124, 266

\bibitem[{{Peng} {et~al.}(2010{\natexlab{a}}){Peng}, {Ho}, {Impey}, \&
  {Rix}}]{peng:2010:galfit}
---. 2010{\natexlab{a}}, \aj, 139, 2097

\bibitem[{{Peng} {et~al.}(2011){Peng}, {Lilly}, {Renzini}, \&
  {Carollo}}]{peng:2011}
{Peng}, Y., {Lilly}, S.~J., {Renzini}, A., \& {Carollo}, M. 2011, ArXiv
  e-prints

\bibitem[{{Peng} {et~al.}(2010{\natexlab{b}}){Peng}, {Lilly}, {Kova{\v c}},
  {Bolzonella}, {Pozzetti}, {Renzini}, {Zamorani}, {Ilbert}, {Knobel},
  {Iovino}, {Maier}, {Cucciati}, {Tasca}, {Carollo}, {Silverman}, {Kampczyk},
  {de Ravel}, {Sanders}, {Scoville}, {Contini}, {Mainieri}, {Scodeggio},
  {Kneib}, {Le F{\`e}vre}, {Bardelli}, {Bongiorno}, {Caputi}, {Coppa}, {de la
  Torre}, {Franzetti}, {Garilli}, {Lamareille}, {Le Borgne}, {Le Brun},
  {Mignoli}, {Perez Montero}, {Pello}, {Ricciardelli}, {Tanaka}, {Tresse},
  {Vergani}, {Welikala}, {Zucca}, {Oesch}, {Abbas}, {Barnes}, {Bordoloi},
  {Bottini}, {Cappi}, {Cassata}, {Cimatti}, {Fumana}, {Hasinger}, {Koekemoer},
  {Leauthaud}, {Maccagni}, {Marinoni}, {McCracken}, {Memeo}, {Meneux}, {Nair},
  {Porciani}, {Presotto}, \& {Scaramella}}]{peng:2010}
{Peng}, Y.-j., {et~al.} 2010{\natexlab{b}}, \apj, 721, 193

\bibitem[{{Pickles}(1998)}]{pickles:1998}
{Pickles}, A.~J. 1998, \pasp, 110, 863

\bibitem[{{Pozzetti} {et~al.}(2010){Pozzetti}, {Bolzonella}, {Zucca},
  {Zamorani}, {Lilly}, {Renzini}, {Moresco}, {Mignoli}, {Cassata}, {Tasca},
  {Lamareille}, {Maier}, {Meneux}, {Halliday}, {Oesch}, {Vergani}, {Caputi},
  {Kova{\v c}}, {Cimatti}, {Cucciati}, {Iovino}, {Peng}, {Carollo}, {Contini},
  {Kneib}, {Le F{\'e}vre}, {Mainieri}, {Scodeggio}, {Bardelli}, {Bongiorno},
  {Coppa}, {de la Torre}, {de Ravel}, {Franzetti}, {Garilli}, {Kampczyk},
  {Knobel}, {Le Borgne}, {Le Brun}, {Pell{\`o}}, {Perez Montero},
  {Ricciardelli}, {Silverman}, {Tanaka}, {Tresse}, {Abbas}, {Bottini}, {Cappi},
  {Guzzo}, {Koekemoer}, {Leauthaud}, {Maccagni}, {Marinoni}, {McCracken},
  {Memeo}, {Porciani}, {Scaramella}, {Scarlata}, \& {Scoville}}]{pozzetti:2010}
{Pozzetti}, L., {et~al.} 2010, \aap, 523, A13

\bibitem[{{Press} {et~al.}(1992){Press}, {Teukolsky}, {Vetterling}, \&
  {Flannery}}]{press:1992}
{Press}, W.~H., {Teukolsky}, S.~A., {Vetterling}, W.~T., \& {Flannery}, B.~P.
  1992, {Numerical recipes in FORTRAN. The art of scientific computing}, ed.
  {Press, W.~H., Teukolsky, S.~A., Vetterling, W.~T., \& Flannery, B.~P. }

\bibitem[{{Renzini}(2009)}]{renzini:2009}
{Renzini}, A. 2009, \mnras, 398, L58

\bibitem[{{Rodighiero} {et~al.}(2011){Rodighiero}, {Daddi}, {Baronchelli},
  {Cimatti}, {Renzini}, {Aussel}, {Popesso}, {Lutz}, {Andreani}, {Berta},
  {Cava}, {Elbaz}, {Feltre}, {Fontana}, {F{\"o}rster Schreiber},
  {Franceschini}, {Genzel}, {Grazian}, {Gruppioni}, {Ilbert}, {Le Floch},
  {Magdis}, {Magliocchetti}, {Magnelli}, {Maiolino}, {McCracken}, {Nordon},
  {Poglitsch}, {Santini}, {Pozzi}, {Riguccini}, {Tacconi}, {Wuyts}, \&
  {Zamorani}}]{rodighiero:2011}
{Rodighiero}, G., {et~al.} 2011, \apjl, 739, L40

\bibitem[{{Rousselot} {et~al.}(2000){Rousselot}, {Lidman}, {Cuby}, {Moreels},
  \& {Monnet}}]{rousselot:2000}
{Rousselot}, P., {Lidman}, C., {Cuby}, J., {Moreels}, G., \& {Monnet}, G. 2000,
  \aap, 354, 1134

\bibitem[{{Salim} {et~al.}(2007){Salim}, {Rich}, {Charlot}, {Brinchmann},
  {Johnson}, {Schiminovich}, {Seibert}, {Mallery}, {Heckman}, {Forster},
  {Friedman}, {Martin}, {Morrissey}, {Neff}, {Small}, {Wyder}, {Bianchi},
  {Donas}, {Lee}, {Madore}, {Milliard}, {Szalay}, {Welsh}, \&
  {Yi}}]{salim:2007}
{Salim}, S., {et~al.} 2007, \apjs, 173, 267

\bibitem[{{S{\'a}nchez-Bl{\'a}zquez} {et~al.}(2006){S{\'a}nchez-Bl{\'a}zquez},
  {Peletier}, {Jim{\'e}nez-Vicente}, {Cardiel}, {Cenarro},
  {Falc{\'o}n-Barroso}, {Gorgas}, {Selam}, \&
  {Vazdekis}}]{sanchezblazquez:2006:miles}
{S{\'a}nchez-Bl{\'a}zquez}, P., {et~al.} 2006, \mnras, 371, 703

\bibitem[{{Sanders} {et~al.}(2007){Sanders}, {Salvato}, {Aussel}, {Ilbert},
  {Scoville}, {Surace}, {Frayer}, {Sheth}, {Helou}, {Brooke}, {Bhattacharya},
  {Yan}, {Kartaltepe}, {Barnes}, {Blain}, {Calzetti}, {Capak}, {Carilli},
  {Carollo}, {Comastri}, {Daddi}, {Ellis}, {Elvis}, {Fall}, {Franceschini},
  {Giavalisco}, {Hasinger}, {Impey}, {Koekemoer}, {Le F{\`e}vre}, {Lilly},
  {Liu}, {McCracken}, {Mobasher}, {Renzini}, {Rich}, {Schinnerer}, {Shopbell},
  {Taniguchi}, {Thompson}, {Urry}, \& {Williams}}]{sanders:2007}
{Sanders}, D.~B., {et~al.} 2007, \apjs, 172, 86

\bibitem[{{Saracco} {et~al.}(2009){Saracco}, {Longhetti}, \&
  {Andreon}}]{saracco:2009}
{Saracco}, P., {Longhetti}, M., \& {Andreon}, S. 2009, \mnras, 392, 718

\bibitem[{{Saracco} {et~al.}(2010){Saracco}, {Longhetti}, \&
  {Gargiulo}}]{saracco:2010}
{Saracco}, P., {Longhetti}, M., \& {Gargiulo}, A. 2010, \mnras, 408, L21

\bibitem[{{Saracco} {et~al.}(2011){Saracco}, {Longhetti}, \&
  {Gargiulo}}]{saracco:2011}
---. 2011, \mnras, 412, 2707

\bibitem[{{Scarlata} {et~al.}(2007){Scarlata}, {Carollo}, {Lilly}, {Feldmann},
  {Kampczyk}, {Renzini}, {Cimatti}, {Halliday}, {Daddi}, {Sargent},
  {Koekemoer}, {Scoville}, {Kneib}, {Leauthaud}, {Massey}, {Rhodes}, {Tasca},
  {Capak}, {McCracken}, {Mobasher}, {Taniguchi}, {Thompson}, {Ajiki}, {Aussel},
  {Murayama}, {Sanders}, {Sasaki}, {Shioya}, \& {Takahashi}}]{scarlata:2007}
{Scarlata}, C., {et~al.} 2007, \apjs, 172, 494

\bibitem[{{Schiavon}(2007)}]{schiavon:2007}
{Schiavon}, R.~P. 2007, \apjs, 171, 146

\bibitem[{{Scoville} {et~al.}(2007){Scoville}, {Aussel}, {Brusa}, {Capak},
  {Carollo}, {Elvis}, {Giavalisco}, {Guzzo}, {Hasinger}, {Impey}, {Kneib},
  {LeFevre}, {Lilly}, {Mobasher}, {Renzini}, {Rich}, {Sanders}, {Schinnerer},
  {Schminovich}, {Shopbell}, {Taniguchi}, \& {Tyson}}]{scoville:2007}
{Scoville}, N., {et~al.} 2007, \apjs, 172, 1

\bibitem[{{Suzuki} {et~al.}(2008){Suzuki}, {Tokoku}, {Ichikawa}, {Uchimoto},
  {Konishi}, {Yoshikawa}, {Tanaka}, {Yamada}, {Omata}, \&
  {Nishimura}}]{suzuki:2008:moircs}
{Suzuki}, R., {et~al.} 2008, \pasj, 60, 1347

\bibitem[{{Taylor} {et~al.}(2009){Taylor}, {Franx}, {van Dokkum}, {Bell},
  {Brammer}, {Rudnick}, {Wuyts}, {Gawiser}, {Lira}, {Urry}, \&
  {Rix}}]{taylor:2009}
{Taylor}, E.~N., {et~al.} 2009, \apj, 694, 1171

\bibitem[{{Trujillo} {et~al.}(2007){Trujillo}, {Conselice}, {Bundy}, {Cooper},
  {Eisenhardt}, \& {Ellis}}]{trujillo:2007}
{Trujillo}, I., {Conselice}, C.~J., {Bundy}, K., {Cooper}, M.~C., {Eisenhardt},
  P., \& {Ellis}, R.~S. 2007, \mnras, 382, 109

\bibitem[{{Valentinuzzi} {et~al.}(2010){Valentinuzzi}, {Poggianti}, {Saglia},
  {Arag{\'o}n-Salamanca}, {Simard}, {S{\'a}nchez-Bl{\'a}zquez}, {D'onofrio},
  {Cava}, {Couch}, {Fritz}, {Moretti}, \& {Vulcani}}]{valentinuzzi:2010}
{Valentinuzzi}, T., {et~al.} 2010, \apjl, 721, L19

\bibitem[{{van de Sande} {et~al.}(2011){van de Sande}, {Kriek}, {Franx}, {van
  Dokkum}, {Bezanson}, {Whitaker}, {Brammer}, {Labb{\'e}}, {Groot}, \&
  {Kaper}}]{vandesande:2011}
{van de Sande}, J., {et~al.} 2011, \apjl, 736, L9

\bibitem[{{van der Wel} {et~al.}(2009){van der Wel}, {Bell}, {van den Bosch},
  {Gallazzi}, \& {Rix}}]{vanderwel:2009:quench}
{van der Wel}, A., {Bell}, E.~F., {van den Bosch}, F.~C., {Gallazzi}, A., \&
  {Rix}, H.-W. 2009, \apj, 698, 1232

\bibitem[{{van Dokkum} {et~al.}(2009){van Dokkum}, {Kriek}, \&
  {Franx}}]{vandokkum:2009}
{van Dokkum}, P.~G., {Kriek}, M., \& {Franx}, M. 2009, \nat, 460, 717

\bibitem[{{van Dokkum} {et~al.}(2008){van Dokkum}, {Franx}, {Kriek}, {Holden},
  {Illingworth}, {Magee}, {Bouwens}, {Marchesini}, {Quadri}, {Rudnick},
  {Taylor}, \& {Toft}}]{vandokkum:2008}
{van Dokkum}, P.~G., {et~al.} 2008, \apjl, 677, L5

\bibitem[{{Vazdekis} {et~al.}(2010){Vazdekis}, {S{\'a}nchez-Bl{\'a}zquez},
  {Falc{\'o}n-Barroso}, {Cenarro}, {Beasley}, {Cardiel}, {Gorgas}, \&
  {Peletier}}]{vazdekis:2010}
{Vazdekis}, A., {S{\'a}nchez-Bl{\'a}zquez}, P., {Falc{\'o}n-Barroso}, J.,
  {Cenarro}, A.~J., {Beasley}, M.~A., {Cardiel}, N., {Gorgas}, J., \&
  {Peletier}, R.~F. 2010, \mnras, 404, 1639

\bibitem[{{Vergani} {et~al.}(2010){Vergani}, {Zamorani}, {Lilly}, {Lamareille},
  {Halliday}, {Scodeggio}, {Vignali}, {Ciliegi}, {Bolzonella}, {Bondi},
  {Kova{\v c}}, {Knobel}, {Zucca}, {Caputi}, {Pozzetti}, {Bardelli}, {Mignoli},
  {Iovino}, {Carollo}, {Contini}, {Kneib}, {Le F{\`e}vre}, {Mainieri},
  {Renzini}, {Bongiorno}, {Coppa}, {Cucciati}, {de la Torre}, {de Ravel},
  {Franzetti}, {Garilli}, {Kampczyk}, {Le Borgne}, {Le Brun}, {Maier}, {Pello},
  {Peng}, {Perez Montero}, {Ricciardelli}, {Silverman}, {Tanaka}, {Tasca},
  {Tresse}, {Abbas}, {Bottini}, {Cappi}, {Cassata}, {Cimatti}, {Guzzo},
  {Koekemoer}, {Leauthaud}, {Maccagni}, {Marinoni}, {McCracken}, {Memeo},
  {Meneux}, {Oesch}, {Porciani}, {Scaramella}, {Capak}, {Sanders}, {Scoville},
  \& {Taniguchi}}]{vergani:2010}
{Vergani}, D., {et~al.} 2010, \aap, 509, A42

\bibitem[{{Whitaker} {et~al.}(2012){Whitaker}, {Kriek}, {van Dokkum},
  {Bezanson}, {Brammer}, {Franx}, \& {Labb{\'e}}}]{whitaker:2012}
{Whitaker}, K.~E., {Kriek}, M., {van Dokkum}, P.~G., {Bezanson}, R., {Brammer},
  G., {Franx}, M., \& {Labb{\'e}}, I. 2012, \apj, 745, 179

\bibitem[{{Whitaker} {et~al.}(2011){Whitaker}, {Labb{\'e}}, {van Dokkum},
  {Brammer}, {Kriek}, {Marchesini}, {Quadri}, {Franx}, {Muzzin}, {Williams},
  {Bezanson}, {Illingworth}, {Lee}, {Lundgren}, {Nelson}, {Rudnick}, {Tal}, \&
  {Wake}}]{whitaker:2011}
{Whitaker}, K.~E., {et~al.} 2011, \apj, 735, 86

\bibitem[{{Williams} {et~al.}(2009){Williams}, {Quadri}, {Franx}, {van Dokkum},
  \& {Labb{\'e}}}]{williams:2009}
{Williams}, R.~J., {Quadri}, R.~F., {Franx}, M., {van Dokkum}, P., \&
  {Labb{\'e}}, I. 2009, \apj, 691, 1879

\bibitem[{{Worthey} {et~al.}(1994){Worthey}, {Faber}, {Gonzalez}, \&
  {Burstein}}]{worthey:1994}
{Worthey}, G., {Faber}, S.~M., {Gonzalez}, J.~J., \& {Burstein}, D. 1994,
  \apjs, 94, 687

\bibitem[{{Worthey} \& {Ottaviani}(1997)}]{worthey:1997}
{Worthey}, G., \& {Ottaviani}, D.~L. 1997, \apjs, 111, 377

\bibitem[{{Wuyts} {et~al.}(2007){Wuyts}, {Labb{\'e}}, {Franx}, {Rudnick}, {van
  Dokkum}, {Fazio}, {F{\"o}rster Schreiber}, {Huang}, {Moorwood}, {Rix},
  {R{\"o}ttgering}, \& {van der Werf}}]{wuyts:2007}
{Wuyts}, S., {et~al.} 2007, \apj, 655, 51

\bibitem[{{Yoshikawa} {et~al.}(2010){Yoshikawa}, {Akiyama}, {Kajisawa},
  {Alexander}, {Ohta}, {Suzuki}, {Tokoku}, {Uchimoto}, {Konishi}, {Yamada},
  {Tanaka}, {Omata}, {Nishimura}, {Koekemoer}, {Brandt}, \&
  {Ichikawa}}]{yoshikawa:2010}
{Yoshikawa}, T., {et~al.} 2010, \apj, 718, 112

\end{thebibliography}

\end{document}